\documentclass[aps, 12pt, preprint, preprintnumbers,  
aps, amsmath,amssymb,nofootinbib,superscriptaddress,hyperref
]{revtex4-1} 
\usepackage{amssymb}
\usepackage{slashed}
\usepackage{amsmath}
\usepackage{braket}
\usepackage{float}
\usepackage[utf8]{inputenc}
\usepackage{graphicx}
\usepackage{subfigure}
\usepackage{latexsym}
\usepackage{epic}
\usepackage{eepic}
\usepackage{epsfig}
\usepackage{color}
\usepackage{natbib}
\usepackage[colorlinks=true,citecolor=darkred,urlcolor=darkred, pdfborder={0 0 0}]{hyperref}
\usepackage[normalem]{ulem}

\definecolor{darkred}{rgb}{0.6,0,0}

\newcommand{\AddrAHEP}{
	AHEP Group, Institut de F\'{i}sica Corpuscular,
	CSIC/Universitat de Val\`{e}ncia, Parc Cient\'ific de Paterna.\\
	C/ Catedr\'atico Jos\'e Beltr\'an, 2 E-46980 Paterna (Valencia), Spain}
\newcommand{\AddrHBNI}{
	Homi Bhabha National Institute, BARC Training School Complex, Anushakti Nagar, Mumbai 400094, India }
\newcommand{\AddrIOP}{
	Institute of Physics, Sachivalaya Marg, Bhubaneswar 751005, India}

\definecolor{linkcolor}{rgb}{0,0,0.5}

\begin{document}
 \title{\boldmath Interference effect in lepton number violating and conserving meson decays for a left-right symmetric model}
 
\author{Rohini M. Godbole}\email{rohini@iisc.ac.in}
\affiliation{Centre for High Energy Physics, Indian Institute of Science, Bengaluru - 560012, India}

\author{Siddharth P. Maharathy}\email{siddharth.m@iopb.res.in}
\affiliation{\AddrIOP}
\affiliation{\AddrHBNI}

\author{Sanjoy Mandal}\email{smandal@ific.uv.es}
\affiliation{\AddrAHEP}

\author{Manimala Mitra}\email{manimala@iopb.res.in}
\affiliation{\AddrIOP}
\affiliation{\AddrHBNI}

\author{Nita Sinha}\email{nita@imsc.res.in}
\affiliation{\AddrHBNI}
\affiliation{The Institute of Mathematical Sciences, C.I.T Campus, Taramani, Chennai 600 113, India}

\bibliographystyle{unsrt} 
\begin{abstract}
We study the effect of interference on the lepton number violating~(LNV) and lepton number conserving~(LNC) three-body meson decays $M_1^{+}\to l_i^{+} l_j^{\pm}\pi^{\mp}$,  that arise in a TeV scale Left Right Symmetric model~(LRSM) with degenerate or nearly degenerate right handed~(RH) neutrinos. LRSM contains three RH neutrinos and a RH gauge boson.
The RH neutrinos with masses in the  range of $M_N \sim$ (MeV -  few GeV)  can give resonant enhancement in the semi-leptonic LNV and LNC meson decays. In the case,  where only one RH neutrino contributes to these  decays, the predicted  new physics branching ratio of semi-leptonic LNV and LNC meson decays  $M_1^{+}\to l_i^{+} l_j^{+}\pi^{-}$ and $M_1^{+}\to l_i^{+} l_j^{-}\pi^{+}$ are equal. We find that with at least  two  RH neutrinos contributing to the process, the LNV and LNC decay rates can differ.
Depending on the neutrino mixing angles and $CP$ violating phases, the branching ratios of LNV and LNC decay channels mediated by the heavy neutrinos can  be either enhanced or suppressed, and the ratio of these two rates can differ from unity.
\end{abstract}
\maketitle
\section{Introduction} \label{Introduction}
\hspace{0.5cm} The observation of light neutrino masses and mixings clearly indicates the existence of beyond standard model~(BSM) physics. A number of models exist in the literature, that have been successful in explaining small neutrino masses and their mixings. One of the most interesting model among them is the LRSM~\cite{Pati:1974yy}, which not only  explains the small light neutrino masses, but also addresses parity violation in the Standard Model~(SM). The model contains three RH neutrinos, and two Higgs triplet fields,  that   generate the Majorana mass terms for light neutrinos via Type-I and Type-II seesaw mechanism. The RH neutrinos in this model are Majorana in nature.  The Majorana masses  violate lepton number and hence these neutrinos can directly induce LNV processes. The Majorana nature of the light and heavy neutrinos, can be tested via the lepton number violating neutrinoless double beta decay~($0\nu\beta\beta$)~\cite{Agostini:2018tnm,Racah:1937qq,Furry:1939qr,Schechter:1981bd,Tello:2010am,Dev:2014xea}. Their LNV nature can also be probed at the colliders through direct searches~\cite{Keung:1983uu,Chen:2013foz,Mitra:2016kov,Nemevsek:2018bbt, Nemevsek:2016enw, Mandal:2019vhf}, as well as through the rare LNV decays of mesons and tau lepton~\cite{Atre:2009rg,Aaij:2014aba,Cvetic:2010rw,Cvetic:2020lyh,Shuve:2016muy, Chun:2019nwi,Mandal:2016hpr,Mandal:2017tab,Milanes:2016rzr,Abada:2017jjx,Helo:2010cw,Abada:2019bac}. The  $\sqrt{s}=13$ TeV LHC search in the same-sign di-lepton and di-jet channel has so far ruled out  RH neutrino masses in the $M_N \sim 100$ GeV upto few TeV mass range, and  $M_{W_R} < 4.7$ TeV~\cite{Aaboud:2018spl,Sirunyan:2018pom}. The boosted RH neutrino search for LRSM also places strong constraint on the RH gauge boson mass $M_{W_R} > 4.8$ TeV \cite{Aaboud:2019wfg,Mitra:2016kov}  for the RH neutrino  and RH gauge boson mass hierarchy $\mathcal{O}(0.1)$. The nature of couplings of additional charged gauge boson with leptons has been studied  in \cite{Ruiz:2013a}. The helicity inversion effects on LNV transition rates at LHC has been studied in the context of Type-1 see-saw model in \cite{Ruiz:2020b}. For complementarity between $Z^\prime$ and $W_R$ searches in LRSM, see~\cite{Lindner:2016lpp}. While LHC searches are mostly sensitive to $M_N \sim \mathcal{O}(100)$ GeV-few TeV, and $M_{W_R} $ upto few TeV, the rare LNV and LNC semi-leptonic meson decays on the other hand are sensitive to a much smaller  RH neutrino mass range $M_N \sim$ (MeV -  few GeV), and to a much higher value of $W_R$ gauge boson mass. These searches are thus complementary to LHC  searches. 
It is well known that for very light and heavy neutrino masses,  the rates of these LNV meson and tau decays are extremely suppressed~\cite{Helo:2010cw,Mandal:2017tab}, well below the sensitivity reach of any future experiment. This changes dramatically, if there exists a heavy neutrino, in the MeV-GeV mass domain, which can be produced on-shell in the parent meson decay.  This can lead to large resonant enhancement of these processes. Various ongoing  experiments including NA62, LHCb, Belle-II are searching for the LNV meson decays. The LHCb experiment looked for the process $B^- \to \mu^- \mu^- \pi^+$, and has constrained the light neutrino-heavy neutrino mixing \cite{Aaij:2014aba}. 
Due to the order of magnitude increase in the  number of meson and tau flux, and long detector distance in experiments, such as SHiP, MATHUSLA and others under discussion, one expects to achieve better sensitivity for  LNV meson and tau decays in future. Even non-observations can set tight limits on the relevant parameter space such as on the heavy neutrino mass $M_N$ and RH gauge boson mass $M_{W_R}$~\cite{Mandal:2017tab,Hirsch:2020klk}. 

In this work, we study the three body LNV and LNC meson decays -  $M_1^{+}\to l_i^{+}l_j^{+}M^{-}_2$ and $M_1^{+}\to l_i^{+}l_j^{-}M^{+}_2$ for  LRSM, in particular focussing on  implications of  possible interference effect, that may occur due to degenerate or nearly degenerate RH neutrino states. With only a single heavy neutrino, the rate of LNV and LNC meson decays, via RH neutrino mediation are predicted to be the same, irrespective of any $CP$ violating phase present in the RH neutrino mixing matrix. This scenario changes dramatically,  if more than one heavy neutrino state contributes in these processes with non-trivial RH neutrino mixing matrix. In this case,  the predictions for LNV and LNC meson decay rates can widely differ due to the interference amongst the contributions of different RH neutrinos. This leads to a  change in  the interpretation of data in LRSM compared to the case of single heavy neutrino. For the study of interference effect in semi-leptonic meson decays for a pure sterile neutrino without any additional gauge extension, see \cite{Abada:2019bac}. The CP violation in semi-leptonic decays of charged mesons with nearly degenerate heavy neutrinos have been extensively studied in \cite{Cvetic:2013eza,Cvetic_2015naa,Cvetic_2015ura,Moreno:2016cfz} and for CP violation in rare $\tau^\pm$ decays with nearly equal heavy neutrino mass, see~\cite{Zamora-Saa:2016ito} . A comprehensive  study of  heavy neutrino oscillation in rare W- decays as well as tau decays with  degenerate heavy neutrinos has been performed in \cite{Cvetic:2018elt,Cvetic:2019rms,Tapia:2019coy}  .The interference effect in LRSM, relevant for collider searches has been discussed in \cite{Gluza:2015goa}. The interference effects have also been studied in type-I and generalized inverse seesaw models in~\cite{Das:2017hmg}. The enhancement of CP-asymmetry at LHC has been studied  for heavy degenerate neutrinos in \cite{pilaftsis:2007sj} and for degenerate scalars in~\cite{pilaftsis:199761}.

To quantify the  interference effect in meson decays, we consider  $K$ and $B$-meson LNV and LNC semileptonic three body decays with a pion in the final state as  illustrative examples. We develop the generic theory framework with two degenerate or nearly degenerate RH neutrino states that contribute significantly in the LNV and LNC semileptonic three body meson decays. Using this we then evaluate the analytic results for the partial decay widths, and branching ratios in the presence of interference terms in the amplitude. 
We  focus on the final states,  that contain electrons and muons. 
We consider  two different mass ranges of the two RH neutrino states $0.14\,\text{GeV}< M_N< 0.49$~GeV and  $0.14\,\text{GeV}< M_N< 5$~GeV,  relevant for    $K^{+} \to e^{+}e^{+} \pi^{-}/e^{+}\mu^{\pm}\pi^{\mp} $ and  $B^{+} \to e^{+}e^{+} \pi^{-}/e^{+}\mu^{\pm}\pi^{\mp} $ meson decays, respectively. These decay modes including the  flavor violating LNC modes: $K^{+} \to e^{+}\mu^{-} \pi^{+}$ and  $B^{+} \to e^{+}\mu^{-} \pi^{+} $ are absent in the SM, hence serve as a clear indication of new physics. 

The paper is organized as follows: In Sec.~\ref{model}, we first review the basic features of LRSM, following which in Sec.~\ref{LNC and LNV meson Decays} we discuss in detail the RH neutrino contributions in LNV and LNC meson decays. In Sec.~\ref{Two Generation Scenario}, we discuss our results with the assumption, that  only two RH neutrinos are contributing with an effective $2\times 2$ RH neutrino mixing matrix. In Sec.~\ref{Three generation case}, we discuss the interference effects with two RH neutrinos considering the full $3\times 3$ RH neutrino mixing matrix. Finally we summarise our most important findings in the conclusion.
\section{Left-Right Symmetric Model \label{model}}
LRSM is a simple extension of the Standard Model~(SM), where both the left and right chiral fermions are treated on an equal footing. The model is based on the gauge group $SU(3)_c\otimes SU(2)_L \otimes SU(2)_R \otimes U(1)_{B-L} $, where the left and right chiral fermions are doublets of $SU(2)_L$ and $SU(2)_R$, respectively. The model necessarily contains three RH neutrinos ($N_{Ri}$), which are part of the three  right-handed lepton doublets. The electric charge generator $Q$ is related with third component of weak isospins $I_{3L}$ and $I_{3R}$ as $Q=I_{3L}+I_{3R}+(B-L)/2$. The scalar sector is also enlarged because of the extra symmetry. The LRSM contains one  bi-doublet $\Phi$ and two scalar triplets $\Delta_R$ and $\Delta_L$ under  $SU(2)_R$ and $SU(2)_L$, respectively. The particle content is given below:
\begin{equation}
	l_L=\begin{bmatrix} \nu_{Li}\\e_{Li}\end{bmatrix} ,\, l_R=\begin{bmatrix} N_{Ri}\\e_{Ri}\end{bmatrix}
\end{equation}
\begin{equation}
	Q_L=\begin{bmatrix} u_{Li}\\d_{Li}\end{bmatrix} ,\, Q_R=\begin{bmatrix} u_{Ri}\\d_{Ri}\end{bmatrix}
\end{equation}
\begin{equation}
	\Phi=\begin{bmatrix} \phi_1^0 &\phi_2^+ \\ \phi_1^- &\phi_2^0\end{bmatrix} ,\,\Delta_{L/R}=\begin{bmatrix} \Delta_{L/R}^+/\sqrt{2} &\Delta_{L/R}^{++} \\ \Delta_{L/R}^0 & -\Delta_{L/R}^+/\sqrt{2} \end{bmatrix}
\end{equation}
The $SU(2)$ doublets $l_L$ and $l_R$ have the charges $(1,2,1,-1)$ and $(1,1,2,-1)$, while the Higgs multiplets have the charges $\Phi \sim (1,2,2,0)$, $\Delta_L \sim (1,3,1,+2)$ and $\Delta_R \sim (1,1,3,+2)$. 
The bi-doublet being neutral under $B-L$, additional Higgs triplets are required to break the left-right symmetric gauge group to the SM gauge group $SU(2)_L \otimes U(1)_Y$.
The neutral component of $\Delta_R$ takes vacuum expectation value~(VEV) $v_R$ and breaks the gauge group $SU(2)_R \otimes U(1)_{B-L}$ to $U(1)_Y$. In the next step, VEV of bi-doublet $\Phi$ breaks the SM gauge group to $U(1)_{Q}$. The VEV of the bi-doublet is denoted as: 
$ \braket{\Phi}$ =  $\text{Diag}\big(\frac{\kappa_1}{\sqrt{2}},\frac{\kappa_2}{\sqrt{2}}\big)$.  Due to the strong constraint on $\rho$ parameter, the VEV of $\Delta_L$ should be very small $v_L < 5 $ GeV \cite{Gunion:1989ci}. Hence, the different VEVs of $SU(2)$ triplets and bi-doublet follow the hierarchy $v_L\ll \kappa_{1,2}\ll v_R$.
The Yukawa Lagrangian responsible for generating the lepton masses has the following form
\begin{eqnarray}
	-\mathcal{L}_Y &=& y\bar{l}_L \Phi {l}_R + \Tilde{y}\bar{l}_L \Tilde{\Phi} {l}_R + y_L l^T_LC^{-1}i \sigma^2  \Delta_L  l_L 
	\nonumber  \\ && + y_R l^T_R C^{-1}i\sigma^2 \Delta_R l_R + \text{H.C.}, 
	\label{eq:laglrsmyu}
\end{eqnarray}
where $C$ is the charge-conjugation operator, $C=i\gamma^2 \gamma^0$  and $\Tilde{\Phi}=\sigma^2\Phi^*\sigma^2$. Here $\gamma^\mu$ and $\sigma^i$ are the Dirac and Pauli matrices, and  $y,\,\Tilde{y}$, $y_L$ and $y_R$ are the Yukawa couplings, respectively. After spontaneous symmetry breaking the neutral lepton   mass matrix is obtained as, 
\begin{equation}
	\mathcal{M}_{\nu}=
	\begin{bmatrix}
		M_L & M_{D}\\M_{D}^T & M_{R}
	\end{bmatrix}
	\label{eqn:massmatrix}
\end{equation}
In the above, the Dirac mass matrix $M_D = \frac{1}{\sqrt{2}}(y \kappa_1 + \Tilde{y} \kappa_2)
=y_D \kappa_s $, and $M_{L,R}$ are given by  $M_L=\sqrt{2}v_L y_L$ and $M_R=\sqrt{2}v_R y_R$. The Higgs triplets $\Delta_R$ and $\Delta_L$ generate  Majorana masses of heavy and light neutrinos, respectively. The parameter $\kappa_s $ is the Electroweak VEV, and is related to $\kappa_{1,2}$ as  $\kappa_s=\sqrt{\kappa_1^2 +\kappa_2^2}$. The light and heavy neutrino masses can be calculated by using the seesaw approximation $M_L\ll M_D\ll M_R$. This leads to the following light and heavy neutrino mass matrix,
\begin{eqnarray}
	M_\nu & \sim & M_L - M_D M_R^{-1}M_D^T+\mathcal{O}(M_R^{-2}) \\ \nonumber
	&& \sim \sqrt 2 v_L y_L - \frac{\kappa^2_s}{\sqrt 2 v_R} y_D y_R^{-1} y_D^{\sf T}
\end{eqnarray}
and
\begin{eqnarray}
	M_N & \sim  & M_R + \mathcal{O}(M_R^{-1}) \nonumber \\  &&   \sim \sqrt{2}  y_R v_R
\end{eqnarray}
The  mass matrix $\mathcal{M}_{\nu}$  in Eq.~\ref{eqn:massmatrix} can be diagonalized by a unitary transformation,
\begin{eqnarray}
	\mathcal{V}^T \begin{bmatrix}
		M_L & M_{D}\\M_{D}^T & M_{R}
	\end{bmatrix} \mathcal{V} = \begin{bmatrix}
		\widetilde{\mathcal{M}}_{{\nu}}  
		& 0\\0 & \widetilde{\mathcal{M}}_{{R}}
	\end{bmatrix}
\end{eqnarray} 
where $\widetilde{\mathcal{M}}_{{\nu}}=\text{Diag}(m_1,m_2,m_3)$ and $\widetilde{\mathcal{M}}_{{R}}=\text{Diag}(M_1,M_2,M_3)$.
Up to $\mathcal{O}(M_R^{-2})$, the mixing matrix $\mathcal{V}$ has the following form
\begin{align}
	\mathcal{V}=\begin{bmatrix}
		U_{\text{PMNS}} & S^{\dagger} \\
		T & K_R^{\dagger}
	\end{bmatrix},
\end{align}
where $S,\,T\approx M_D M_R^{-1}$. In the above, $K_R$ is the diagonalization matrix for the RH neutrino mass matrix $M_R$. We will neglect the effect of $S,\,T$ in our subsequent discussions as, $S,T\sim \mathcal{O}(10^{-5})$ for $M_\nu\sim\mathcal{O}(0.1)\,\text{eV}$ and $M_N\sim\mathcal{O}(1\,\text{GeV})$.

\subsection{Gauge Sector}
In addition to the SM gauge bosons $W_L$ and $Z$, this model also has RH gauge boson $W_R$ and an additional  neutral gauge boson $Z^{\prime}$. The left and right handed charged gauge bosons ($W_L , W_R$) will mix and the mixing angle can be approximated to be
\begin{equation}
	\zeta \simeq \frac{\kappa_1 \kappa_2}{v_R^2} \simeq 2\frac{\kappa_2}{\kappa_1}\Big(\frac{M_{W_L}}{M_{W_R}}\Big)^2
\end{equation}
Due to this small mixing between the charged gauge bosons, the masses of the gauge bosons can be approximated as
\begin{equation}
	M_{W_L}\simeq M_{W_1}\simeq \frac{g \kappa_1}{\sqrt{2}} ,   \hspace{0.5cm}      M_{W_R}\simeq M_{W_2}\simeq g v_R 
\end{equation}
Note that, throughout our calculation, we assume $g\equiv g_L = g_R$, which is justified, as  we consider parity as a symmetry in LRSM. The mass of neutral gauge boson $Z^\prime$ for this choice becomes  $M_{Z^{\prime}}\sim 1.7 M_{W_R}$.

\subsection{Charged and neutral current Lagrangian}
\begin{figure}[b]
	\centering
	\includegraphics[height=3.5cm,width=0.45\textwidth]{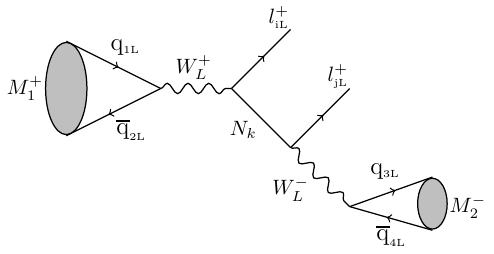}
	\includegraphics[height=3.5cm,width=0.45\textwidth]{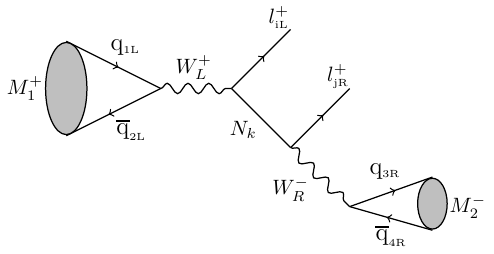}
	\includegraphics[height=3.5cm,width=0.45\textwidth]{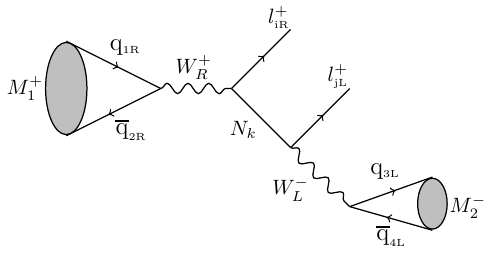}
	\includegraphics[height=3.5cm,width=0.45\textwidth]{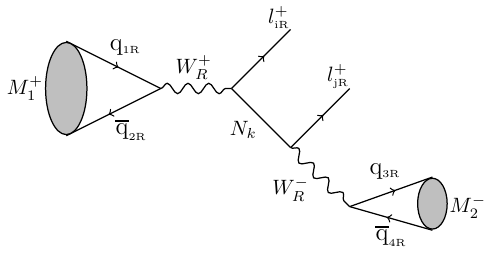}
	\caption{\footnotesize{The Feynman diagrams for LNV meson decays.  See text for details.} }
	\label{LNV process}
\end{figure}

\begin{figure}[h]
	\centering
	\includegraphics[height=3.5cm,width=0.45\textwidth]{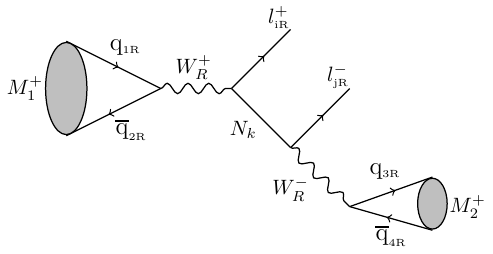}
	\caption{\footnotesize{The Feynman diagram for LNC meson decay, mediated  via  RH-neutrino.}}
	\label{LNC process}
\end{figure}

\hspace{0.5cm} The charged current Lagrangian for the quark sector has the following form
\begin{eqnarray}
	\mathcal{L}^q_{cc} &= & \frac{g}{\sqrt{2}}\sum_{i,j}\bar{u}_i V^{\text{CKM}}_{ij}W_{L\mu}^{+}\gamma^{\mu}P_L d_j \\ \nonumber 
	&& +\frac{g}{\sqrt{2}}\sum_{i,j}\bar{u}_i V^{\text{R-CKM}}_{ij}W_{R\mu}^{+}\gamma^\mu P_R d_j+\text{H.c.},
	\label{eq:ccinteractiopn}
\end{eqnarray}
where $P_L=\frac{1}{2}(1-\gamma_5)$ and $P_R=\frac{1}{2}(1+\gamma_5)$.
In our analysis, we consider $V^{\text{R-CKM}}$ to be the same as $V^{\text{CKM}}$. This holds naturally if parity is realised as a symmetry in LRSM together with the phase of the  bi-doublet Higgs VEV considered to be  zero \cite{Maiezza:2010ic,Senjanovic:2014pva}.
The charge current Lagrangian for lepton-neutrino sector is given by
\begin{align}
	&\mathcal{L}^{l}_{CC} = \frac{g}{\sqrt{2}}\sum_{i}\bar{l}_{L_i}W_{L\mu}^-\gamma^{\mu}P_L\Big(U_{\text{PMNS}}\nu_{L}+S^{\dagger}N^c\Big)_i \nonumber\\
	&+ \frac{g}{\sqrt{2}}\sum_{i}\bar{l}_{R_i}W_{R\mu}^-\gamma^\mu P_R\Big(K_R^{T} N+T^{*}\nu_{L}^c\Big)_i+\text{H.c.}
	\label{eq:cc}
\end{align}
Note that, in few of the decay channels of $N_i$, neutral current will also contribute. The neutral current for LRSM has the following form \cite{Mohapatra:1977be,Tello}: 
\begin{eqnarray}
	\mathcal{L}_{\text{NC}}=\frac{g}{\cos\theta_w}\Big(Z_\mu J_Z^\mu+\frac{\cos^2\theta_w}{\sqrt{\cos 2\theta_w}}Z'_{\mu}J_{Z'}^{\mu}\Big)
\end{eqnarray}
where,
\begin{align}
	J_Z^{\mu} =  \sum_i\bar{f}_i\gamma^{\mu}(I_{3L}P_L-Q\sin^2\theta_w)f_i \\ 
	J_{Z'}^{\mu} = \sum_i\bar{f}_i\gamma^\mu(I_{3R}P_R-\tan^2\theta_w(Q-I_{3L}))f_i. 
\end{align}

As emphasized before, for the particular choice of the neutrino mixing matrix $\mathcal{V}$, we neglect interaction terms proportional to the mixing matrices $S$ and $T$.

Note that, masses of both the RH gauge bosons and RH neutrinos  are proportional to $SU(2)_R$ breaking scale $v_R$. However,  since the RH neutrino masses  also depend on the Yukawa couplings of $\Delta_R$ with the heavy neutrinos, one can choose to have a wide  splitting between the two. 
In this paper we  consider the masses of the heavy neutrino in the MeV-GeV range, in particular, in between $0.14\,\text{GeV}< M_{N_i}< 5\,\text{GeV}$, so that the decay of the mesons can produce on-shell RH neutrinos. Semi-leptonic meson decays,  such as $M_1^{+}\to l_i^{+}l_{j}^{\pm}M_2^{\mp}$ will be then resonantly enhanced due to the on-shell production of the RH neutrinos. 

\section{LNV and LNC meson decays \label{LNC and LNV meson Decays}} 

\hspace{0.5cm} Lepton number is broken in LRSM due to $SU(2)_R \times U(1)_{B-L}$ symmetry breaking. The heavy neutrinos being Majorana particles, can result in LNV as well as LNC meson decay processes:
\begin{eqnarray}
	\text{LNV} :M_1^+(p)\rightarrow l_i^+(k_1) +l_j^+(k_2) +M_2^-(k_3)   \\ 
	\text{LNC} :M_1^+(p)\rightarrow l_i^+(k_1) +l_j^-(k_2) +M_2^+(k_3).
\end{eqnarray}
In the above, $M_1$ is a pseudoscalar meson and $M_2$ can be either a pseudoscalar or a vector meson. Here we consider only the case of pseudoscalar meson $M_2$. The LNV mode for all  flavors of  final state leptons arises entirely from  RH neutrino mediation. Also for LNC mode with  different lepton flavors (such as,  $e^+ \mu^-$ combination), the contribution  arises  solely from RH neutrino mediation. These processes are absent in the SM, and hence serve as a distinct signature of new physics. On the other hand, for LNC mode with same  lepton flavors ($e^+e^-,\mu^+\mu^-$), virtual photon and virtual $Z$ diagrams (one loop penguin diagrams) will also contribute in addition to the RH neutrino contribution
with a substantial branching ratio $\sim 10^{-7}/10^{-8}$ for $K^{+} \to e^+ e^- \pi^-/\mu^{+} \mu^{-} \pi^+$ \cite{Li:2014uha, Littenberg:1993qv, Cirigliano:2011ny}. In our subsequent discussions, the contribution to such LNC processes coming  from diagrams involving RH neutrino mediation will be referred to new physics (NP) contribution.  
In rest of the paper, we  focus mainly on the  RH neutrino contribution to the LNV and LNC meson decays, as the main focus of this paper is on the interference effect of RH neutrino states.  

{For our later discussion on the RH neutrino contribution in LNV and LNC semi-leptonic meson decays, the main subject matter of this paper,} we assume  that there are atleast two RH
neutrinos with masses in the range  $100$ MeV$-5$ GeV,
mediating  these meson decays. The Feynman diagrams for the LNV process are shown in Fig.~\ref{LNV process}. The different contributions are mediated through $W_L-N_k-W_L$, $W_L-N_k-W_R$, $W_R-N_k-W_L$ and $W_R-N_k-W_R$, respectively. Note that, while $W_R-N_k-W_R$ diagram completely depends on the mixing matrix in the RH neutrino sector, the other diagrams also depend on the light-heavy neutrino mixing. Throughout this work, we consider  the contribution from $W_R-N_k-W_R$ diagram only, as the light-heavy neutrino mixing angle which comes from the off-diagonal blocks ($S,T$) of mixing matrix $\mathcal{V}$ is very small. Considering the heavy neutrinos to be $\mathcal{O}$(MeV), the RH neutrinos can be produced on-shell and the semi-leptonic meson decay will be resonantly enhanced. In addition, there can also be contribution from $W_L-W_R$ mixing in one of the legs, but these are suppressed due to small mixing angle $\zeta$. The contributions from the light neutrino mediated process will be much smaller due to mass-suppression. 
Hence we do not consider all of  these other contributions in our analysis. In Fig.~\ref{LNC process}, we have shown the Feynman diagram for LNC process. 

The contribution from heavy neutrinos $N_a$ to the decay amplitude of the LNV process $M_1^{+}(p)\to l_{i}^+(k_1) l_{j}^+(k_2)M_{2}^-(k_3)$ can be written as,
\begin{align}
	\mathcal{M}^{\text{LNV},a}_{ij}=(\mathcal{M}_{\text{lep}}^{\mu\nu})_{ij}^a\mathcal{M}^{\text{had}}_{\mu\nu}
\end{align}
where 
\begin{align}
	\mathcal{M}^\text{had}_{\mu\nu}&=\frac{G_F}{\sqrt{2}}M_{W_L}^2 V^{\text{CKM}}_{M_1} V^{\text{CKM}}_{M_2}\braket{0|\bar{q}_2\gamma_\mu\gamma^5 q_1|M_1^{+}(p)} \nonumber \braket{M_2^{+}(k_3)|\bar{q}_3\gamma_\nu\gamma^5|0} \nonumber \\
	&=\frac{G_F}{\sqrt{2}}M_{W_L}^2 V^{\text{CKM}}_{M_1} V^{\text{CKM}}_{M_2} f_{M_1} f_{M_2} p_{\mu} k_{3\nu}.
\end{align}
In the above, $G_F$ is the Fermi coupling constant, $V_{M_1}^{\text{CKM}}$( $V_{M_2}^{\text{CKM}}$) are the Cabbibo-Kobayashi-Maskawa (CKM) matrix elements at the decay (production) vertex of the meson $M_1$($M_2$), $f_{M_1}$, $f_{M_2}$ are the decay constants of $M_1$, $M_2$.
The relevant leptonic matrix element for ($\Delta L$=2) LNV process is given by the following product of two charged currents
\begin{align}
	\mathcal{M}_{\text{lep}}^{\mu\nu} &\propto(\bar{N}\gamma^\mu P_{R}l)(\bar{N}\gamma^\nu P_{R}l)\nonumber \\
	&=(\bar{N}_a (K_R^*)_{ai} \gamma^{\mu} P_{R} l_i)(\bar{N}_a (K_R^*)_{aj} \gamma^{\nu} P_{R} l_j)\nonumber\\
	&=(\bar{l^c_i} {K_R^*}_{ai}\gamma^{\mu}P_{L}N_a)(\bar{N_a}{K_R^*}_{aj}\gamma^{\nu}P_{R}l_j), 
\end{align}

where we have used the fact that massive neutrinos are Majorana type ($N_a^c=N_a$). We can now write the leptonic part of the amplitude as
\begin{align}
	(\mathcal{M}_{\text{lep}}^{\mu\nu})_{ij}^a=2\sqrt{2}G_F\frac{M_{W_L}^2}{M_{W_R}^4}(K_R^*)_{ai}(K_R^*)_{aj}M_{N_a}\frac{\bar{u}(k_1)\slashed{p}\slashed{k}_3 P_R v(k_2)}{q^2-M_{N_a}^2+i\Gamma_{N_a}M_{N_a}},
\end{align}
where $q=p-k_1$.
The $1/M^4_{W_R}$ term appears due to the two $W_R$ gauge boson propagators in the lower-most panel of Fig.~\ref{LNV process}. Finally, we can write the individual contribution from heavy neutrino $N_a$ to the amplitude as   
\begin{equation}
	\mathcal{M}^{\text{LNV},a}_{ij} = 2G_F^2 V^{\text{CKM}}_{M_1}V^{\text{CKM}}_{M_2}f_{M_1}f_{M_2}\Big(\frac{M_{W_L}}{M_{W_R}}\Big)^4 (K_R^*)_{ai}(K_R^*)_{aj}M_{N_a}\frac{\bar{u}(k_1)\slashed{p} \slashed{k_3} P_R v(k_2)}{q^2-M_{Na}^2+i\Gamma_{Na}M_{Na}},
\end{equation}
where $\Gamma_{N_a}$ is the decay width of heavy neutrino $N_a$, obtained by summing over all accessible final states. Adding the contributions from all heavy neutrinos, we can write the full amplitude as
\begin{align}
	\mathcal{M}^{\text{LNV}}=\sum_{a=1}^{3}\Big(\mathcal{M}^{\text{LNV},a}_{ij}+\mathcal{M}^{\text{LNV},a}_{ji}\Big)
\end{align}
where the second contribution is coming from the exchange of two leptons. Finally the total amplitude square, $|\mathcal{M}^{\text{LNV}}|^2$ can be written as
\begin{eqnarray}
	|\mathcal{M}^{\text{LNV}}|^2  &=&  \sum_{a,b=1;\, b> a}^{3}\Bigg(|\mathcal{M}_{ij}^{\text{LNV},a}|^2+|\mathcal{M}_{ji}^{\text{LNV},a}|^2 \nonumber \\
	&& +2\text{Re}\Big[\big(\mathcal{M}^{\text{LNV},a}_{ij}\big)^{\dagger}\big(\mathcal{M}^{\text{LNV},b}_{ij}\big)\Big]  \nonumber \\
	&& +2\text{Re}\Big[\big(\mathcal{M}^{\text{LNV},a}_{ji}\big)^{\dagger}\big(\mathcal{M}^{\text{LNV},b}_{ji}\big)\Big]\Bigg)
\end{eqnarray}
The explicit form of these squared matrix elements are provided in Appendix~\ref{Details of LNV and LNC calculations}.\, 
\\

\begin{figure}
	\centering
	\includegraphics[height=1.8in]{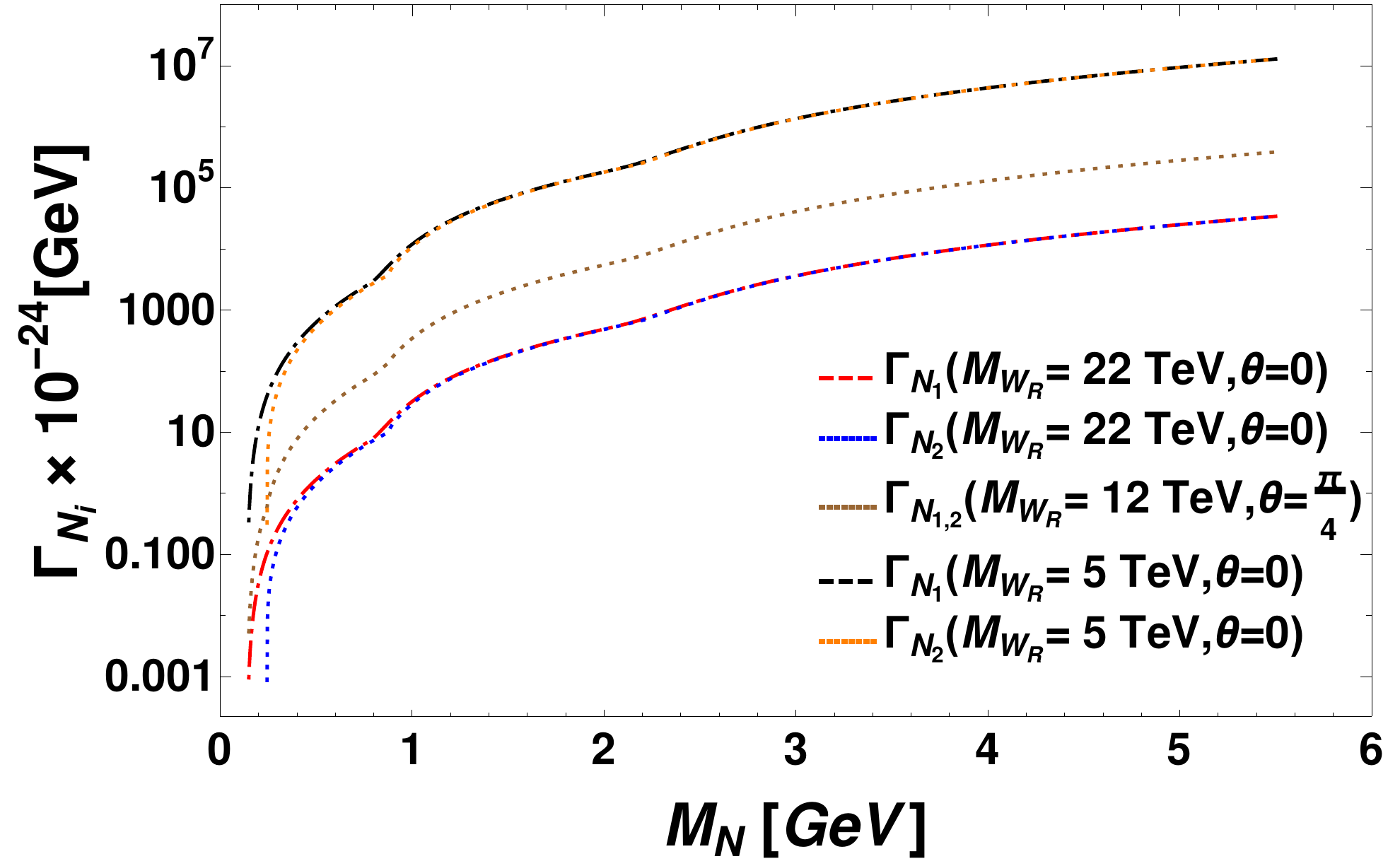}
	\includegraphics[height=1.8in]{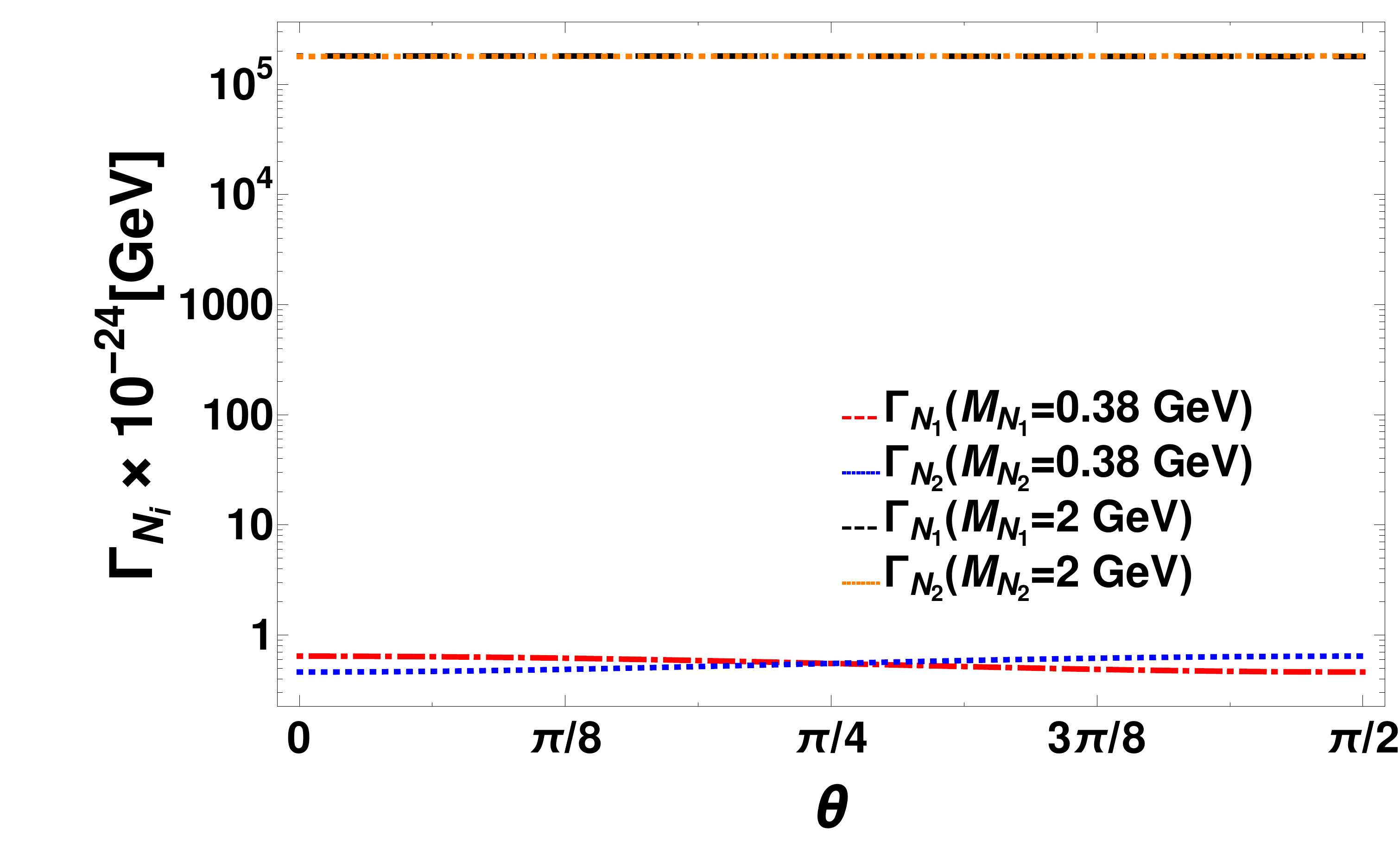}
	\caption{Left panel: variation of the decay widths of the RH neutrino states $N_{1,2}$ with the masses of RH neutrinos for different values of $\theta$ and RH gauge boson mass $M_{W_R}$. Right panel: variation of the decay widths of $N_{1,2}$ with  $\theta$. For $M_{N_{1,2}}=0.38 (2)$ GeV, we consider $M_{W_R}=22 (5)$ TeV, respectively.}
	\label{N decay width}
\end{figure}

{\bf{Decay widths of RH neutrino:}} The RH neutrino state $N_{i}$ of mass MeV to few GeV can decay to various final states, such as, $l^{\pm} V^{\mp}, l^{\pm} P^{\mp}, \nu_l V^0, \nu_l P^0$, where $V,P$ are the vector and pseudoscalar mesons, respectively. Choice of $K_R$ plays a crucial role  in determining the different possible decay modes of RH neutrino. Following the parametrization given in Eq.~\ref{KRmatrix}, which depends on only one angle, and one phase,  we show the decay width of the RH neutrino state $N_{1,2}$ in Fig.~\ref{N decay width}. The left panel represents the variation of the decay widths of $N_{1,2}$ with the masses of the RH neutrinos, and right panel represents the variation w.r.t the mixing angle $\theta$, where $\theta$ parametrises the mixing between $N_1$ and $N_2$ in the  two generation case. For detailed discussion, see Eq.~\ref{KRmatrix}. We find that, for smaller values of $M_{N_{1,2}}$, the dependency of decay width on mixing angles are more prominent,  whereas for higher values of $M_{N_{1,2}}$, the $\theta$ dependency is negligible. This is clearly evident from the right panel, where for $M_N=2$ GeV, both the decay widths $\Gamma_{N_{1,2}}$ coincide.
In the left panel, that represents decay widths for various $\theta$, the decay widths of $N_{1,2}$ show some difference for smaller masses, and  for $\theta \neq \pi/4$, while for larger masses $M_{N_{1,2}}> 1.5$ GeV, both the decay widths become same.  This will have an impact on the estimated branching ratios for $K$ and $B$ mesons, which we will discuss in the subsequent sections.  From Fig.~\ref{N decay width}, it is clear, that the decay width of RH neutrino is indeed very small for our chosen mass range. Hence, we can safely use the narrow width approximation, 
\begin{align}
	|\mathcal{M}_{ij}^{\text{LNV},a}|^2 & \propto  \frac{1}{\big((p-k_1)^2-M_{N_a}^2\big)^2+\Gamma_{N_a}^2M_{N_a}^2}  \nonumber \\ & =\frac{\pi}{M_{N_a}\Gamma_{N_a}}\delta((p-k_1)^2-M_{N_a}^2)
\end{align}
 We have verified that for  the parametrisation given in Eq~\ref{eq:3KR}, the narrow width approximation is valid too. Note that, we are neglecting terms like $(\mathcal{M}_{ij}^{\text{LNV},a})^{\dagger}\mathcal{M}_{ji}^{\text{LNV},b}$ as
\begin{align}
	(\mathcal{M}_{ij}^{\text{LNV},a})^{\dagger}\mathcal{M}_{ji}^{\text{LNV},b}&\propto \frac{1}{\Big((p-k_1)^2-M_{N_a}^2-i\Gamma_{N_a}M_{N_a}\Big)\Big((p-k_2)^2-M_{N_b}^2+i\Gamma_{N_b}M_{N_b}\Big)} \\\nonumber
	&\neq \frac{\pi}{M_{N_a}\Gamma_{N_a}}\delta((p-k_1)^2-M_{N_a}^2)
\end{align}
Hence, contributions from these terms will not be resonantly enhanced and can be safely neglected compared to other terms, see Ref.~\cite{Helo:2010cw}.
Finally, the LNV decay rate can be written as
\begin{align}
	d\Gamma^{\text{LNV}}=\bigg(1-\frac{\delta_{ij}}{2}\bigg)\frac{1}{2m}|\mathcal{M}^{\text{LNV}}|^2d_{3}(\text{PS};p\to k_1 k_2 k_3)
	\label{Decay Width LNV}
\end{align}
where $d_3(\text{PS};p\to k_1 k_2 k_3)$ is the three-body phase space which can be written in terms of product of two two-body phase space as follows
\begin{align}
	d_3(\text{PS};p\to k_1 k_2 k_3)=d_2(\text{PS};p\to k_1 q)dq^2 d_2(\text{PS};q\to k_2 k_3)
\end{align}
The full analytical expression for the LNV decay width and three-body phase space are given in Appendix.~\ref{Details of LNV and LNC calculations}.
\\
Similarly the LNC process can also be mediated by the heavy neutrinos, and the relevant   leptonic part of matrix element:
\begin{equation}
	\mathcal{M}_{\text{lep}}^{\mu\nu}\propto(\bar{l}\gamma^\mu P_{R}N)(\bar{N}\gamma^\nu P_{R}l)
	=(\bar{l}_i  \gamma^{\mu} P_{R} (K_R^T)_{ia}N_a)(\bar{N}_a (K_R^*)_{aj} \gamma^{\nu} P_{R} l_j).
\end{equation}
The amplitude coming from individual contribution of heavy neutrino $N_a$ is given by
\begin{equation}
	\mathcal{M}^{\text{LNC},a}_{ij} = 2G_F^2V^{\text{CKM}}_{M_1}V^{\text{CKM}}_{M_2}f_{M_1}f_{M_2}\Big(\frac{M_{W_L}}{M_{W_R}}\Big)^4 (K_R)_{ai}(K_R^*)_{aj}\frac{\bar{u}(k_1)\slashed{p}\slashed{q} \slashed{k_3} P_R v(k_2)}{q^2-M_{N_a}^2+i\Gamma_{N_a}M_{N_a}}
\end{equation}
Following the same approach as LNV case, one can also  derive the partial decay width for LNC process. The  details are given in Appendix.~\ref{Details of LNV and LNC calculations}. In the subsequent sections, we will consider a simplified scenario, where both the RH neutrino states $N_{1,2}$  can give resonantly enhanced contributions in  the LNV and LNC processes, and we will quantify the effect of the interference.

Before presenting our analysis on the interference effect,  we make few remarks about the bound appearing  on such RH neutrino states from cosmology. Note that,  the RH neutrinos of mass in the range MeV to $\mathcal{O}(\textrm{GeV})$
can  also be constrained  from  cosmological considerations. A RH neutrino with a mixing angle  $\theta$ with the active neutrino, can decay to leptonic and hadronic final states. If the decay happens around the time of Big Bang Nucleosynthesis~(BBN)  $\tau \geq 1$ sec, this can alter the prediction of  light element abundance in the Universe. Constraints from BBN  on the MeV scale RH neutrino  have been discussed in detail in \cite{Dolgov:2000jw}, with the assumption of a  pure sterile neutrino. See also \cite{Kawasaki:2004qu,Kawasaki:2017bqm} for a generic discussion on  BBN constraints for a late decaying massive  state 
$X$. Similar constraints would also be applicable for LRSM, where the RH neutrino decays via off-shell $W_R$ gauge bosons, and lead to semi-leptonic final states. We estimate the lifetime of the RH neutrino states for 380 MeV and 2 GeV, and the  RH gauge boson masses as 22 TeV and 5 TeV, respectively. We find that, the RH neutrino lifetime  varies in  between $10^{-6 }$ sec to $1$ sec. For these   RH neutrino masses, a $W_R$ gauge boson heavier than 22 TeV will be tightly constrained from BBN. 
Note that, the decay width/lifetime of the RH neutrino states can be made larger/smaller, if additional channels for RH neutrino decay open up. For, example, with a sizeable active-sterile mixing $S, T$ in Eq.~\ref{eq:cc}, the decay lifetime can be made sufficiently smaller, and hence BBN constraints can be avoided. In that scenario, additional diagrams $L-R, L-L$ shown in Fig.~\ref{LNV process} will also contribute in three body LNV and LNC meson decays. A detailed evaluation of  these processes  is beyond the scope of this paper, and will be pursued elsewhere.
\begin{figure}[b!]
	\centering
	\includegraphics[width=0.45\textwidth]{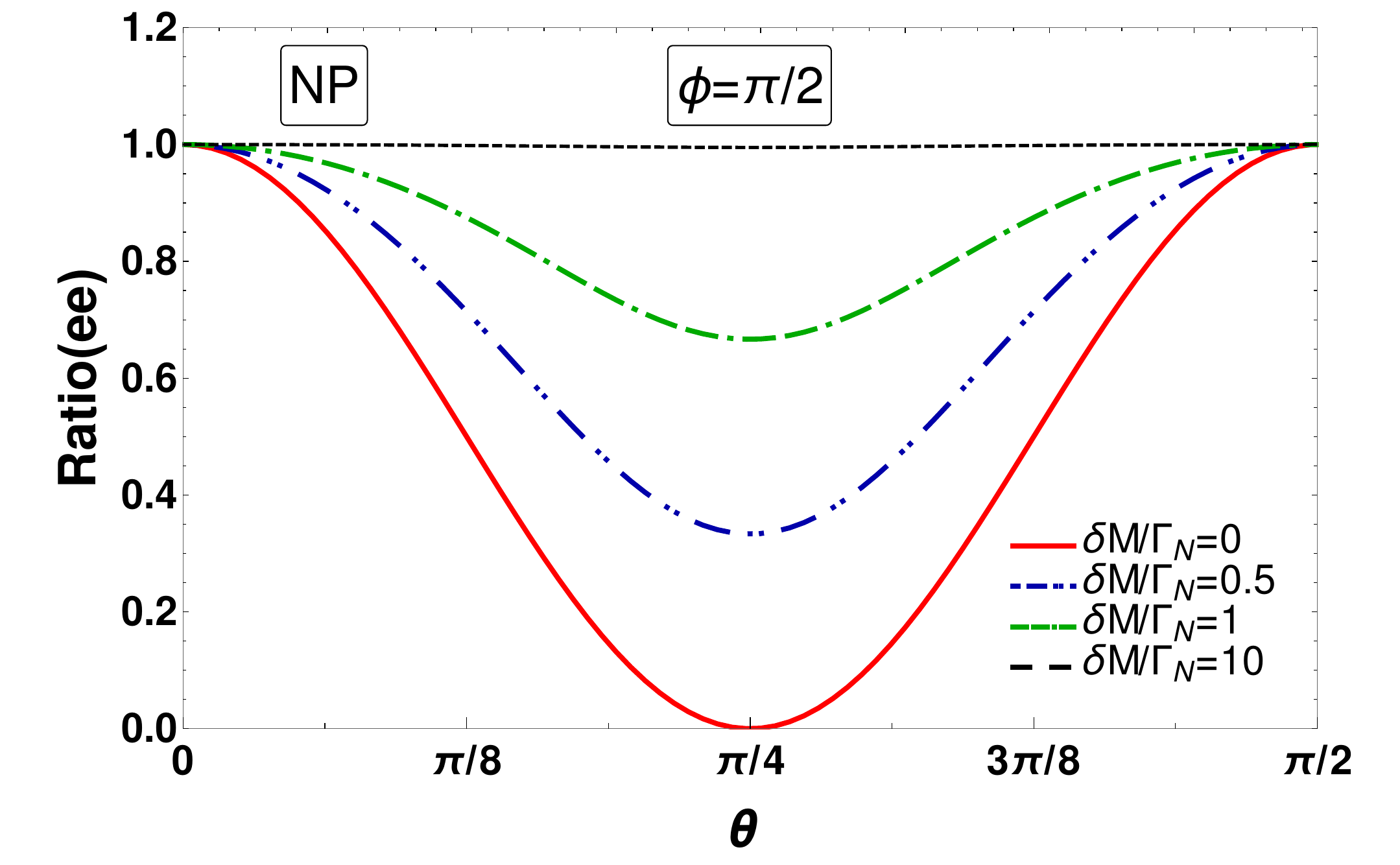}
	\includegraphics[width=0.45\textwidth]{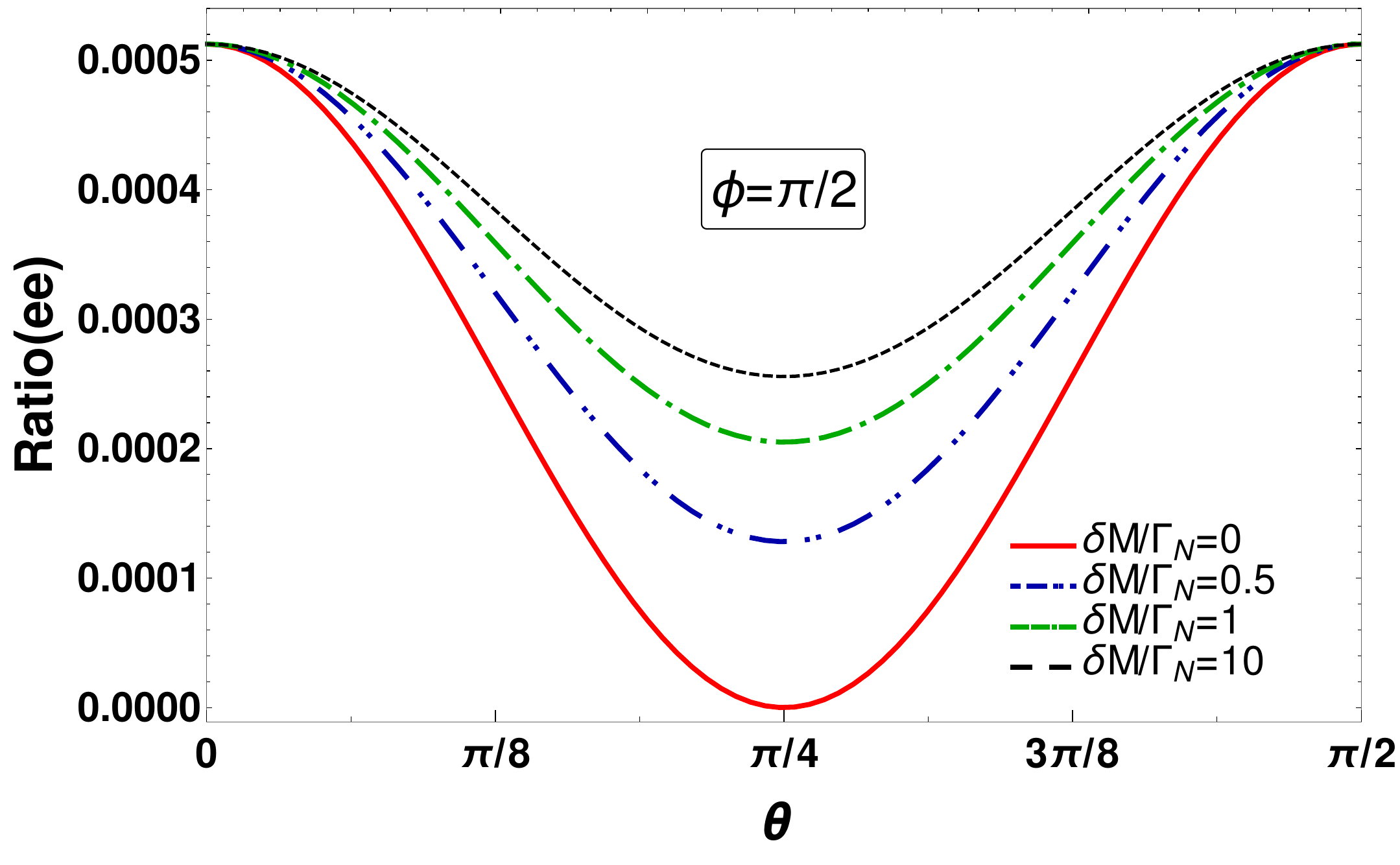}
	\caption{Variation of the ratio of LNV and LNC branching ratios of $K^{+} \to e^{+} e^{\pm} \pi^{\mp}$ with  the mixing angle $\theta$. The red solid, blue dashed-double-dotted, green dashed-dotted, and black dashed lines represent four different $\delta M/\Gamma$ ratios (0,0.5,1,10). The RH neutrino mass has been set at $M_{N} \sim 0.38$ GeV. Left panel represents only RH neutrino mediated contribution. 
		In the right panel we present a conservative estimate by including the 'polluting'  SM contribution maximally. } 
	\label{fig:interferencesimple}
\end{figure}

\section{Analysis for the case of two generations}
\label{Two Generation Scenario}

\hspace{0.5cm} We consider the case where two of the heavy neutrino states are degenerate or nearly degenerate i.e. $M_{N_1} \sim M_{N_2}$ and $M_{N_3}$ is very heavy. The two degenerate states mediate the LNV and LNC meson decays of $K$, and $B$ mesons, and give resonance enhancement in the branching ratios, being in the MeV-GeV mass domain. The choice of the RH neutrino masses can be justified, as  we have free parameters $y_D$ and $y_R$ in the Lagrangian Eq.~\ref{eq:laglrsmyu}, that can be adjusted to get eV light neutrino mass,  along with two nearly degenerate RH neutrinos. The matrix $K_R$ is in general  a 3$\times$3 unitary matrix with few angles, and phases. However, to present the effect of interference in a more simplified  way, we choose  the following parameterisation of $K_R$, 
\begin{equation}
	K_{R}=\begin{bmatrix}
		\cos\theta & -\sin\theta & 0\\
		e^{-i\phi}\sin\theta & e^{-i\phi}\cos\theta & 0 \\
		0 & 0 & 1
	\end{bmatrix},
	\label{KRmatrix}
\end{equation}
The above matrix  $K_R$ is just the product of an orthogonal matrix with angle $\theta$,  and a diagonal phase matrix.
By the choice of such parameterisation, we are interested only at mixing between the two flavors ($N_1,N_2$) which are assumed to have degenerate or nearly degenerate masses. The parameterisation of the mixing matrix $K_R$ enables  the three body decay of meson into  $e$, and $\mu$ lepton flavors, and suppresses any final state with  tau. In the  subsections Sec.~\ref{lnv}, and Sec.~\ref{lnc},  we explicitly demonstrate the impact of this angle and phase on LNV and LNC meson decay rates.

Note that, in addition to the mixing angle, and phase, the contributions  of $N_{1}$ and $N_{2}$ states in LNV and LNC decays also depend on the mass difference of the RH neutrino states $N_{1,2}$, along with  their decay widths $\Gamma_{N_{1,2}}$. For degenerate or almost nearly degenerate masses of RH neutrinos, 
the states $N_{1,2}$ will both be resonantly produced in the $K, B$ meson decays. 
Depending upon the angle and phases of the mixing matrix, the contributions of the $N_{1,2}$ states can either interfere  constructively or destructively. For a very large mass difference between the $N_{1,2}$ states, the interference effect would fade away.  Therefore, for a large  mass splitting between the RH neutrinos,  the LNV and LNC meson decays are similar to the one generation case, that has been studied in detail in \cite{Helo:2010cw,Atre:2009rg,Chun:2019nwi,Mandal:2017tab}.   To demonstrate the effect of the mass splitting $\delta M$, and the decay width $\Gamma$=$\frac{\Gamma_{N_1}+\Gamma_{N_2}}{2}$ on the interference effect, in Fig.~\ref{fig:interferencesimple}, we show the ratio of LNV and LNC branching ratios for different values of the $\delta M/\Gamma$, where we consider the $K^{+}\to e^{+} e^{\pm} \pi^{\mp}$ channel as an illustrative example. The left panel shows  new physics contribution, mediated via RH neutrinos (referred as NP in the plot). In the right panel, we  show a conservative estimate of the LNV and LNC ratio. The ratio of mass difference between two states $N_{1,2}$ and the decay width $\Gamma$ has a large impact on the interference effect. Without any interference effect, the LNV and LNC decay branching ratio would be same, leading  the ratio to be identity.   For a very small value of $\delta M/\Gamma$, the ratio deviates significantly from identity.  As is evident from the figure, increasing  $\delta M/\Gamma$ ratio, the interference between $N_{1,2}$ state tends to become less prominent, but  still has a visible effect for $\delta M \sim \Gamma$. For much larger values, such as,  $ \delta M/ \Gamma =10 $ the interference  effect and the oscillatory behaviour fade away, leading to the LNV and LNC branching ratios to be equal. While we present the discussion on the $ee$ final state, it is to be noted that this is not the most optimal channel due to large SM contribution in LNC process. However, left panel of Fig.~\ref{fig:interferencesimple} can be used to demonstrate the dependency of interference effect on $\delta M/\Gamma_N$. We have verified that this feature is present in $\mu \mu$ channel, and in $e\mu$ channel (for non-degenerate RHN masses) with the   $K_R$ matrix given in Eq.~\ref{KRmatrix}, and also in  $e\mu$ channel for the parametrisation of $K_R$ discussed in Section.~\ref{Three generation case}

\begin{figure*}[t!]
	\centering
	\includegraphics[width=0.40\textwidth]{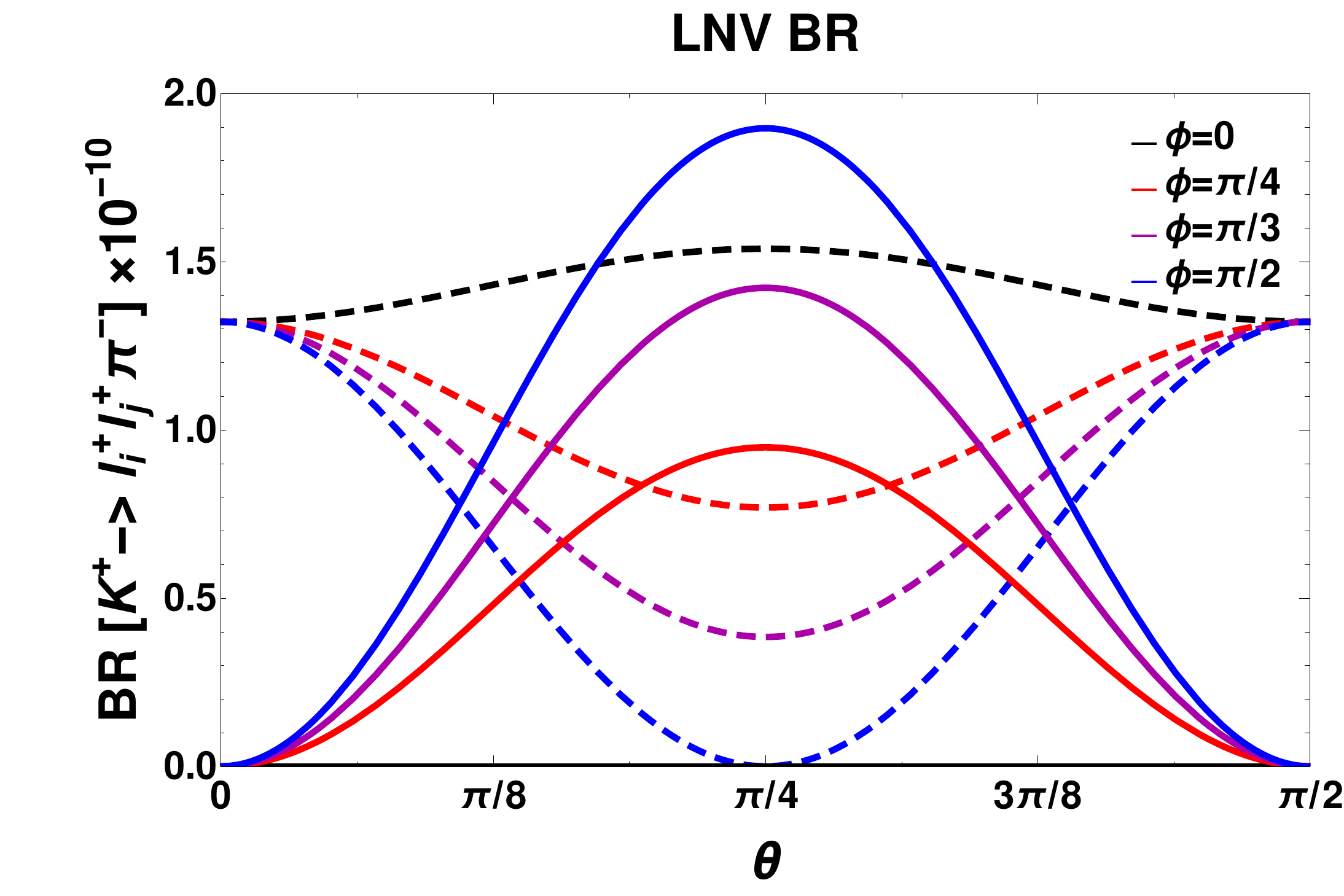}
	\includegraphics[width=0.40\textwidth]{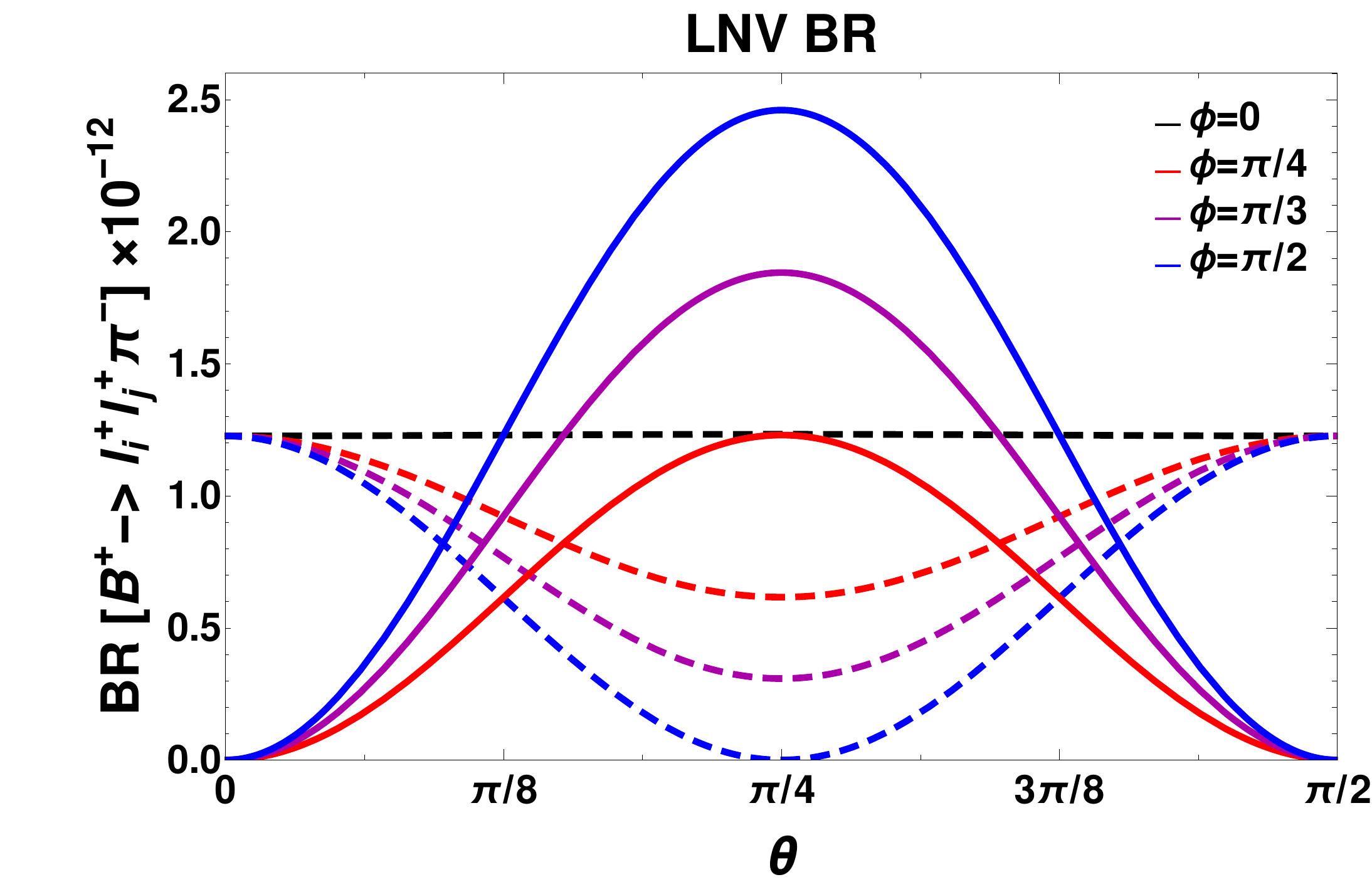}
	\caption{Branching ratio of LNV  meson decay to same flavor $(e^+,e^+)$ and different  flavor $(e^+,\mu^+)$ final state along with a pion ($\pi^+$). The  plot in the left panel is for kaon ($K^+$),  and the plot in the right panel is for B-meson($B^+$) decay, respectively. The BR is not constant rather shows constructive and destructive interference effect for different values of $\theta$ and $\phi$. The dotted line and solid line correspond to $e^+e^+$ and $e^+\mu^+$ mode respectively. For $K^+$ decay we have considered $M_{N} \simeq 0.38\,  \rm{GeV}$ and $M_{W_R} = 22\,
		\rm{TeV}$,  and for $B^+$-meson decay we have considered $M_{N} \simeq 2\, \rm{ GeV}$ and $M_{W_R} = 5\, \rm{TeV}$, respectively.}
	\label{fig:brratioBKLNV}
\end{figure*}
\subsection{LNV \label{lnv}}
We give the complete list of decay channels of $N_i$ for general light-heavy mixing matrix in Appendix~\ref{decay width of N}. We list only the channels that are kinematically allowed for RH neutrino masses in the range $0.14\,\text{GeV}< M_{N_i}< 5\,\text{GeV}$. Note that, for our chosen mass range of RH neutrino states and with the off-diagonal mixing elements $S,T \sim 0$, decay modes such as $N \to l^{\pm}P^{\mp}, l^{\mp} V^{\pm}$, $\nu_l P^0$ and $\nu_l V^0$ will only be allowed. With the  mixing matrix $K_R$ given in Eq.~\ref{KRmatrix}, RH neutrino decay width will also depends on the mixing angle $\theta$. One can write down a generic expression for the total decay width of degenerate RH neutrinos as,  
\begin{align}
	\Gamma_{N_1}=\frac{1}{M_{W_R}^4}\Big(A(M_N)+B(M_N)\cos 2\theta\Big), \label{N1 width} \\ \Gamma_{N_2}=\frac{1}{M_{W_R}^4}\Big(A(M_N)-B(M_N)\cos 2\theta\Big)
	\label{N1 and N2 decay width}
\end{align}
Note that, these above expressions  can be obtained by substituting $K_R$ into the expressions of  the decay widths given in the Appendix~\ref{decay width of N}, where we considered $S,T  \sim 0$. For the choice of $M_{N_1} \sim M_{N_2} \sim M_{N}=0.38$ GeV, that we have considered for $K^{+}$ meson decay,  the RH neutrinos will have only decay modes with a charged pion in the final state, i.e. $N_j\rightarrow l^{\pm}_i+\pi^{\mp}$. The expressions of the two functions, $A(M_N)$ and $B(M_N)$ have a simpler form, and have been given  in the Appendix~\ref{decay width of N} (see Eq.~\ref{AMN} and Eq.~\ref{BMN}). We note that, for  relatively larger mass  $M_{N_i}> 1.5$ GeV of the RH neutrino states, decay widths are nearly independent of the mixing angle $\theta$ (Fig.~\ref{N decay width}). This can be understood easily as follows. We can neglect the final state lepton masses  in  evaluating the decay widths for RH neutrinos with masses $M_N  \sim $ few GeV.  The unitarity relation for the mixing matrix $K_R$ makes the total decay widths of $N_{1,2}$ nearly independent of $\theta$. In this case, one can write down the expression for  the total decay width as,
\begin{align}
	\Gamma_{N_1}\sim\frac{1}{M_{W_R}^4} A(M_{N_1}),\,\,\Gamma_{N_2}\sim\frac{1}{M_{W_R}^4} A(M_{N_2}).
	\label{N1 and N2 decay width set1}
\end{align}
Here $A(M_N)$ is function of the mass $M_N$, and can be derived from different decay modes,  given in the Appendix~\ref{decay width of N}. Therefore, for degenerate masses ($M_{N_1} = M_{N_2}\equiv M_{N}$), the above two decay widths will be nearly the same. 
\begin{figure*}
	\centering
	\includegraphics[height=2.50in]{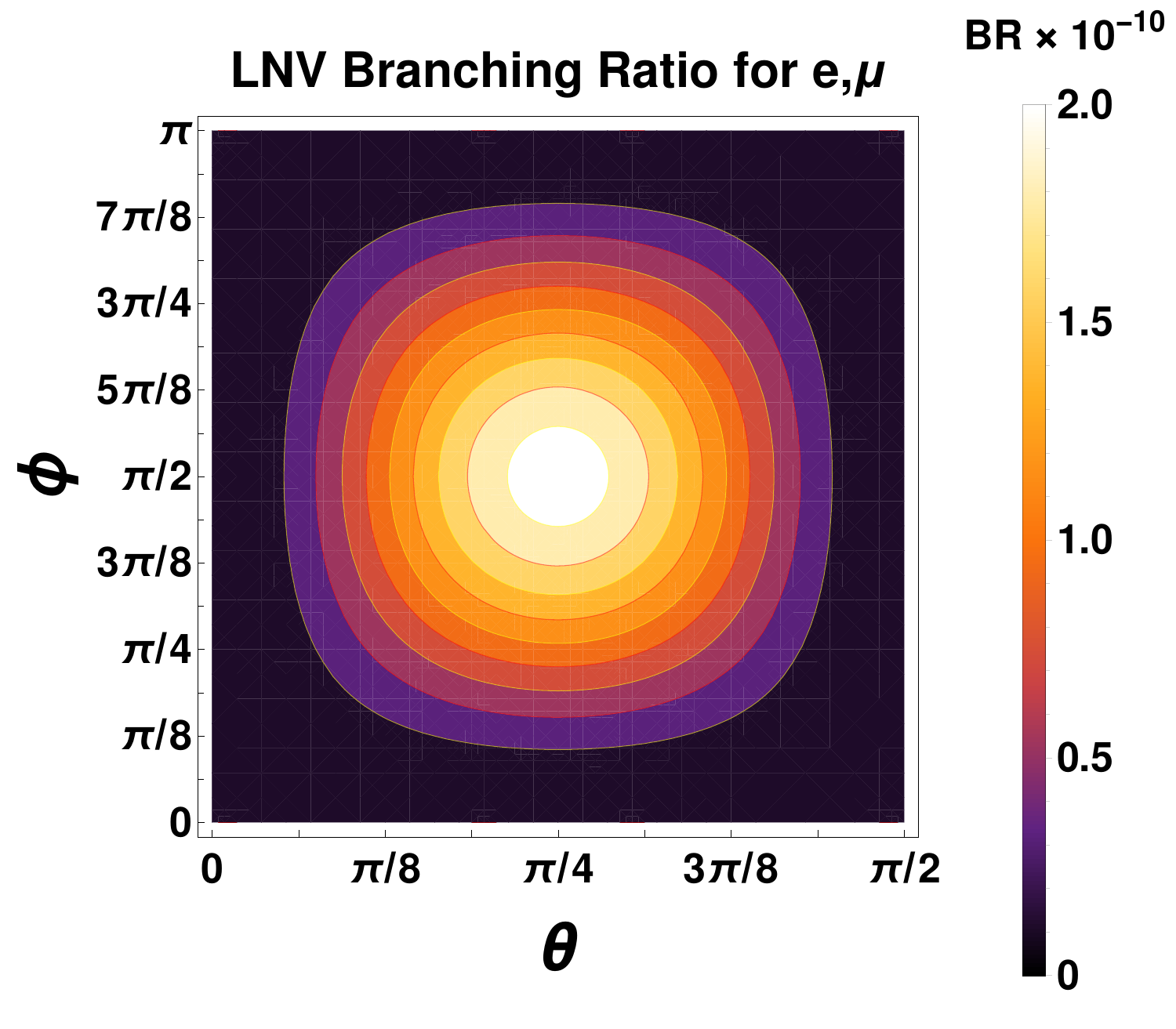}
	\includegraphics[height=2.50in]{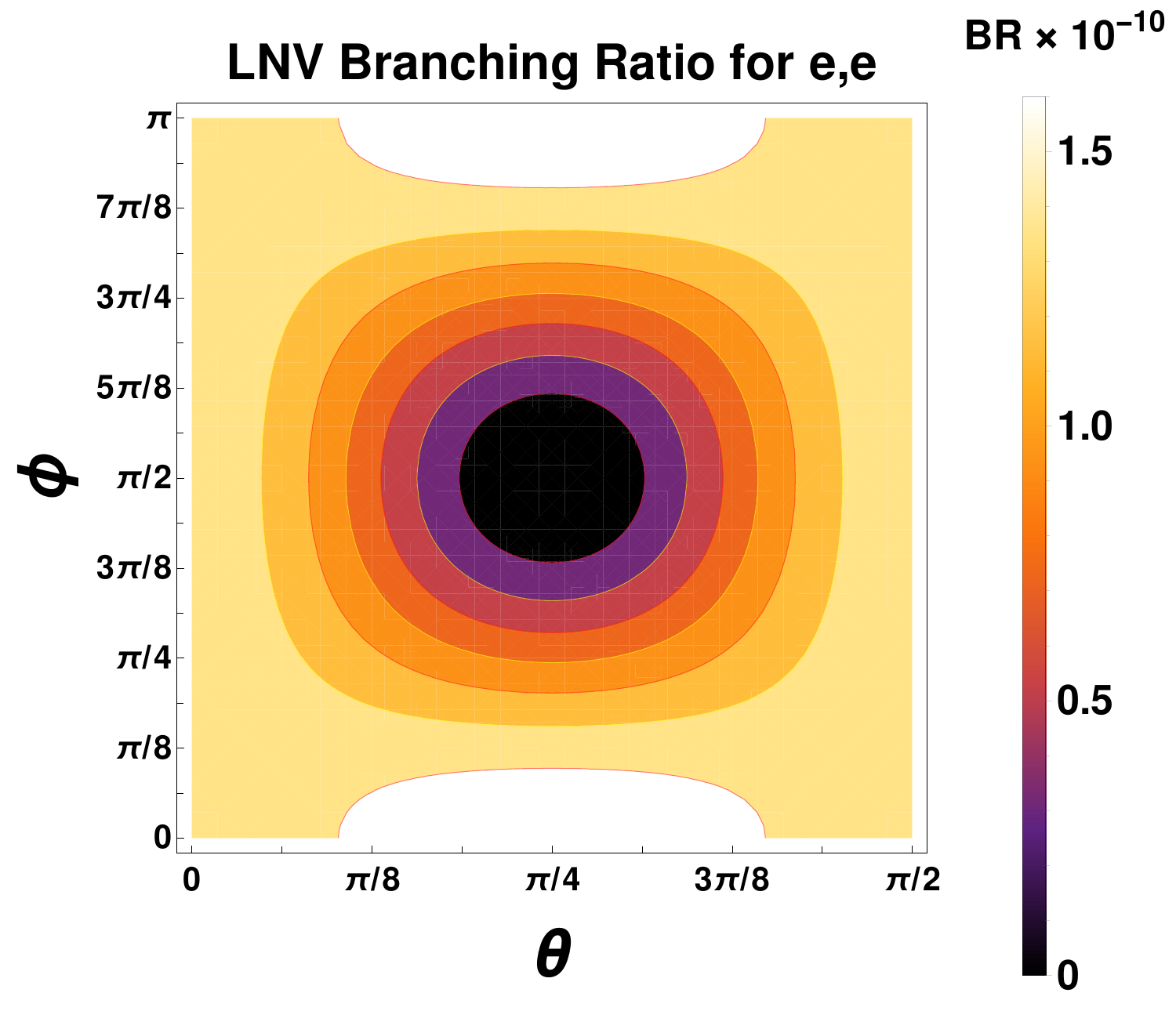}
	\caption{Left panel:  variation of the branching ratio of LNV kaon decay $K^+$ $\rightarrow$  $e^+$ $\mu^+ \pi^-$ with the variation of the angle and phase $(\theta,\phi$). The final states leptons are of same charge, and different flavors. Right panel: the same, however for final states with same charge and same lepton flavors $K^+$ $\rightarrow$  $e^+$ $e^+ \pi^-$. The masses of RH neutrino and RH gauge bosons are   $M_{N_1}\simeq M_{N_2}\simeq 0.38\, \rm{GeV}$, and $M_{W_R} = 22\, \rm{TeV}$, respectively. The maximum value of the branching ratios are at $\mathcal{O}$($10^{-10}$), consistent with the current experimental bounds}. 
	\label{fig:countourKemu}
\end{figure*}

Using the general  form of the heavy $N$ decay width given in  Eq.~\ref{N1 width}  and Eq.~\ref{N1 and N2 decay width}, one can write down the LNV decay rate of the parent meson $M_1$ into same flavor final state leptons and it leads to a rather complicated expression. The explicit expressions for the $ee$ and $\mu\mu$ case are as follows:
\begin{align}
	\label{eq:lnvpartialee}
	&\Gamma^{\text{LNV}}_{ee} =   \pi\beta M_N\Big(\frac{\cos^4\theta}{A(M_N)+B(M_N)\cos 2\theta} +\frac{\sin^4\theta}{A(M_N)-B(M_N)\cos 2\theta} \\ \nonumber
	&+\frac{1}{4}\cos 2\phi\sin^2 2\theta   \Big[\frac{1}{A(M_N)+B(M_N)\cos 2\theta}+\frac{1}{A(M_N)-B(M_N)\cos 2\theta}\Big]\Big)f^{ee}(M_N)
\end{align}

\begin{align}
	\label{eq:lnvpartialmumu}
	&\Gamma^{\text{LNV}}_{\mu\mu}=  \pi\beta M_N \Big(\frac{\sin^4\theta}{A(M_N)+B(M_N)\cos 2\theta}+\frac{\cos^4\theta}{A(M_N)-B(M_N)\cos 2\theta}\\ \nonumber
	& +\frac{1}{4}\cos 2\phi\sin^2 2\theta\Big[\frac{1}{A(M_N)+B(M_N)\cos 2\theta}+\frac{1}{A(M_N)-B(M_N)\cos 2\theta}\Big]\Big)f^{\mu\mu}(M_N)
\end{align}

The partial decay width for different flavor final state leptons is given by
\begin{align}
	\label{eq:lnvpartialemu}
	&\Gamma^{\text{LNV}}_{e\mu}=\pi\beta M_N\Big(\frac{\sin^2 2\theta}{4(A(M_N)+B(M_N)\cos 2\theta)}+\frac{{\sin^2 2\theta}}{4(A(M_N)-B(M_N)\cos 2\theta)} \\ \nonumber
	& -\frac{1}{4}\cos 2\phi \sin^2 2\theta\Big[\frac{1}{A(M_N)+B(M_N)\cos 2\theta}+\frac{1}{A(M_N)-B(M_N)\cos 2\theta}\Big]\Big)f^{e\mu}(M_N)
\end{align}

where $\beta=\frac{1}{128\pi^3}\frac{M_{W_L}^8}{M_{W_R}^4}G_F^2\Big(V_{M_1}^{\text{CKM}}V_{M_2}^{\text{CKM}}\Big)^2f_{M_1}^2f_{M_2}^2$, and we consider $M_{N_1}=M_{N_2}=M_N$. The functions $f^{ee}$, $f^{\mu\mu}$ and $f^{e\mu}$ can be identified from the Eq.~\ref{LNV Calculation}. The 
first two terms are due to the individual $N_{1,2}$ contributions in the decay amplitude. The term in the partial decay width expressions proportional to $\cos 2\phi \sin^2 2\theta$ is due to the interference  of the $N_{1,2}$ contributions. 

These above complicated expressions   simplify for  RH neutrinos with  mass $M_N \sim $ few GeV range, where the mass of the final state leptons can be ignored in evaluating the total decay width of RH neutrinos.  Using Eq.~\ref{N1 and N2 decay width set1}, and Eq.~\ref{eq:lnvpartialee}, Eq.~\ref{eq:lnvpartialmumu}, Eq.~\ref{eq:lnvpartialemu}, we  find  that, for RH neutrinos with masses $\sim$ few GeV,  LNV partial width of the parent meson $M_1$~(such as $B$ meson) has the following simplified dependencies upon the mixing angle and phase,
\begin{equation}
	\Gamma^{\text{LNV}}_{ij}\propto\left\{
	\begin{array}
		{cl}
		(1-\sin^{2}2\theta \sin^{2}\phi),&\mbox{for $i=j$  }\\\\
		(\sin^{2}2\theta \sin^{2}\phi),& \mbox{for $i\neq j$}
	\end{array}\right.
\end{equation}
In the above equation $i,j$ can either be $e$ or $\mu $. Similar expressions for the $lljj$ production cross-section at LHC has been obtained in Ref.~\cite{Gluza:2015goa}.

The branching ratios of the different LNV modes of the parent meson $M_1$ are then given by
\begin{equation}
	\textrm{Br}( M^{+}_1 \to l^{+}_i l^{+}_j M^{-}_2)=\frac{\Gamma^{\text{LNV}}_{ij}}{\Gamma_{M^+_1}}
	\label{eq:brlnv}
\end{equation}
where $\Gamma_{M^+_1}$ is the total decay width of the parent meson. In Fig.~\ref{fig:brratioBKLNV},  we show the  branching ratios of LNV meson decay $K^{+}/B^{+} \to e^{+} e^{+} \pi^-$, and  $K^{+}/B^{+} \to e^{+} \mu^{+} \pi^-$ vs mixing angle $\theta$ for different values of $\phi$. The left panel corresponds to the kaon decay. We consider the  mass of the RH gauge boson as  $M_{W_R}$ = 22 TeV, and mass of heavy neutrino as $M_N$ $\simeq$ 0.38 GeV. In the right panel,  we show  the LNV decay of $B^+$-meson. For this case, we consider  the RH gauge boson mass to be $M_{W_R}$ = 5 TeV, and mass of heavy neutrino is $M_N$ $\simeq$ 2 GeV. This choice of  RH neutrino and RH gauge boson mass, which is relevant for  $B^+$ study,  is unconstrained from $K^{+}$ meson decays. The branching ratios derived for these benchmark mass points are consistent with the experimental limits $\textrm{Br}{(K^+ \to e^+ e^+\pi^-, e^+\mu^+  \pi^-) < (2.2, 5.0) \times 10^{-10}}$  and    \textrm{Br}$(B^+ \to e^+ e^+\pi^-, e^+ \mu^+ \pi^-) < (2.3, 15.0) \times 10^{-8}$, respectively \cite{Tanabashi:2018oca}. The figures confirm the angular dependencies of the branching ratios of LNV decay process. The solid and dashed lines correspond to $e^{+} \mu^{+}$, and $e^+ e^+$ channels. Note that,  the $e^+e^+$ and $e^+\mu^+$ channels have complimentary nature w.r.t the angular variables. This can further be highlighted by a contour plot in the $\theta-\phi$ plane. Fig.~\ref{fig:countourKemu} 
shows the variation of LNV branching ratios of  $K^+ \to e^{+} e^{+} \pi^{-}$, and $K^{+} \to e^{+} \mu^{+} \pi^-$ processes for different values of mixing angle $\theta$ and phase $\phi$. With two degenerate heavy neutrinos, the branching ratio of the above processes  exhibit constructive and destructive interferences, as is clearly evident from the figures.  The white region in the right panel of Fig.~\ref{fig:countourKemu},  occurs as for $\theta=\pi/4$, and $\phi \sim 0, \pi$, the LNV branching ratio to $e^{+} e^{+}$ channel shows a maxima (see left panel of Fig.~\ref{fig:brratioBKLNV} ). A different value of the RH gauge boson mass will simply give an overall scaling in the branching ratios and will not change the nature of Fig.~\ref{fig:brratioBKLNV}. For  $B^{+}$ meson, the plots are very similar. Hence,  we do not show them explicitly.   

\subsection{LNC \label{lnc}}
\hspace{0.5cm} The parent meson can decay via LNC processes  $M^{+}_1 \to l^{+}_i l^{-}_j M^{+}_2$. The RH neutrino states $N_{1,2}$ will mediate these processes. 
Considering the general form of $N_i$ decay width as given in Eq.~\ref{N1 width} and Eq.~\ref{N1 and N2 decay width}, LNC decay rate can be calculated. For $ee$ and $\mu\mu$ channel the $N$ mediated decay rate is given by
\begin{eqnarray}
	\begin{aligned}
		\Gamma^{\text{LNC}}_{ee} & =\pi\frac{\beta}{M_N}\Bigg(\frac{\cos^4\theta}{A(M_N)+B(M_N)\cos 2\theta}+\frac{\sin^4\theta}{A(M_N)-B(M_N)\cos 2\theta}+\frac{1}{4}\sin^2 2\theta\\ 
		& \Big[\frac{1}{A(M_N)+B(M_N)\cos 2\theta}+\frac{1}{A(M_N)-B(M_N)\cos 2\theta}\Big]\Bigg)h^{ee}(M_N)
	\end{aligned}
	\label{eq:LNCee}
\end{eqnarray}
\begin{eqnarray}
	\begin{aligned}
		\Gamma^{\text{LNC}}_{\mu\mu}&=\pi\frac{\beta}{M_N}\Bigg(\frac{\sin^4\theta}{A(M_N)+B(M_N)\cos 2\theta}+\frac{\cos^4\theta}{A(M_N)-B(M_N)\cos 2\theta}+\frac{1}{4}\sin^2 2\theta\\ 
		&\Big[\frac{1}{A(M_N)+B(M_N)\cos 2\theta}+\frac{1}{A(M_N)-B(M_N)\cos 2\theta}\Big]\Bigg)h^{\mu\mu}(M_N)
	\end{aligned}
	\label{eq:LNCmumu}
\end{eqnarray}
and for $e\mu$ channel, the decay rate has the following form,
\begin{eqnarray}
	\begin{aligned}
		\Gamma^{\text{LNC}}_{e\mu}&=\pi\frac{\beta}{M_N}\Bigg(\frac{\sin^2 2\theta}{4(A(M_N)+B(M_N)\cos 2\theta)}+\frac{\sin^2 2\theta}{4(A(M_N)-B(M_N)\cos 2\theta)}-\frac{1}{4}\sin^2 2\theta\\ 
		&\Big[\frac{1}{A(M_N)+B(M_N)\cos 2\theta}+\frac{1}{A(M_N)-B(M_N)\cos 2\theta}\Big]\Bigg)h^{e\mu}(M_N)
	\end{aligned}
	\label{eq:LNCemu}
\end{eqnarray}
\begin{figure*}
	\centering
	\includegraphics[width=0.45\textwidth]{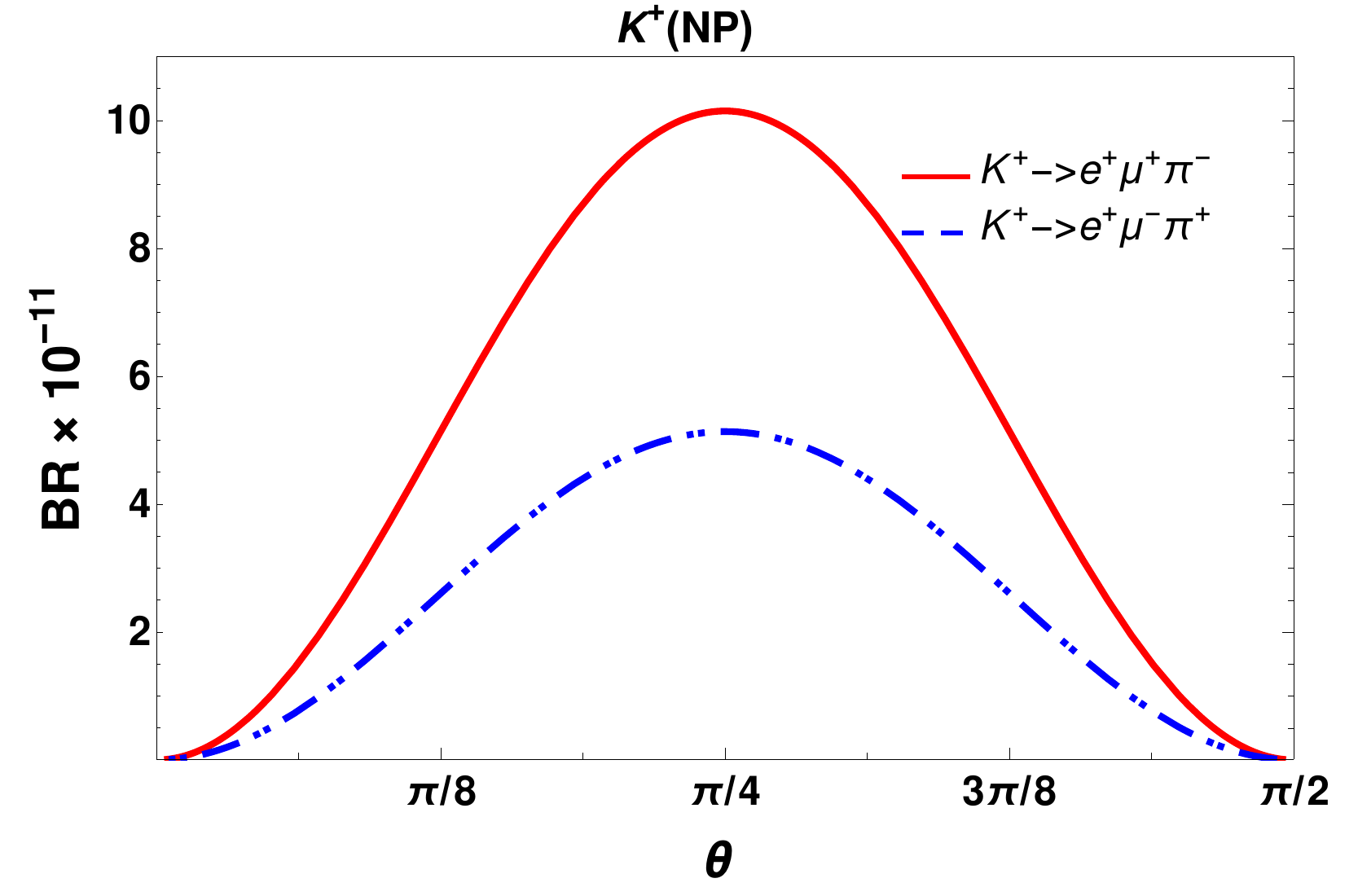}
	\includegraphics[width=0.45\textwidth]{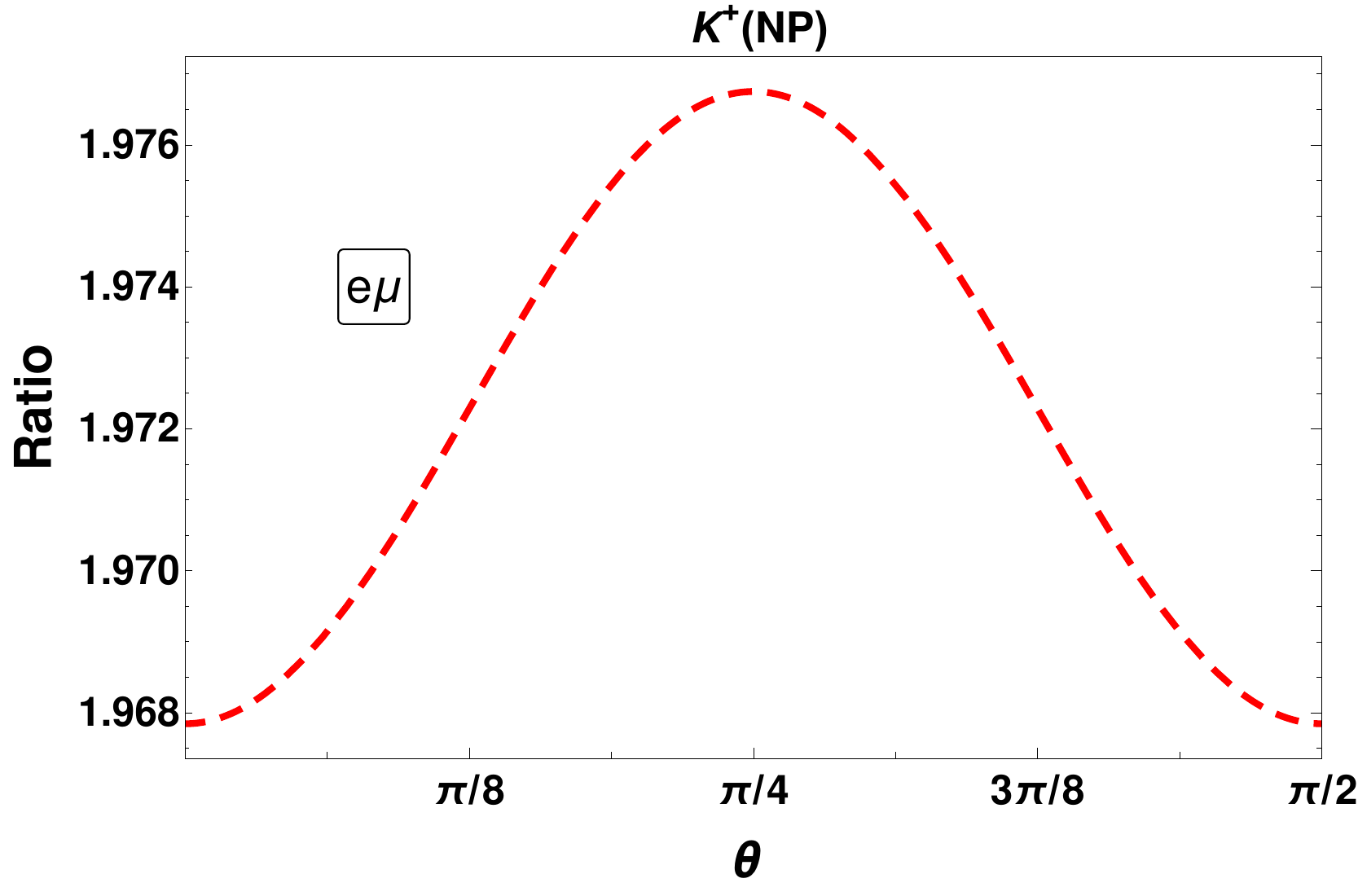}
	\caption{Left panel: Branching ratio of $K^+ \to e^+ \mu^+ \pi^-$, and $e^+ \mu^- \pi^+$ vs angle $\theta$ for non-degenerate RH neutrino masses considering $\frac{\delta M}{\Gamma_{N}} \sim 10$. Right panel: variation of the ratio of the two branching ratios vs mixing angle $\theta$. The results for $B^+$ decays are similar, hence, we do not show them explicitly.}
	\label{fig:brratioBKLNC}	
\end{figure*}

 The functions $h^{ee}$, $h^{\mu\mu}$ and $h^{e\mu}$ can be identified from Eq.~\ref{LNC Calculation}. Note that,  the functions $h^{ij}$ in the above expressions  are related with $f^{ij}$ of the previous section, as $M_N^2 f^{ij}=h^{ij}$. We use the same values for RH neutrino and RH gauge boson masses, that we considered for LNV processes. Using Eq.~\ref{eq:LNCemu}, and after simplification, LNC partial decay width for different lepton flavor $e\mu$ becomes zero for degenerate RHN masses $M_{N_1}=M_{N_2}$. This however becomes non-zero, if we consider non-degenerate masses of RH neutrinos. In the left panel of Fig.~\ref{fig:brratioBKLNC}, we show the branching ratio of $K^+ \to e^+\mu^+\pi^- ,e^+\mu^-\pi^+ $ for $\frac{\delta M}{\Gamma_N}$ $\sim$ 10. In the right panel of Fig.~\ref{fig:brratioBKLNC}, we show the ratio of the two branching ratios for the  $e\mu$ channel. As can be seen due to the interference between two RH neutrino states, the ratio of LNV and LNC branching ratios differ from unity. Using Eq.~\ref{eq:LNCee} and Eq.~\ref{eq:LNCmumu},  the partial  width for  same flavor leptons turns out to be large, and is almost   independent of the mixing angle $\theta$. We find that the contribution of RHN mediated diagrams in the branching ratio of same flavor LNC decay mode of $K^+$($K^+ \to e^+ e^- \pi^+$) and $B^+$($B^+ \to e^+ e^- \pi^+$) mesons are $\mathcal{O}(10^{-10})$ and $\mathcal{O}(10^{-12})$, respectively. As these same flavor LNC decay modes do also contain  large SM contributions (BR $\sim$ $10^{-7}/10^{-8}$) along with the order of magnitude suppressed new physics contribution, we do not show them here. In Section.~\ref{Three generation case}, we elaborate on the large LNC branching ratio for $e\mu$ case, while using a $3\times 3$   $K_R$ matrix.

\begin{figure*}
	\centering
	\includegraphics[height=1.8in]{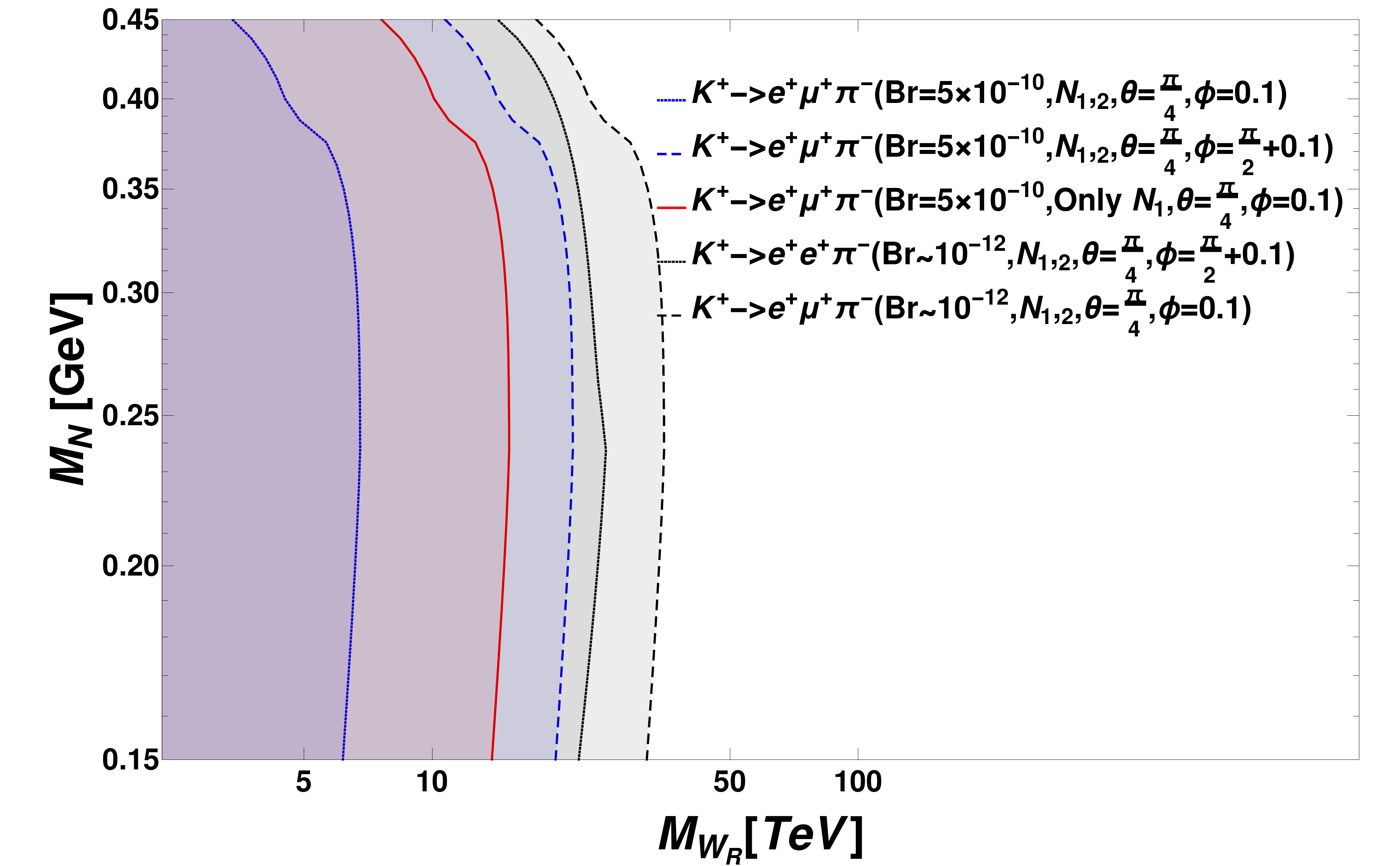}
	\includegraphics[height=1.9in]{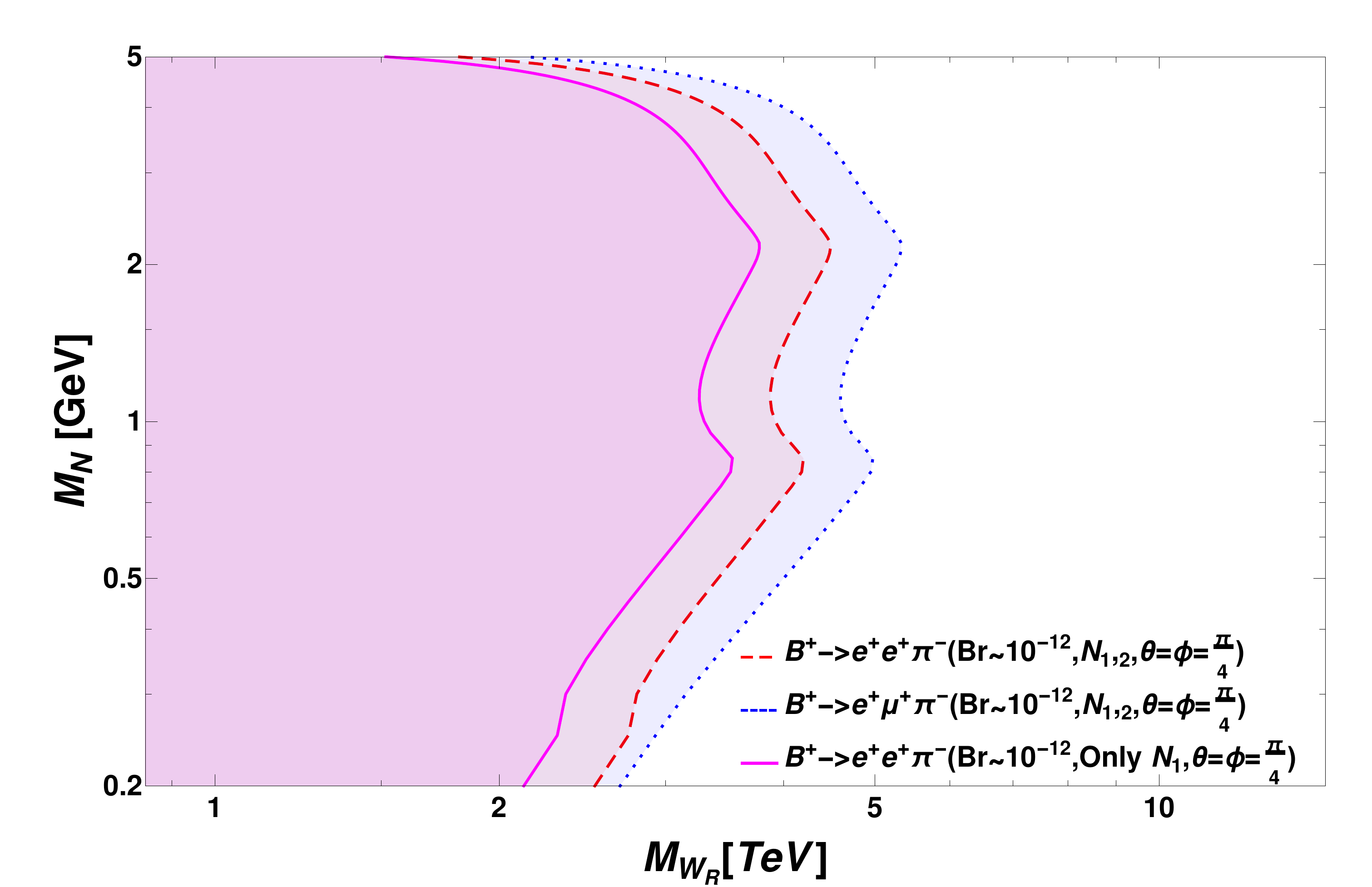}
	\caption{Sensitivity of the RH neutrino mass $M_N$ and the RH gauge boson mass $M_{W_R}$ from the LNV processes $K^+ \to  e^+ e^+ \pi^-$, $K^+ \to  e^+ \mu^+ \pi^-$ (left panel) and $B^+ \to  e^+ \mu^+ \pi^-$, $e^+ e^+ \pi^-$ (right panel). For $K^+$ decay mode in the upper panel, the red solid line corresponds to one generation scenario, blue dashed  line represents two generation scenario with constructive interference $\theta=\pi/4 $, $\phi \sim 0, \pi/2$. The blue solid  line represents two generation scenario with destructive interference. The black solid and dashed lines represent  $Br \sim 10^{-12}$. The figure in the right panel indicates the future sensitivity of LNV  $B^+$ meson decay with $\text{Br} \sim 10^{-12}$.}
	\label{fig:mnmwr}
\end{figure*}

{\bf  Effect of interference  for RH neutrino and RH gauge boson masses}-In Fig.~\ref{fig:mnmwr}, we represent the effect of interference in the $M_N$-$M_{W_R}$ mass plane, considering LNV modes. We consider two different decay modes $K^{+}/B^+ \to e^{+} \mu^{+} \pi^{-}$, and $K^{+}/B^+ \to e^{+} e^{+} \pi^-$.  The solid red line in the left panel has been derived by assuming only one generation RH neutrino state $N_1$. This corresponds to  the present experimental limit of the  branching ratio $K^+ \to e^{+} \mu^{+} \pi^{-}$, which is $\textrm{Br} (K^+ \to e^{+} \mu^{+} \pi^{-})< 5\times 10^{-10}$~\cite{Tanabashi:2018oca}. The blue dashed and blue solid lines correspond to the two generation RH neutrino scenario where both $N_{1,2}$ can be produced on-shell  from parent meson decay. The subsequent decays of $N_{1,2}$  lead to the same final state $e^{+} \mu^{+} \pi^{-}$ with the same value of the branching ratio. 
We have set the mixing angle, and phase as $\theta=\pi/4, \phi = \pi/2+0.1$ for blue dashed line. The solid blue line corresponds to the two generation scenario, with $\theta=\pi/4, \phi=0.1$, and again corresponds the experimental limit on the branching ratio~\footnote{We have not considered $\phi$ to be exactly $\pi/2$ or 0 as the theoretical branching ratio will be exactly zero and hence it is not possible to derive any kind of bound in $M_N-M_{W_R}$ plane.}.  The black dashed line and black solid line correspond to a value of $10^{-12}$ for branching ratios of $K^{+} \to e^{+} \mu^{+} \pi^-$, and $K^{+} \to e^{+} e^{+} \pi^-$. 
In the right panel, we show the result  for $B$ meson decays $B^+ \to e^{+} e^{+} \pi^-/e^{+} \mu^{+} \pi^-$. The two kinks occur, as new decay modes of the RH neutrinos into a lepton associated with a  $\rho$ and $D$ meson open up in these mass ranges. The solid pink line in the right panel has been again derived by assuming only one generation RH neutrino state $N_1$, where as blue dotted and red dashed lines stand for two generation RH neutrino scenario. For $B$ meson case, we have considered the future sensitivity on the branching ratio to be $\mathcal{O}(10^{-12})$\footnote{The current bounds for the $B$ meson LNV semileptonic decays are: $\text{BR}(B^+\to e^+e^+\pi^-)\leq 2.3\times 10^{-8}$ and $\text{BR}(B^+\to e^+\mu^+\pi^-)\leq 1.5\times 10^{-7}$~\cite{Tanabashi:2018oca}.}.  
Note that, the limit on $W_R$ can be significantly lowered in the presence of destructive interference as it leads to a natural suppression of the LNV branching ratio.  Non-observation of any LNV signal for LRSM could signify the interference effects due to  more than one generation of RH neutrinos.

\subsection{Asymmetries in   LNV and LNC Processes}

The non-trivial phase $\phi$, along with the  interference effect between $N_{1,2}$ contributions result in  asymmetries in LNV and LNC processes.  Before discussing the different asymmetries, and the effect of interference, we first  quantify  the effect of  small mass splitting on the LNV and LNC meson decays. The obtained expressions will be used in deriving the expressions of  CP asymmetries.  We represent  the average mass of the two RH neutrinos as  $M_N$,  and mass splitting as  $\delta M$. Therefore,  we follow the convention,  $M_{N_1}=M_N-\frac{\delta M}{2}$ and $M_{N_2}=M_N+\frac{\delta M}{2}$. We assume, the following  conditions are realised:
\begin{align}
	\delta M\ll M_N\,\,\text{and}\,\,\delta M < \Gamma_{N_1}, \Gamma_{N_2},
\end{align}
With the above set of approximations, the LNV and LNC meson decay rates for  a non-trivial mixing matrix $K_R$  can be written as follows,
\begin{align}
	\Gamma^{\text{LNV},++}_{ij}&=\bigg(1-\frac{\delta_{ij}}{2}\bigg)\pi\beta\Big(|(K_R)_{1i}|^2 |(K_R)_{1j}|^2 \frac{\big(M_N-\frac{\delta M}{2}\big)}{\Gamma_{N_1}} f_{ij}\big(M_N-\frac{\delta M}{2}\big) + \nonumber \\
	& |(K_R)_{2i}|^2 |(K_R)_{2j}|^2 \frac{\big(M_N+\frac{\delta M}{2}\big)}{\Gamma_{N_2}} f_{ij}\big(M_N+\frac{\delta M}{2}\big) + \frac{(M_N-\frac{\delta M}{2}) (M_N+\frac{\delta M}{2})}{M_N}\nonumber \\
	& \text{Re}\big[(K_R)_{1i}(K_R)_{1j}(K_R)_{2i}^* (K_R)_{2j}^* \left(\frac{1-4\mathrm{i} \frac{\delta M}{\Gamma_{N_1}}}{\Gamma_{N_1} \big(1+4 \frac{\delta M^2}{\Gamma_{N_1}^2}\big)} + \frac{1-4\mathrm{i} \frac{\delta M}{\Gamma_{N_2}}}{\Gamma_{N_2} \big(1+4 \frac{\delta M^2}{\Gamma_{N_2}^2}\big)}\right)\big] f_{ij}(M_N) \nonumber \\
	& + \left(i \Leftrightarrow j\right) \Big)
	\label{LNV plusplus}
\end{align}
The charged conjugate process $M_1^-\to\ell_i^-\ell_j^-\pi^+$ can be obtained  by replacing $K_R$ to $K_R^*$ in Eq.~\ref{LNV plusplus}:
\begin{align}
	\Gamma^{\text{LNV},--}_{ij}=\Gamma^{\text{LNV},++}_{ij}\left(K_R \to K_R^*\right)
	\label{LNV minusminus}
\end{align}
For  LNC,  this  takes the following form,
\begin{align}
	\Gamma^{\text{LNC}}_{ij}&=\pi\beta\Big(|(K_R)_{1i}|^2 |(K_R)_{1j}|^2 \frac{1}{\Gamma_{N_1} \big(M_N-\frac{\delta M}{2}\big)} h_{ij}\big(M_N-\frac{\delta M}{2}\big) + 
	|(K_R)_{2i}|^2 |(K_R)_{2j}|^2 \nonumber \\
	& \frac{1}{\Gamma_{N_2} \big(M_N+\frac{\delta M}{2}\big)} h_{ij}\big(M_N+\frac{\delta M}{2}\big) + \frac{1}{M_N} \text{Re}\bigg[(K_R)_{1i}^*(K_R)_{1j}(K_R)_{2i} (K_R)_{2j}^* \nonumber \\
	&\left(\frac{1-4\mathrm{i} \frac{\delta M}{\Gamma_{N_1}}}{\Gamma_{N_1} \big(1+4 \frac{\delta M^2}{\Gamma_{N_1}^2}\big)} + \frac{1-4\mathrm{i} \frac{\delta M}{\Gamma_{N_2}}}{\Gamma_{N_2} \big(1+4 \frac{\delta M^2}{\Gamma_{N_2}^2}\big)}\right)\bigg] h_{ij}(M_N) \Big)
	\label{LNC simplified}
\end{align}
The interference between the two RH neutrino $N_1$ and $N_2$ contributions, along with a non-zero CP phase $\phi$  can lead to a difference between the LNV process $M_1^+\to\ell_i^+\ell_j^+\pi^-$ and charge conjugate process $M_1^-\to\ell_i^-\ell_j^-\pi^+$. This can be quantified by the following definition, 
\begin{align}
	A_{CP}^{ij}=\frac{\Gamma^{\text{LNV},++}_{ij}-\Gamma^{\text{LNV},--}_{ij}}{\Gamma^{\text{LNV},++}_{ij}+\Gamma^{\text{LNV},--}_{ij}}\hspace{0.5cm},(i,j = e,\mu)
	\label{ACP}
\end{align}
Using Eqs.~\ref{LNV plusplus}, \ref{LNV minusminus} and \ref{KRmatrix},  and using nearly-degenerate limit of the masses and decay widths,  $M_{N_1}\approx M_{N_2}\equiv M_N$~\footnote{The mass difference $\delta M\ne 0$, however,  $\delta M\ll M_N$.}, $\Gamma_{N_1}\approx\Gamma_{N_2}\equiv\Gamma_N$, $A_{CP}^{ij}$ takes the following simple forms,
\begin{align}
	&A_{CP}^{ii}\approx \frac{4x \sin 2\phi \sin^2 2\theta}{(1+4x^2)(2\cos^4\theta+2\sin^4\theta)+\cos 2\phi\sin^2 2\phi}\,\, \text{with}\,i=e,\mu. \\
	& A_{CP}^{ij}\approx \frac{-4x \sin 2\phi }{(1+4x^2-\cos 2\phi)}, \,\,\text{where}\,i\ne j. 
\end{align}
In the above,  $x=\frac{\delta M}{\Gamma_N}$. For a  fixed $\theta$ and $x$, the above equations lead to maximum value of  $A_{CP}$, if   $\phi=\frac{\pi}{4}$. 
\begin{figure*}
	\centering
	\includegraphics[height=2.5in]{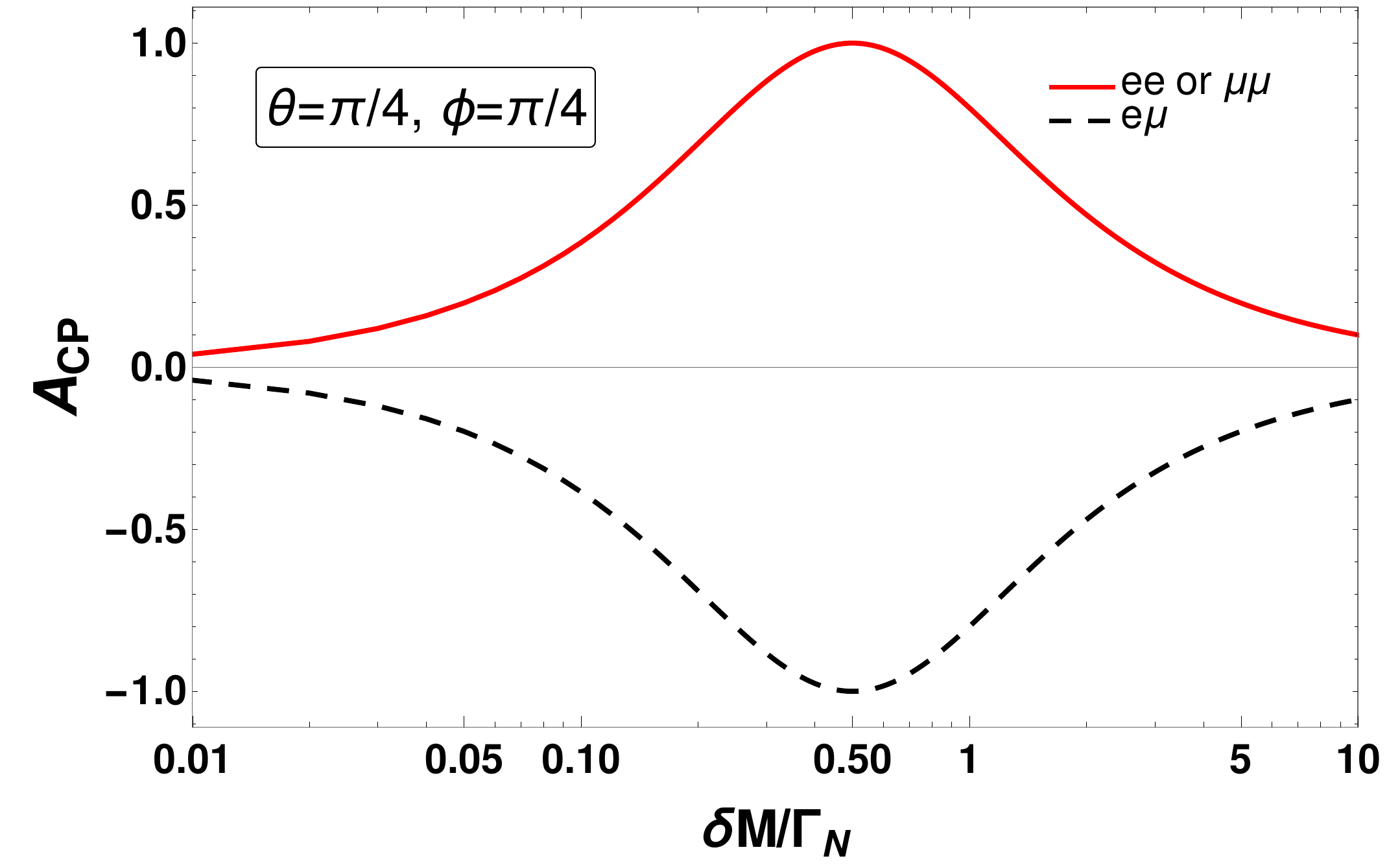}
	\caption{Variation  of $A_{\text{CP}}$ as a function of $\frac{\delta M}{\Gamma_N}$, for $\theta=\frac{\pi}{4}$ and $\phi=\frac{\pi}{4}$. Red solid line represents  $e^+e^+$ or $\mu^+\mu^+$ channel, whereas black dashed line represents  $e^+\mu^+$ channel. For this figure, we consider $K^+$ as parent meson.  For $B^+$ meson,  the figure  is similar, and hence, we do not show this explicitly.  }
	\label{fig:ACP}
\end{figure*} 
In Fig.~\ref{fig:ACP},  we show the variation  of the LNV - CP asymmetry as a function of $\frac{\delta M}{\Gamma_N}$ for the  channels $e^+e^+$, $\mu^+\mu^+$ and $e^+\mu^+$,  where we fix $\theta=\frac{\pi}{4}$ and $\phi=\frac{\pi}{4}$. We see from this figure,  that the $A_{CP}^{ee\,\text{or}\,\mu\mu}$ and $A_{CP}^{e\mu}$ have a complementary behaviour as a function of  $\frac{\delta M}{\Gamma_N}$.   
In order to discuss the impact of the $CP$ phase and mixing angle on the LNV width, we further define one more asymmetry  $\tilde{R}$ as,
\begin{align}
	\tilde{R}_{ij}&=\frac{\Gamma^{\text{LNV}}_{K^{+}\to \ell_i^+\ell_i^+\pi^-}-\Gamma^{\text{LNV}}_{K^{+}\to \ell_i^+\ell_j^+\pi^-}}{\Gamma^{\text{LNV}}_{K^{+}\to \ell_i^+\ell_i^+\pi^-}+\Gamma^{\text{LNV}}_{K^{+}\to \ell_i^+\ell_j^+\pi^-}}\,\,\text{where}\,\,i\ne j.
	\label{eq:R1}
\end{align}
\begin{figure*}
	\centering
	\includegraphics[height=2.5in]{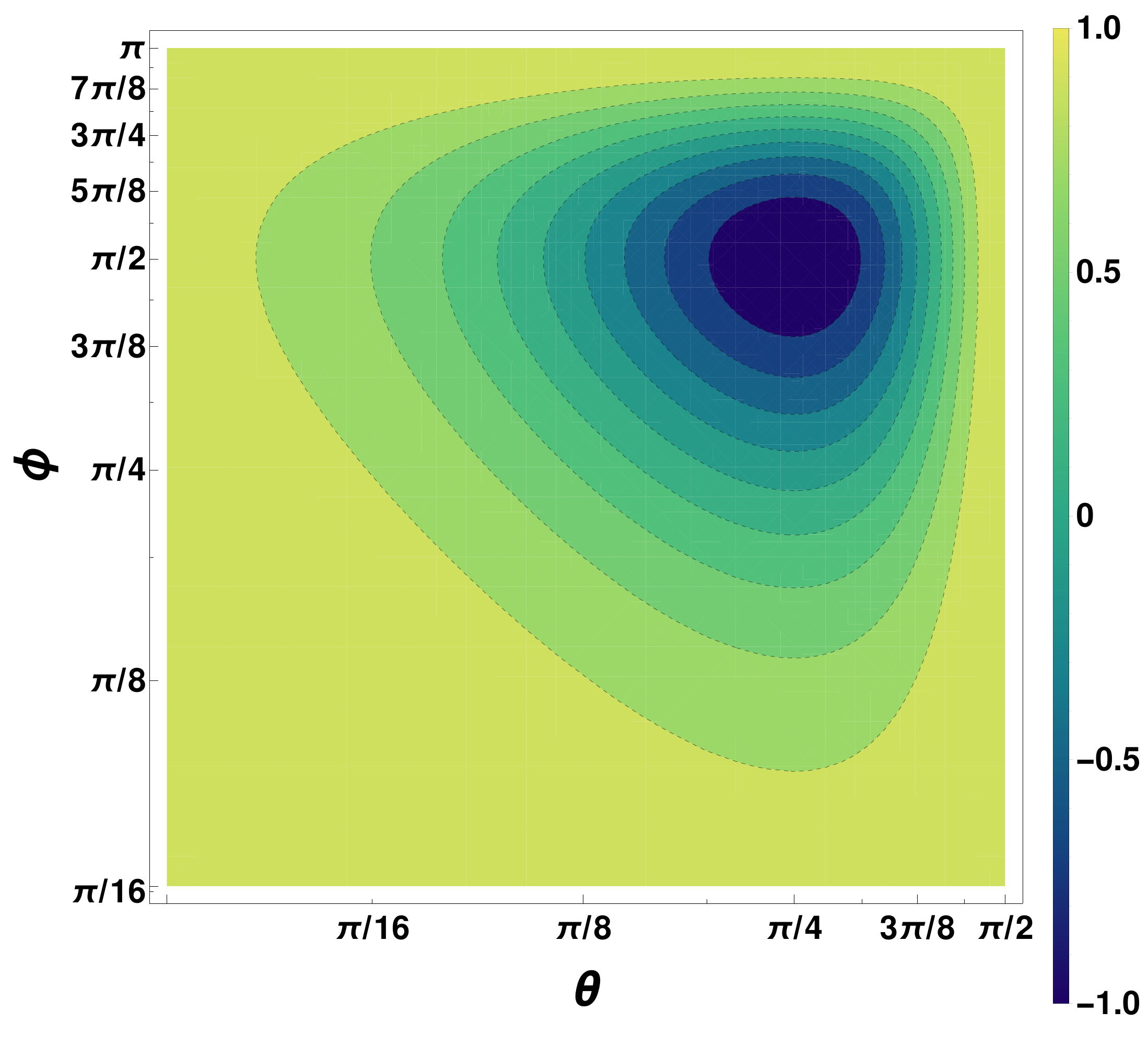}
	\caption{Variation of 
		$\tilde{R}_{e\mu}$~ with the variation of angle and phase~($\theta$,$\phi$). $\tilde{R}$ varies between  $[-1:1]$.}
	\label{fig:asym}
\end{figure*} 
Fig.~\ref{fig:asym} shows the variation of  $\tilde{R}_{e\mu}$~ for different values of mixing angle $\theta$ and phase $\phi$. It is evident from this figure that  $\tilde{R}_{e\mu}$ varies between  $[-1:1]$. Note that  $|\tilde{R}_{e\mu}|$ have complimentary nature w.r.t the angular variables {and their analytical expressions can be found in Appendix~\ref{app:R}}. Similar results can also be obtained for $B$ decays.  This is to note that,  the two asymmetries  $A_{CP}$, and $\tilde{R}_{e\mu}$ do not contain any SM contribution and hence, serve as  a clear indication of  new physics.


\section{Result for two generation  with the full $3\times 3$ mixing matrix $K_R$}
\label{Three generation case}
\hspace{0.5cm} 
In the previous section, we considered a simple form of the $K_R$ matrix, which contains only one angle and one phase.  In general the mixing matrix $K_R$ is a $3\times3$ unitary matrix, with more number of parameters. We consider a special scenario, where the $K_R$ matrix is identical with the PMNS mixing matrix in the light neutrino sector. The additional CP phases can give sizeable contributions in both the LNV  and the LNC  meson decays, in particular give a different result in the LNC $e\mu$ mode, as compared to the previous scenario, where we considered  a simple form  for $K_R$. 
We consider  the following parameterisation of $K_R$,

\begin{equation}
	K_{R}=\begin{bmatrix}
		c_{12} c_{13}  &  -c_{23} s_{12} - e^{-i\delta} c_{12} s_{13} s_{23}  &  -e^{-i\delta} c_{12} c_{23} s_{13} + s_{12} s_{23}  \\
		e^{-i\alpha_1} c_{13} s_{12}  & e^{-i\alpha_1} c_{12} c_{23} - e^{-i(\alpha_1+\delta)} s_{12} s_{13} s_{23}  & -e^{-i(\alpha_1+\delta)} c_{23} s_{12} s_{13} - e^{-i\alpha_1} c_{12} s_{23}\\
		e^{-i\alpha_2} s_{13}  &  e^{-i(\alpha_2+\delta)} c_{13} s_{23}  &   e^{-i(\alpha_2+\delta)} c_{13} c_{23}
	\end{bmatrix}
	\label{eq:3KR}
\end{equation} 
where $s_{12},s_{13}, s_{23}$ refer to $\sin \theta_{12},\sin \theta_{13}, \sin \theta_{23}$, respectively, and $(\alpha_1,\alpha_2 )\in (0,\pi)$, $\delta\in (0,2\pi)$. Considering such a $3\times3$ matrix the number of parameters increase and the result is much more complicated as compared to the former scenario.
\begin{figure*}
	\centering
	\includegraphics[width=0.49\textwidth]{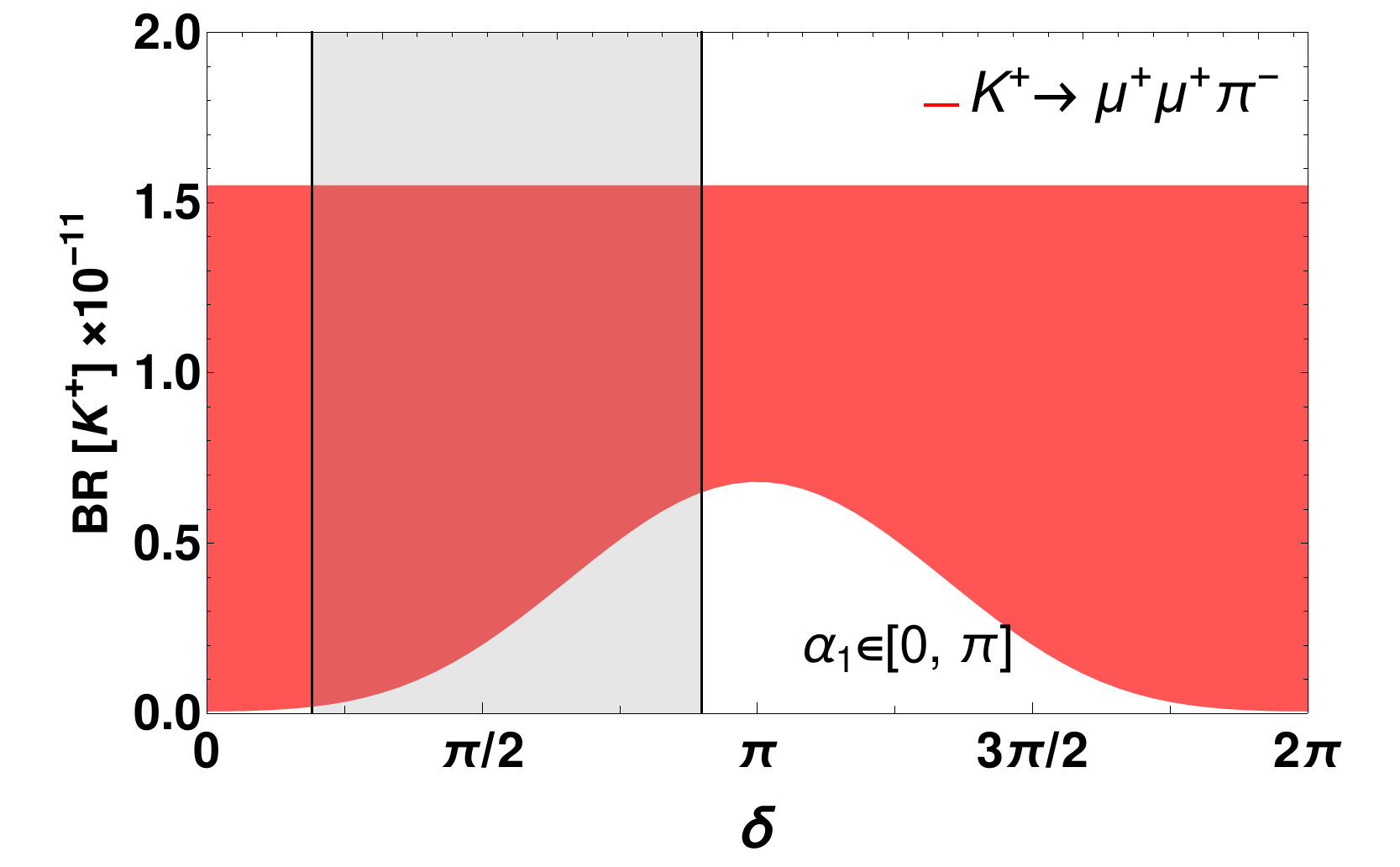}
	\includegraphics[width=0.45\textwidth]{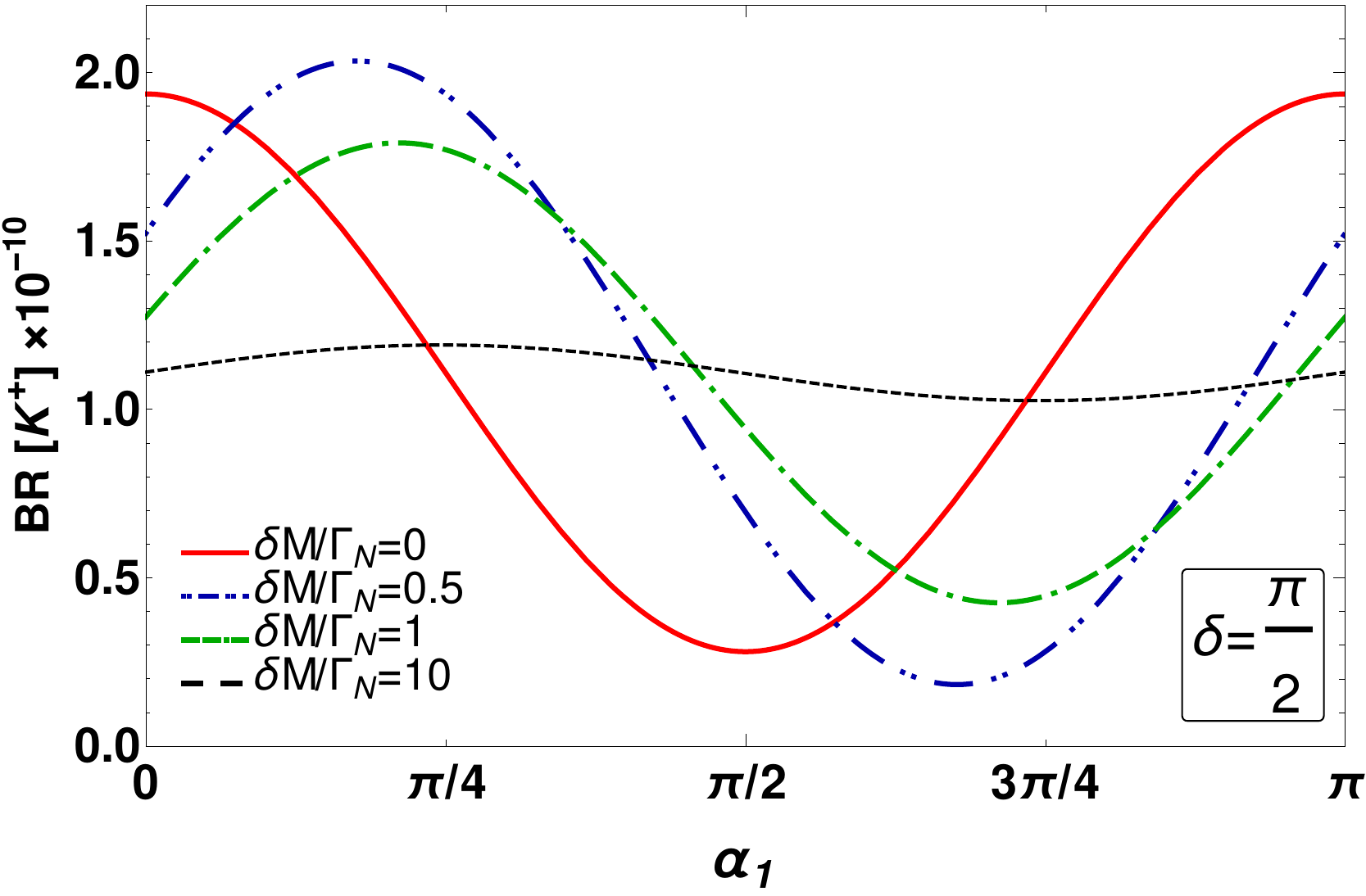}
	\caption{Left panel: variation of the branching ratio of $K^+ \to  \mu^+ \mu^+ \pi^- $ decay mode with the variation of the CP violating phase $\delta$. The Majorana phases $\alpha_{1}$ have been varied in the denoted range. 
		The shaded region is disallowed from the recent results from T2K. Note that, excluded region from T2K for $\delta$ is given with the convention $[-\pi:\pi]$, which we have translated in the range $[0:2\pi]$,  to be consistent with our convention. See text for more details. Right panel: variation of the branching ratio of $K^+ \to  e^+ e^+ \pi^- $ decay mode with the variation of the Majorana phase $\alpha_1$. The red solid, blue dashed-double-dotted, green dashed-dotted, and black dashed
		lines represent four different $\frac{\delta M}{\Gamma_N}$ ratios (0,0.5,1,10). }
	\label{lnv3}
\end{figure*} 

Note that, from Eq.~\ref{eq:3KR}, one would obtain Eq.~\ref{KRmatrix}, by considering $\theta_{13},\theta_{23}=0$, $\alpha_1=\delta$ and $\alpha_2=0$. In Fig.~\ref{lnv3}, we show the variation of the LNV  branching ratio $K^+ \to \mu^+ \mu^+ \pi^-$   with the variation of the Dirac CP phase $\delta$. We have checked the RH neutrino contribution in the LNC branching ratio $K^+ \to \mu^+ \mu^- \pi^+$ is independent of the Dirac phase $\delta$, and hence we do not show that explicitly. We consider best fit values of the mixing angles $\theta_{12}, \theta_{13}, \theta_{23}$ for inverted mass hierarchy~\cite{deSalas:2020pgw}, and  vary the Majorana phases $\alpha_{1}$ in between $0-\pi$. As an illustrative example, we consider  the mass splitting $\delta M=0$ and $M_N=0.38$ GeV. We also show the  regions of $\delta$, disallowed by the T2K neutrino oscillation experiment for inverted mass hierarchy. The red region represents the LNV branching ratio,  which shows clearly a non-trivial variation w.r.t $\alpha_1$ and $\delta$. This is true that LNC mode should be independent of any Majorana phase\footnote{Note that although the LNC decay modes do not depend on the Majorana phases, they can in principle depend on the Dirac phase $\delta$. We have verified that with our chosen $K_R$ matrix, there is no dependency of $K^{+}\to \mu^{+}\mu^{-}\pi^{+}$ on the Dirac CP phase,  for $\delta M=0$.}. 
We further find that, the decay mode $K^{+}\to e^+ e^+\pi^-$   depends  only on Majorana phase $\alpha_1$, and not on the Dirac phase $\delta$. The LNC mode with the same lepton flavor is also independent of the Dirac CP phase, which happens due to the chosen form of $K_R$ in Eq.~\ref{eq:3KR}. This can be verified  substituting  the explicit form of the $K_R$ matrix~(Eq.~\ref{eq:3KR}) in Eq.~\ref{LNV plusplus}. 
In the right  panel of Fig.~\ref{lnv3} we show the variation of the LNV  branching ratio $K^+ \to e^+ e^+ \pi^-$  with respect to Majorana phase $\alpha_1$ for different $\frac{\delta M}{\Gamma_N}$ ratios. The RH neutrino contribution in the LNC branching ratio $K^+ \to e^+ e^- \pi^+$ is independent of Majorana phase, and hence we do not show that explicitly. 
\\\\
\begin{figure*}
	\centering
	\includegraphics[width=0.45\textwidth]{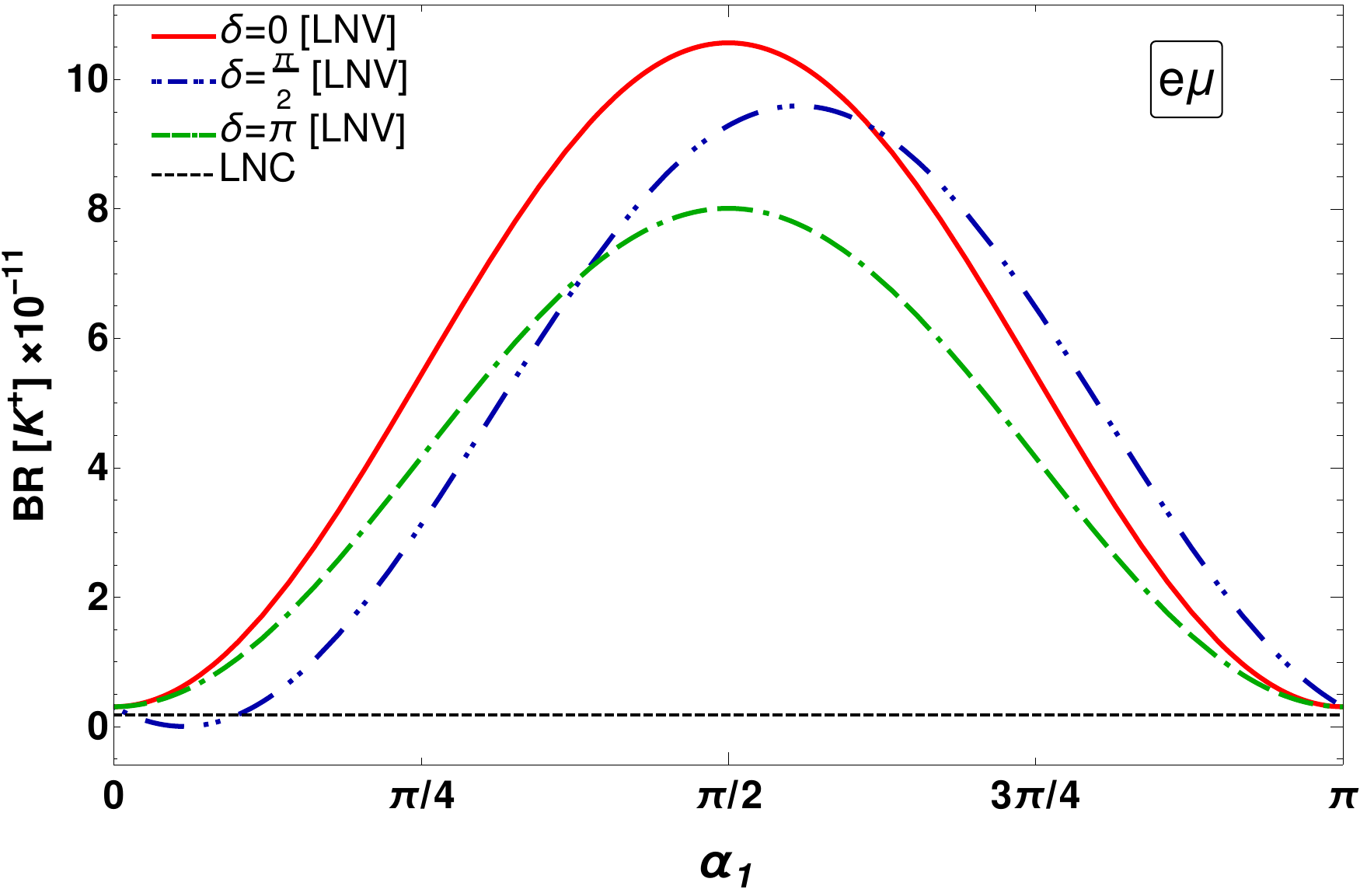}
	\includegraphics[width=0.49\textwidth]{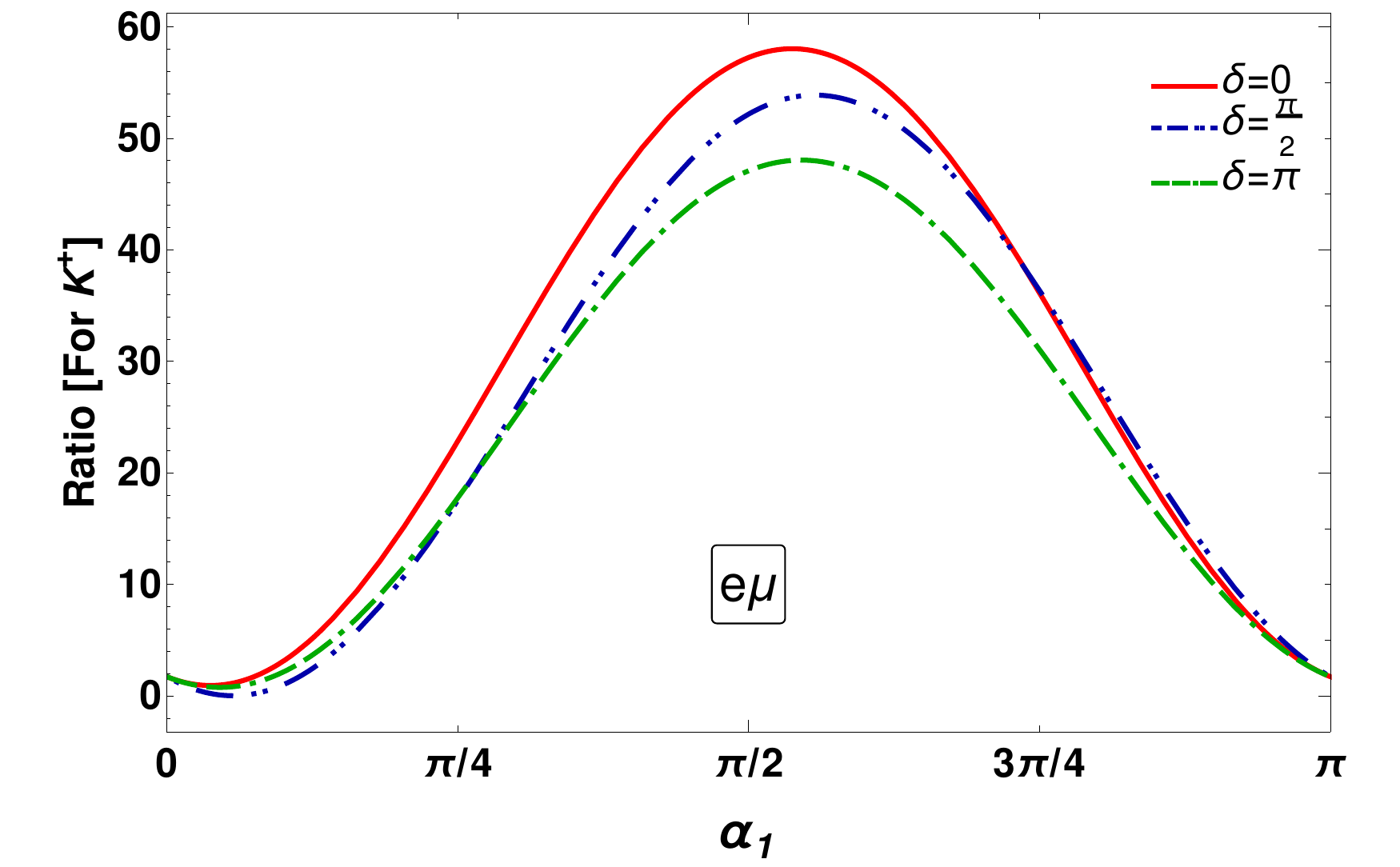}
	\caption{Left panel: variation of the branching ratio of $K^+ \to  e^+ \mu^+ \pi^- $, and $K^+ \to e^+ \mu^- \pi^+$ decay modes, with the variation of the Majorana phase $\alpha_1$ and for different Dirac phases $\delta=\frac{\pi}{4}$~(solid red), $\frac{\pi}{2}$~(blue dashed-double-dotted) and $\frac{3\pi}{4}$~(green dashed-dotted). Right panel: the ratio of LNV and LNC branching ratios in the $e\mu$ mode.   }
	\label{lnv3ee}
\end{figure*} 
Finally, we elaborate on the LNC $e\mu$ channel, which we found to be vanishing  for a simple form of $K_R$, shown in Eq.~\ref{KRmatrix}, and for degenerate RH neutrino masses.  With the approximation $\delta M=0$ and $\Gamma_{N_1}=\Gamma_{N_2}\equiv \Gamma_N$, Eq.~\ref{LNC simplified} can be simplified and the LNC decay rate $\Gamma_{e\mu}^{\text{LNC}}$ is proportional to the following expression, 
\begin{align}
	\tilde{\Gamma}_{e\mu}= |(K_R)_{11}|^2 |(K_R)_{12}|^2 + |(K_R)_{21}|^2 |(K_R)_{22}|^2 + 2\text{Re}\big[(K_R)_{11}^* (K_R)_{12} (K_R)_{21} (K_R)_{22}^*\big]
\end{align}
This is to note that,  for the preceding scenario, our choice of  $2\otimes 2$ sub-block of $K_R$ matrix~(Eq.~\ref{KRmatrix}) as  a unitary matrix leads to $\tilde{\Gamma}_{e\mu}=0$. This occurs due to the  unitary properties  of  $K_R$.  Hence, for this simplified scenario,  the ratio $\Gamma_{e\mu}^{\text{LNV}}/\Gamma_{e\mu}^{\text{LNC}}$ is ill-defined. However,  this changes if one considers  full $3\otimes 3$ matrix,  as in this case $2\otimes 2$ sub-block itself is not unitary. We find that, for $3 \times 3$ case, and using Eq.~\ref{eq:3KR},  for degenerate RH neutrinos, $\tilde{\Gamma}_{e\mu}=c_{13}^2 s_{13}^2 s_{23}^2$. This certainly is   non-zero for  the best fit values of PMNS mixing angles. We stress that, in this case it is possible to define the ratio $\Gamma_{e\mu}^{\text{LNV}}/\Gamma_{e\mu}^{\text{LNC}}$ and the ratio can be large as the best fit value of $s_{13}$ is small.
In the right panel of Fig.~\ref{lnv3ee} we show the variation of the LNV and LNC branching ratio $K^+ \to e^+ \mu^+ \pi^-$ and $K^+ \to e^+ \mu^- \pi^+$  with respect to Majorana phase $\alpha_1$ for different choices of Dirac CP phase $\delta$. In the right panel,  we show the ratio of LNV and LNC mode. We see that LNV mode $K^+\to e^+ \mu^+ \pi^-$  depends on both Majorana and Dirac phases even for $\delta M=0$. For degenerate RH neutrinos, the LNC $e\mu$ mode is  independent of Dirac phase (left panel). Contrary to the results obtained in Section.~\ref{lnc2}, the  LNC $e\mu$ mode in this scenario has  a large branching ratio $\sim 10^{-11}$. The ratio of LNV and LNC mode in this case is order of magnitude large ($\sim 40-60$) compared to the earlier scenario,  due to relative suppression of LNC rate for a small $\theta_{13}$.  

\section{Conclusion}
The LRSM is one of the most appealing models, that accommodate the embedding of RH neutrinos in a natural way.  The model contains RH neutrinos, RH gauge bosons, and other BSM states that can offer distinctive experimental signatures. In this work, we explore the LNV and LNC semi-leptonic meson decays, mediated by the RH neutrinos and RH gauge bosons, and  quantify the effect of interference in these decays due to the presence of  at least two degenerate/quasi-degenerate RH neutrino states. The RH neutrinos of masses in the range of MeV- few GeV can give a  resonance enhancement in these rates. We consider few specific decay modes of $K^+$ and $B^{+}$ meson to $K^{+}/B^{+} \to e^{+} e^{+} \pi^{-}$ and $K^{+}/B^{+} \to e^{+} \mu^{+} \pi^{-}, e^{+} \mu^{-} \pi^{+}$, and analyse the effect of interference in detail.  We consider two RH neutrinos to be degenerate/nearly  degenerate in mass with their  masses in the range 
$M_{N_1} \sim M_{N_2} \sim 380 $ MeV, and $M_{N_1} \sim M_{N_2} \sim 2 $ GeV, relevant for the LNV and LNC $K^{+}$ and $B^{+}$ meson decays. In case of a single generation of RH neutrino,  the LNV and LNC decay contributions  are the same, leading to the ratio of these two to be unity. We find that,  in the presence of interference between the two RH neutrino contributions in the amplitude, the contributions of the RH neutrinos in the LNV and LNC decay rates can differ widely, which leads to the ratio to be different than unity.

We first consider a  simplistic scenario, where two RH neutrinos contribute to  meson decay, with a simple  RH neutrino mixing matrix  consisting of one angle $\theta$ and one phase $\phi$. We show that the LNV branching ratio, in particular the interference term, depends on both the angle and phase. The RH neutrino contribution in the LNC mode  only depends on the angle, however, is  insensitive to the phase. Few comments are in order:
\begin{itemize}
	\item The  channels $K^{+} \to e^{+} e^{+} \pi^{-}$ and $K^{+} \to e^{+} \mu^{+} \pi^{-}$   offer a complimentary nature  in their predicted branching ratios.  This  holds true for other meson decays as well. 
	\item We find that overall, the LNV branching ratios $K^{+}/B^{+} \to e^{+} e^{+} \pi^{-}$ have a large variation w.r.t the variation in $\theta$ and $\phi$. The decay rates are highly suppressed due to destructive interference at $\phi \sim \pi/2$, and $\theta \sim \pi/4$. For both the processes, the predicted branching ratios can reach maximum $ \mathcal{O}(10^{-10}),  \mathcal{O}(10^{-12})$ range with a RH gauge boson mass $22$ TeV and $5 $ TeV, respectively. 
	\item For the different lepton flavor in the final states, $K^{+} \to e^{+} \mu^{-} \pi^{+}$ mode, the LNC branching ratio is Br $\sim 10^{-11}$ for non-degenerate RH neutrinos with 
	$\frac{\delta M}{\Gamma_N}$ $\sim$ 10 .  The branching ratio for the LNV decay mode,  $K^+ \rightarrow e^+ \mu^+\pi^-$, can be as high as ${\cal O} (10^{-10})$.
	
\end{itemize}

We also explore a scenario where the mixing matrix in the RH neutrino sector has the same form as the PMNS mixing in the neutrino sector. We consider the best fit values of neutrino mixing parameters, and vary the phases. We find that, the Majorana and Dirac phases in this case  have a large impact on the branching ratio for the LNV channel $K^{+} \to \mu^{+} \mu^{+} \pi^{-}$, as well as $K^{+} \to e^{+} \mu^{+} \pi^{-}$. The same LNV decay mode, but  with $e^+ e^+$ in the final state  is independent of the Dirac phase, and depends only on the Majorana phase. For the LNC modes there is no dependency on the Majorana CP phase.   Choosing degenerate RH neutrino masses, we find that the $e^+ \mu^-$ mode also becomes independent of the Dirac CP phase. We further find that, in this case, branching ratio of $K^{+} \to e^{+}\mu^- \pi^+$ can be sizeable, even for RH neutrinos with degenerate masses. 
The ratio between LNV and LNC decays now takes values over a wider range.  
Any future measurement of the ratio, different than unity will indicate possible interference effect. Furthermore,  observations of the LNV signature $K^{+} \to e^+ \mu^+ \pi^-$ together with LNC signature $K^{+} \to e^+ \mu^- \pi^-$ will indicate  a non-trivial form of $K_R$ matrix.  
Hence,  the LNV mode as well as the LNC mode with different lepton flavors can serve as a smoking gun signal for underlying new physics.   


\section*{Acknowledgements}
SM acknowledges the support from Spanish grant FPA2017-85216-P (AEI/FEDER, UE), PROMETEO/2018/165 (Generalitat Valenciana) and the Spanish Red Consolider MultiDark FPA2017-90566-REDC. The work of RMG is supported by the Department of Science and Technology, India under Grant No. SR/S2/JCB-64/2007. MM and RMG thank the {\it Indo-French Centre for the Promotion of Advanced Research} for the  support  (project no: 6304-2). 
\section*{Appendix} 
\section{Details of LNV and LNC calculations}
\label{Details of LNV and LNC calculations}
\subsection{LNV}
Amplitude from $N_a$ contribution:
\begin{equation}
	\mathcal{M}^{\text{LNV},a}_{ij} = 2G_F^2 V^{\text{CKM}}_{M_1}V^{\text{CKM}}_{M_2}f_{M_1}f_{M_2}\Big(\frac{M_{W_L}}{M_{W_R}}\Big)^4 (K_R^*)_{ai}(K_R^*)_{aj}M_{N_a}\frac{\bar{u}(k_1)\slashed{p} \slashed{k_3} P_R v(k_2)}{q^2-M_{Na}^2+i\Gamma_{Na}M_{Na}},
\end{equation}
Amplitude for leptonic exchange diagram:    
\begin{equation}
	\mathcal{M}^{\text{LNV},a}_{ji} = 2G_F^2 V^{\text{CKM}}_{M_1}V^{\text{CKM}}_{M_2}f_{M_1}f_{M_2}\Big(\frac{M_{W_L}}{M_{W_R}}\Big)^4 (K_R^*)_{aj}(K_R^*)_{ai}M_{N_a}\frac{\bar{u}(k_2)\slashed{p} \slashed{k_3} P_R v(k_1)}{q^2-M_{Na}^2+i\Gamma_{Na}M_{Na}},
\end{equation}
\begin{align}
	|\mathcal{M}_{ij}^{\text{LNV},a}|^2=\kappa M_{N_a}^2 (K_R^*)_{ai}(K_R^*)_{aj}(K_R)_{ai}(K_R)_{aj}\frac{Tr[(\slashed{k_2}-m_2)\slashed{k_3}\slashed{p}P_L(\slashed{k_1}+m_1)\slashed{p}\slashed{k_3}P_R]}{(q^2-M_{N_a}^2)^2+\Gamma_{N_a}^2 M_{N_a}^2},
\end{align}
\begin{align}
	|\mathcal{M}_{ji}^{\text{LNV},a}|^2=\kappa M_{N_a}^2 (K_R^*)_{aj}(K_R^*)_{ai}(K_R)_{aj}(K_R)_{ai}\frac{Tr[(\slashed{k_1}-m_1)\slashed{k_3}\slashed{p}P_L(\slashed{k_2}+m_2)\slashed{p}\slashed{k_3}P_R]}{(q^2-M_{N_a}^2)^2+\Gamma_{N_a}^2 M_{N_a}^2},
\end{align}
where $\kappa=4G_F^4 (V^{\text{CKM}}_{M_1})^2(V^{\text{CKM}}_{M_2})^2f_{M_1}^2f_{M_2}^2\Big(\frac{M_{W_L}}{M_{W_R}}\Big)^8$.
The interference terms are,
\begin{align}
	\Big(\mathcal{M}_{ij}^{\text{LNV},a}\Big)^{\dagger}\mathcal{M}^{\text{LNV},b}_{ij}& = \kappa M_{N_a}M_{N_b}(K_R)_{ai}(K_R)_{aj}(K_R^*)_{bi} (K_R^*)_{bj}\\ \nonumber
	&\times\frac{Tr[(\slashed{k_2}-m_2)\slashed{k_3}\slashed{p}P_L(\slashed{k_1}+m_1)\slashed{p}\slashed{k_3}P_R]}{(q^2-M_{N_a}^2-i\Gamma_{N_a}M_{N_a})(q^2-M_{N_b}^2+i\Gamma_{N_b}M_{N_b})}
\end{align}
\begin{align}
	\Big(\mathcal{M}_{ji}^{\text{LNV},a}\Big)^{\dagger}\mathcal{M}^{\text{LNV},b}_{ji}& = \kappa M_{N_a}M_{N_b}(K_R)_{aj}(K_R)_{ai}(K_R^*)_{bj} (K_R^*)_{bi}\\ \nonumber
	&\times\frac{Tr[(\slashed{k_1}-m_1)\slashed{k_3}\slashed{p}P_L(\slashed{k_2}+m_2)\slashed{p}\slashed{k_3}P_R]}{(q^2-M_{N_a}^2-i\Gamma_{N_a}M_{N_a})(q^2-M_{N_b}^2+i\Gamma_{N_b}M_{N_b})}
\end{align}
The traces are
\begin{align}
	Tr_{21}=Tr[(\slashed{k_2}-m_2)\slashed{k_3}\slashed{p}P_L(\slashed{k_1}+m_1)\slashed{p}\slashed{k_3}P_R]=2k_1.k_2 k_3.k_3 p.p -4k_1.p k_2.p k_3.k_3\\ \nonumber
	-4k1.k_3 k_2.k_3 p.p+8k_1.p k_2.k_3 k_3.p
\end{align}
\begin{align}
	Tr_{12} =Tr[(\slashed{k_1}-m_1)\slashed{k_3}\slashed{p}P_L(\slashed{k_2}+m_2)\slashed{p}\slashed{k_3}P_R]=2k_1.k_2 k_3.k_3 p.p -4k_1.p k_2.p k_3.k_3\\ \nonumber
	-4k_1.k_3 k_2.k_3 p.p+8k_1.k_3 k_2.p k_3.p
\end{align}
As the RH neutrinos are produced onshell and having $\Gamma_{N_a}\ll M_{N_a}$, hence in our decay width calculation we can use narrow width approximation,
\begin{align}
	\frac{1}{(q^2-M_{N_a}^2)^2+\Gamma_{N_a}^2 M_{N_a}^2}\longrightarrow \frac{\pi}{\Gamma_{N_a}M_{N_a}}\delta(q^2-M_{N_a}^2)
\end{align}
Using the narrow-width approximation, the product of propagators in the interference term can be written as
\begin{align}
	&\frac{1}{\big[(q^2-M_{N_1}^2)^2-i\Gamma_{N_1} M_{N_1}\big] \big[(q^2-M_{N_2}^2)^2+i\Gamma_{N_2} M_{N_2}\big]}  = -\frac{i (M_{N_2}^2-M_{N_1}^2)\pi \delta (q^2-M_{N_2}^2)}{(M_{N_2}^2-M_{N_1}^2)^2+\Gamma_{N_1}^2 M_{N_1}^2}\nonumber \\
	& + \frac{i (M_{N_1}^2-M_{N_2}^2)\pi \delta (q^2-M_{N_1}^2)}{(M_{N_1}^2-M_{N_2}^2)^2+\Gamma_{N_2}^2 M_{N_2}^2} + \frac{\Gamma_{N_1}M_{N_1}\pi\delta (q^2-M_{N_2}^2)}{2\big[(M_{N_2}^2-M_{N_1}^2)^2+\Gamma_{N_1}^2 M_{N_1}^2\big]} + \frac{\Gamma_{N_2}M_{N_2}\pi\delta (q^2-M_{N_1}^2)}{2\big[(M_{N_1}^2-M_{N_2}^2)^2+\Gamma_{N_2}^2 M_{N_2}^2\big]}
\end{align}
Finally we can write the decay width as
\begin{multline}
	d\Gamma^{\text{LNV}}=
	\Big(1-\frac{\delta_{ij}}{2}\Big)\frac{1}{2m}\hspace{0.2cm}\frac{\kappa}{(2\pi)^5}\frac{\pi}{2}\frac{\pi}{4} \hspace{0.2cm}d\cos\theta\times\\ 
	\Bigg(\sum_{a=1}^{2}\Big[ \frac{\pi M_{N_a}}{\Gamma_{N_a}} |{K_R}_{ai}|^2 |{K_R}_{aj}|^2 Tr_{21}\lambda^{1/2}\Big(1,\frac{m_i^2}{m^2},\frac{M_{N_a}^2}{m^2}\Big)\lambda^{1/2}\Big(1,\frac{m_j^2}{M_{N_a}^2},\frac{m_3^2}{M_{N_a}^2}\Big)\Big]+\\
	\pi M_N \text{Re}\Big[{K_R}_{1i}{K_R}_{1j}{K_R^*}_{2i}{K_R^*}_{2j} \bigg(\frac{1-4i\frac{\delta M}{\Gamma_{N_1}}}{\Gamma_{N_1}(1+4\frac{\delta M^2}{\Gamma_{N_1}^2})}+\frac{1-4i\frac{\delta M}{\Gamma_{N_2}}}{\Gamma_{N_2}(1+4\frac{\delta M^2}{\Gamma_{N_2}^2})}\bigg)\Big]Tr_{21}\\
	\lambda^{1/2}\Big(1,\frac{m_i^2}{m^2},\frac{M_{N}^2}{m^2}\Big)\lambda^{1/2}\Big(1,\frac{m_j^2}{M_{N}^2},\frac{m_3^2}{M_{N}^2}\Big)+(i\leftrightarrow j)
	\Bigg)\\
	\label{LNV Calculation}
\end{multline}
\subsection{LNC \label{lnc2}}
Following the similar procedure the decay width for opposite sign leptons can be calculated as,
\begin{align}
	\mathcal{M}^{\text{LNC},a}_{ij} = 2G_F^2V^{\text{CKM}}_{M_1}V^{\text{CKM}}_{M_2}f_{M_1}f_{M_2}\Big(\frac{M_{W_L}}{M_{W_R}}\Big)^4 (K_R)_{ai}(K_R^*)_{aj}\frac{\bar{u}(k_1)\slashed{p}\slashed{q} \slashed{k_3} P_R v(k_2)}{q^2-M_{N_a}^2+i\Gamma_{N_a}M_{N_a}}
\end{align}
\begin{align}
	|\mathcal{M}^{\text{LNC},a}_{ij}|^2 = \kappa (K_R^*)_{ai}(K_R)_{aj}(K_R)_{ai}(K_R^*)_{aj}\frac{Tr\Big[(\slashed{k_2}-m_2)\slashed{k_3}\slashed{q}\slashed{p}P_R(\slashed{k_1}+m_1)\slashed{p}\slashed{q}\slashed{k_3}P_R\Big]}{(q^2-M_{N_a}^2)^2+\Gamma_{N_a}^2M_{N_a}^2} 
\end{align}
where
\begin{multline}
	Tr_{21}^{\text{LNC}}=Tr\Big[(\slashed{k_2}-m_2)\slashed{k_3}\slashed{q}\slashed{p}P_R(\slashed{k_1}+m_1)\slashed{p}\slashed{q}\slashed{k_3}P_R\Big]=\\4k_1.qk_2.qk_3.k_3 p.p-8k_1.qk_2.k_3k_3.qp.p-8k_1.pk_2.qk_3.k_3p.q+16k_1.pk_2.k_3k_3.qp.q+\\4k_1.pk_2.pk_3.k_3q.q-8k1.pk_2.k_3k_3.pq.q+4k_1.k_3k_2.k_3p.pq.q-2k_1.k_2k_3.k_3p.pq.q
\end{multline}
We can write the LNC decay width as
\begin{multline}
	d\Gamma^{\text{LNC}}=\frac{1}{2m}\hspace{0.2cm}\frac{\kappa}{(2\pi)^5}\frac{\pi}{2}\frac{\pi}{4} \hspace{0.2cm}d\cos\theta\times\\ 
	\sum_{a=1}^2\Big[\frac{\pi}{\Gamma_{N_a}M_{N_a}} |{K_R}_{ai}|^2 |{K^*_R}_{aj}|^2 Tr_{21}^{\text{LNC}}\lambda^{1/2}\Big(1,\frac{m_i^2}{m^2},\frac{M_{N_a}^2}{m^2}\Big)\lambda^{1/2}\Big(1,\frac{m_j^2}{M_{N_a}^2},\frac{m_3^2}{M_{N_a}^2}\Big)\Big]+\\
	\frac{\pi}{M_N}\text{Re}\Big[ {K^*_R}_{1i}{K_R}_{1j}{K_R}_{2i}{K_R^*}_{2j} \bigg(\frac{1-4i\frac{\delta M}{\Gamma_{N_1}}}{\Gamma_{N_1}(1+4\frac{\delta M^2}{\Gamma_{N_1}^2})}+\frac{1-4i\frac{\delta M}{\Gamma_{N_2}}}{\Gamma_{N_2}(1+4\frac{\delta M^2}{\Gamma_{N_2}^2})}\bigg)\Big] Tr_{21}^{\text{LNC}}\\
	\lambda^{1/2}\Big(1,\frac{m_i^2}{m^2},\frac{M_{N}^2}{m^2}\Big)\lambda^{1/2}\Big(1,\frac{m_j^2}{M_{N}^2},\frac{m_3^2}{M_{N}^2}\Big)
	\label{LNC Calculation}
\end{multline}
\section{$R_{ee}$ and $\tilde{R}_{e\mu}$}
\label{app:R}
\begin{align}
	R_{ee}&=\frac{\sin^2 2\theta \sin\alpha_1 (\sin\alpha_1-4x\cos\alpha_1)}{(1+4x^2)(2\cos^4\theta+2\sin^4\theta)+\sin^2 2\theta (\cos^2 \alpha_1+2x\sin\alpha_1)}\\
	\tilde{R}_{e\mu}&=\frac{(1+4x^2)(1+3\cos 4\theta)+6\sin^2 2\theta (\cos 2\alpha+4x \sin 2\alpha)}{(1+4x^2) (5-\cos 4\theta)-2 \sin^2 2\theta (\cos 2\alpha+4x\sin 2\alpha)}
\end{align}
\section{Total decay width of heavy Majorana neutrino $N_j$}
\label{decay width of N}
	Here we present the analytic expression of total decay width of $N_j$ for our chosen mass range $0.14\, \text{GeV}\leq M_{N_j}\leq 6\,\text{GeV}$. In addition to
the SM gauge bosons $W_L$ , $Z$, the gauge bosons $W_R$ , $Z'$
will also now contribute in the decays of RH neutrinos via charged current and neutral
current interactions.
The analytical expressions for different two and three body partial decay widths of the RH neutrinos $N_{i}$ are given as:
\subsection{Two-body decays of N}
\begin{align}
	\Gamma (N_{j}\rightarrow{\ell_i}^{-}P^{+})&=\frac{G_{F}^{2}M_{N_{j}}^{3}}{16\pi}f_{p}^{2}\left|V_{q\bar{q}^{'}}\right|^{2}\bigg(\left|S_{ij}\right|^{2}F_{P}\left(x_{\ell_i},x_{P}\right)+\left|K_{R_{ij}}\right|^{2}\xi_{1}^{4}F_{P}
	\left(x_{\ell_i},x_{P}\right)\\ \nonumber
	& +4Re\left[S_{ij}K_{R_{ij}}\right]\xi_{1}^{2}x_{\ell_i}x_{P}^{2}\lambda^{\frac{1}{2}}\left(1,x_{\ell_i}^{2},x_{P}^{2}\right)\bigg),
\end{align}
where $\ell_1=e,\ell_2=\mu,\ell_3=\tau$ and $P^+=\pi^+,K^+,D^+,D_s^+$.
\begin{align}
	\Gamma\left(N_{j}\rightarrow{\ell_i}^{-}V^{+}\right)&=\frac{G_{F}^{2}M_{N_{j}}^{3}}{16\pi}f_{V}^{2}\left|V_{q\bar{q}^{'}}\right|^{2}\bigg(\left|S_{ij}\right|^{2}F_{V}\left(x_{\ell_i},x_{V}\right)+
	\left|K_{R_{ij}}\right|^{2}\xi_{1}^{4}F_{V}\left(x_{\ell_i},x_{V}\right)\\ \nonumber
	&-12Re\left[S_{ij}K_{R_{ij}}\right]\xi_{1}^{2}x_{\ell_i}x_{V}^{2}\lambda^{\frac{1}{2}}\left(1,x_{\ell_i}^{2},x_{V}^{2}\right)\bigg),
\end{align}
where $\ell_1=e,\ell_2=\mu,\ell_3=\tau$ and $V^+=\rho^+,K^{*+},D^{*+},D_s^{*+}$.
\begin{align}
	\Gamma\left(N_{j}\rightarrow\nu_{\ell_i}P^{0}\right)&=\frac{G_{F}^{2}M_{N_{j}}^{3}}{4\pi}f_{P}^{2}\sum_{i}\left|U_{ij}\right|^{2}\left|S_{ij}\right|^{2}\bigg(K_{P}^{2}
	+K_{P}^{'2}\xi_{2}^{4}-2K_{P}K_{P}^{'}\xi_{2}^{2}\bigg)F_{P}\left(x_{\nu_{\ell}},x_{P}\right),
\end{align}
where $\nu_{\ell_i}$ are the flavor eigenstates $\nu_e,\nu_\mu,\nu_\tau$ and $P^0=\pi^0,\eta,\eta',\eta_c$.
\begin{align}
	\Gamma\left(N_{j}\rightarrow\nu_{\ell_i}V^{0}\right)&=\frac{G_{F}^{2}M_{N_{j}}^{3}}{4\pi}f_{V}^{2}\sum_{i}\left|U_{ij}\right|^{2}\left|S_{ij}\right|^{2}\\ \nonumber &\bigg(K_{V}^{2}
	+K_{V}^{'2}\xi_{2}^{4}-2K_{V}K_{V}^{'}\xi_{2}^{2}\bigg)F_{V}\left(x_{\nu_{\ell}},x_{P}\right),
\end{align}
where $\nu_\ell=\nu_e,\nu_\mu,\nu_\tau$ and $V^0=\rho^0,\omega,\phi,J/\psi$. In the above $\xi_{1}=\frac{M_{W_{L}}}{M_{W_{R}}}$, $\xi_{2}=\frac{M_{Z}}{M_{Z'}}$, $x_{i}=\frac{m_{i}}{M_{N}}$ with $m_{i}=m_{\ell}, m_{P^{0}},m_{V^{0}},m_{P^{+}},m_{V}^{+}$.
The kinematical functions required for two-body decay are given by,
\begin{align*}
	F_{P}(x,y)&=\big((1+x^{2})(1+x^{2}-y^{2})-4x^{2}\big)\lambda^{\frac{1}{2}}(1,x^{2},y^{2});\\
	F_{V}(x,y)&=\big((1-x^{2})^{2}+(1+x^{2})y^{2}-2y^{4}\big)\lambda^{\frac{1}{2}}(1,x^{2},y^{2}).
\end{align*}
Neutral current coupling of pseudoscalar mesons are given by,
\begin{eqnarray}
	&K_{\pi^{0}}=-\frac{1}{2\sqrt{2}},\,\,\,K_{\pi^{0}}^{\prime}=\frac{1}{\sqrt{2}}(\frac{1}{2}-\text{sin}^{2}\theta_{w}),\nonumber\\
	&K_{\eta}=-\frac{1}{2\sqrt{6}},\,\,\,K_{\eta}^{\prime}=\frac{1}{\sqrt{6}}(\frac{1}{2}-\text{sin}^{2}\theta_{w}),\nonumber\\
	&K_{\eta^{\prime}}=\frac{1}{4\sqrt{3}},\,\,\,K_{\eta^{\prime}}^{\prime}=\frac{1}{\sqrt{3}}(-\frac{1}{4}+\frac{1}{2}\text{sin}^{2}\theta_{w}),\nonumber\\
	&K_{\eta_{c}}=-\frac{1}{4},\,\,\,K_{\eta_{c}}^{\prime}=(\frac{1}{4}-\frac{1}{2}\text{sin}^{2}\theta_{w}),\nonumber
\end{eqnarray}
Neutral current coupling of vector mesons are given by
\begin{eqnarray}
	&K_{\rho^{0}}=\frac{1}{\sqrt{2}}(\frac{1}{2}-\text{sin}^{2}\theta_{w}),\nonumber\\
	&K_{\omega}=-\frac{1}{3\sqrt{2}}\text{sin}^{2}\theta_{w},\nonumber\\
	&K_{\phi}=(-\frac{1}{4}+\frac{1}{3}\text{sin}^{2}\theta_{w}),\nonumber\\
	&K_{J/\psi}=(\frac{1}{4}-\frac{2}{3}\text{sin}^{2}\theta_{w}).\nonumber\\
\end{eqnarray}
\subsection{Three-body decays of N}
\begin{align}
	\Gamma\left(N_{j}\rightarrow\ell_{i}^{-}\ell_{k}^{+}\nu_{\ell_{k}}\right)&=\frac{G_{F}^{2}M_{N_{j}}^{5}}{16\pi^{3}}\bigg(\left|S_{ij}\right|^{2}\sum_{\kappa}\left|U_{k \kappa}\right|^{2}
	I_{1}\left(x_{\ell_{i}},x_{\nu_{\ell_{k}}},x_{\ell_{k}}\right)+\left|K_{R_{ij}}\right|^{2}\\  \nonumber
	&\sum_{\kappa}\left|T_{k \kappa}\right|^{2}\xi_{1}^{4} I_{1}\left(x_{\ell_{i}},x_{\nu_{\ell_{k}}},x_{\ell_{k}}\right)
	-8Re\big(S_{ij}^{*}V_{ij}^{*}\sum_{\kappa}U_{k \kappa}T_{k \kappa}\big)\xi_{1}^{2}I_{3}\left(x_{\ell_{i}},x_{\nu_{\ell_{k}}},x_{\ell_{k}}\right)\bigg),
\end{align}
where $\ell_i,\ell_j=e,\mu,\tau$, $\ell_i\ne\ell_j$.
\begin{align}
	\Gamma\left(N_{j}\rightarrow\nu_{\ell_{i}}\ell_{i}^{-}\ell_{i}^{+}\right) &=\frac{G_{F}^{2}M_{N_{j}}^{5}}{16\pi^{3}}\bigg(\left|S_{ij}\right|^{2}\sum_{k}\left|U_{ik}\right|^{2}\bigg[I_{1}
	\left(x_{\nu_{\ell_{i}}},x_{\ell_{i}},x_{\ell_{i}}\right)  \\ \nonumber  & +2\left((g_{V}^{\ell})^{2}+(g_{A}^{\ell})^{2}\right)
	I_{1}
	\left(x_{\nu_{\ell_{i}}},x_{\ell_{i}},x_{\ell_{i}}\right)
	+2\left((g_{V}^{\ell})^{2}-(g_{A}^{\ell})^{2}\right)I_{2}
	\left(x_{\nu_{\ell_{i}}},x_{\ell_{i}},x_{\ell_{i}}\right)\\ \nonumber &+2\left((g_{V}^{\prime\ell})^{2}+(g_{A}^{\prime\ell})^{2}\right)\xi_{2}^{4}I_{1}
	\left(x_{\nu_{\ell_{i}}},x_{\ell_{i}},x_{\ell_{i}}\right)+2\left((g_{V}^{\prime\ell})^{2}-(g_{A}^{\prime\ell})^{2}\right)\xi_{2}^{4}I_{2}
	\left(x_{\nu_{\ell_{i}}},x_{\ell_{i}},x_{\ell_{i}}\right)
	\\ \nonumber
	&-4\xi_{2}^{2}\big((g_{V}^{\ell}g_{V}^{\prime\ell}+g_{A}^{\ell}g_{A}^{\prime\ell})I_{1}
	\left(x_{\nu_{\ell_{i}}},x_{\ell_{i}},x_{\ell_{i}}\right)+(g_{V}^{\ell}g_{V}^{\prime\ell}-g_{A}^{\ell}g_{A}^{\prime\ell})I_{2}
	\left(x_{\nu_{\ell_{i}}},x_{\ell_{i}},x_{\ell_{i}}\right)\big)\bigg]
	\\ \nonumber
	&+\left|V_{ij}\right|^{2}\sum_{k}\left|T_{ik}\right|^{2}\xi_{1}^{4}I_{1}
	\left(x_{\nu_{\ell_{i}}},x_{\ell_{i}},x_{\ell_{i}}\right)-8Re\big[S_{ij}^{*}V_{ij}^{*}\sum_{k}U_{ik}T_{ik}\big]\xi_{1}^{2}
	I_{3}\left(x_{\ell_{i}},x_{\nu_{\ell_{i}}},x_{\ell_{i}}\right)\\ \nonumber
	&+2Re\big[\left|S_{ij}\right|^{2}\sum_{k}\left|U_{ik}\right|^{2}\big]
	\bigg[\xi_{2}^{2}(g_{A}^{\prime\ell}-g_{V}^{\prime\ell})I_{1}
	\left(x_{\nu_{\ell_{i}}},x_{\ell_{i}},x_{\ell_{i}}\right)
	-\xi_{2}^{2}(g_{A}^{\prime\ell}+g_{V}^{\prime\ell})I_{2}
	\left(x_{\nu_{\ell_{i}}},x_{\ell_{i}},x_{\ell_{i}}\right)\\ \nonumber
	&-(g_{A}^{\ell}-g_{V}^{\ell})I_{1}
	\left(x_{\nu_{\ell_{i}}},x_{\ell_{i}},x_{\ell_{i}}\right)
	+(g_{A}^{\ell}+g_{V}^{\ell})I_{2}
	\left(x_{\nu_{\ell_{i}}},x_{\ell_{i}},x_{\ell_{i}}\right)\bigg]
	\\ \nonumber
	&-8Re\big[S_{ij}V_{ij}\sum_{k}U_{ik}^{*}T_{ik}^{*}\big]\xi_{1}^{2}\bigg[(g_{V}^{\prime\ell}-g_{A}^{\prime\ell})\xi_{2}^{2}I_{3}\left(x_{\nu_{\ell_{i}}},x_{\ell_{i}},x_{\ell_{i}}\right)
	+\frac{1}{4}(g_{V}^{\prime\ell}+g_{A}^{\prime\ell})\xi_{2}^{2}\\ \nonumber
	&I_{4}\left(x_{\ell_{i}},x_{\ell_{i}},x_{\nu_{\ell_{i}}}\right)+(g_{V}^{\ell}-g_{A}^{\ell})I_{3}\left(x_{\nu_{\ell_{i}}},x_{\ell_{i}},x_{\ell_{i}}\right)
	+\frac{1}{4}(g_{V}^{\ell}+g_{A}^{\ell})I_{4}\left(x_{\ell_{i}},x_{\ell_{i}},x_{\nu_{\ell_{i}}}\right)\bigg]\bigg),
\end{align}
where $\ell_i=e,\mu,\tau$.
\begin{align}
	\Gamma\left(N_{j}\rightarrow\nu_{\ell_{i}}\ell_{k}^{-}\ell_{k}^{+}\right)&=\frac{G_{F}^{2}M_{N_{j}}^{5}}{8\pi^{3}}\left|S_{ij}\right|^{2}\sum_{\kappa}\left|U_{i\kappa}\right|^{2}\bigg[\big((g_{V}^{\ell})^{2}
	+(g_{A}^{\ell})^{2}\big)I_{1} \left(x_{\nu_{\ell_{i}}},x_{\ell_{k}},x_{\ell_{k}}\right)\\ \nonumber
	&+\big((g_{V}^{\ell})^{2}-(g_{A}^{\ell})^{2}\big)I_{2} \left(x_{\nu_{\ell_{i}}},x_{\ell_{k}},x_{\ell_{k}}\right)
	+\big((g_{V}^{\prime\ell})^{2}
	+(g_{A}^{\prime\ell})^{2}\big)\xi_{2}^{4}I_{1} \left(x_{\nu_{\ell_{i}}},x_{\ell_{k}},x_{\ell_{k}}\right)\\ \nonumber
	&+\big((g_{V}^{\prime\ell})^{2}-(g_{A}^{\prime\ell})^{2}\big)\xi_{2}^{4}I_{2} \left(x_{\nu_{\ell_{i}}},x_{\ell_{k}},x_{\ell_{k}}\right)
	-2\xi_{2}^{2}\big[(g_{V}^{\ell}g_{V}^{\prime\ell}+g_{A}^{\ell}g_{A}^{\prime\ell})
	I_{1} \left(x_{\nu_{\ell_{i}}},x_{\ell_{k}},x_{\ell_{k}}\right)\\ \nonumber
	&+(g_{V}^{\ell}g_{V}^{\prime\ell}-g_{A}^{\ell}g_{A}^{\prime\ell})
	I_{2}\left(x_{\nu_{\ell_{i}}},x_{\ell_{k}},x_{\ell_{k}}\right)\big]\bigg].
\end{align}
where $\ell_i,\ell_j=e,\mu,\tau$ and $\ell_{i}\ne\ell_{j}$.
\begin{align}
	\Gamma\left(N_{j}\rightarrow\nu_{\ell_i}\nu\overline{\nu}\right)&=\frac{G_{F}^{2}M_{N_{j}}^{5}}{192\pi^{3}}\left|S_{ij}\right|^{2}\sum_{k}\left|U_{ik}\right|^{2}
	\bigg(1-\text{sin}^{2}\theta_{w}\xi_{2}^{2}\bigg)^{2},\\ \nonumber
\end{align}
The kinematical functions required for three-body decay are given by,
\begin{align*}
	I_{1}(x,y,z)&=\int_{(x+y)^{2}}^{(1-z)^{2}}\frac{ds}{s}(s-x^{2}-y^{2})(1+z^{2}-s)\lambda^{\frac{1}{2}}(s,x^{2},y^{2})\lambda^{\frac{1}{2}}(1,s,z^{2});\\
	I_{2}(x,y,z)&=yz\int_{(y+z)^{2}}^{(1-x)^{2}}\frac{ds}{s}(1+x^{2}-s)\lambda^{\frac{1}{2}}(s,y^{2},z^{2})\lambda^{\frac{1}{2}}(1,s,x^{2});\\
	I_{3}(x,y,z)&=xyz\int_{(x+y)^{2}}^{(1-z)^{2}}\frac{ds}{s}\lambda^{\frac{1}{2}}(s,x^{2},y^{2})\lambda^{\frac{1}{2}}(1,s,z^{2});\\
	I_{4}(x,y,z)&=z\int_{(x+y)^{2}}^{(1-z)^{2}}\frac{ds}{s}\lambda^{\frac{1}{2}}(s,x^{2},y^{2})\lambda^{\frac{1}{2}}(1,s,z^{2});	
\end{align*}
Neutral current couplings of leptons are given by,
\begin{eqnarray}
	&g_{V}^{\ell}=-\frac{1}{4}+\text{sin}^{2}\theta_{w},\,\,\,g_{A}^{\ell}=\frac{1}{4},\nonumber\\
	&g_{V}^{\prime\ell}=-\frac{1}{4}+\text{sin}^{2}\theta_{w},\,\,\,g_{A}^{\prime\ell}=-\frac{1}{4}+\frac{1}{2}\text{sin}^{2}\theta_{w}.\nonumber
\end{eqnarray}
The functions $A(M_N)$ and $B(M_N)$, relevant for $N \to l \pi$ decay mode is given by 
\begin{eqnarray}
	A(M_N)=\frac{G^2_F M^3_N}{16 \pi} f^2_\pi V^2_{ud}\xi_{1}^4\Big[\frac{F_P(x_e,x_\pi)+F_P(x_\mu,x_\pi)}{2}\Big]
	\label{AMN}
\end{eqnarray}
\begin{eqnarray}
	B(M_N)=\frac{G^2_F M^3_N}{16 \pi} f^2_\pi V^2_{ud}\xi_{1}^4\Big[\frac{F_P(x_e,x_\pi)-F_P(x_\mu,x_\pi)}{2}\Big]
	\label{BMN}
\end{eqnarray}
\bibliography{bibitem}

\providecommand{\href}[2]{#2}\begingroup\raggedright\begin{thebibliography}{10}

\bibitem{Pati:1974yy}
J.~C. Pati and A.~Salam, ``{Lepton Number as the Fourth Color},''
  \href{http://dx.doi.org/10.1103/PhysRevD.10.275}{{\em Phys. Rev. D}
  {\bfseries 10} (1974) 275--289}. [Erratum: Phys.Rev.D 11, 703--703 (1975)].

\bibitem{Agostini:2018tnm}
{\bfseries GERDA} Collaboration, M.~Agostini {\em et~al.}, ``{Improved Limit on
  Neutrinoless Double-$\beta$ Decay of $^{76}$Ge from GERDA Phase II},''
  \href{http://dx.doi.org/10.1103/PhysRevLett.120.132503}{{\em Phys. Rev.
  Lett.} {\bfseries 120} no.~13, (2018) 132503},
  \href{http://arxiv.org/abs/1803.11100}{{\ttfamily arXiv:1803.11100
  [nucl-ex]}}.

\bibitem{Racah:1937qq}
G.~Racah, ``{On the symmetry of particle and antiparticle},''
  \href{http://dx.doi.org/10.1007/BF02961321}{{\em Nuovo Cim.} {\bfseries 14}
  (1937) 322--328}.

\bibitem{Furry:1939qr}
W.~Furry, ``{On transition probabilities in double beta-disintegration},''
  \href{http://dx.doi.org/10.1103/PhysRev.56.1184}{{\em Phys. Rev.} {\bfseries
  56} (1939) 1184--1193}.

\bibitem{Schechter:1981bd}
J.~Schechter and J.~Valle, ``{Neutrinoless Double beta Decay in SU(2) x U(1)
  Theories},'' \href{http://dx.doi.org/10.1103/PhysRevD.25.2951}{{\em Phys.
  Rev. D} {\bfseries 25} (1982) 2951}.

\bibitem{Tello:2010am}
V.~Tello, M.~Nemevsek, F.~Nesti, G.~Senjanovic, and F.~Vissani, ``{Left-Right
  Symmetry: from LHC to Neutrinoless Double Beta Decay},''
  \href{http://dx.doi.org/10.1103/PhysRevLett.106.151801}{{\em Phys. Rev.
  Lett.} {\bfseries 106} (2011) 151801},
  \href{http://arxiv.org/abs/1011.3522}{{\ttfamily arXiv:1011.3522 [hep-ph]}}.

\bibitem{Dev:2014xea}
P.~S. Bhupal~Dev, S.~Goswami, and M.~Mitra, ``{TeV Scale Left-Right Symmetry
  and Large Mixing Effects in Neutrinoless Double Beta Decay},''
  \href{http://dx.doi.org/10.1103/PhysRevD.91.113004}{{\em Phys. Rev. D}
  {\bfseries 91} no.~11, (2015) 113004},
  \href{http://arxiv.org/abs/1405.1399}{{\ttfamily arXiv:1405.1399 [hep-ph]}}.

\bibitem{Keung:1983uu}
W.-Y. Keung and G.~Senjanovic, ``{Majorana Neutrinos and the Production of the
  Right-handed Charged Gauge Boson},''
  \href{http://dx.doi.org/10.1103/PhysRevLett.50.1427}{{\em Phys. Rev. Lett.}
  {\bfseries 50} (1983) 1427}.

\bibitem{Chen:2013foz}
C.-Y. Chen, P.~S.~B. Dev, and R.~Mohapatra, ``{Probing Heavy-Light Neutrino
  Mixing in Left-Right Seesaw Models at the LHC},''
  \href{http://dx.doi.org/10.1103/PhysRevD.88.033014}{{\em Phys. Rev. D}
  {\bfseries 88} (2013) 033014},
  \href{http://arxiv.org/abs/1306.2342}{{\ttfamily arXiv:1306.2342 [hep-ph]}}.

\bibitem{Mitra:2016kov}
M.~Mitra, R.~Ruiz, D.~J. Scott, and M.~Spannowsky, ``{Neutrino Jets from
  High-Mass $W_R$ Gauge Bosons in TeV-Scale Left-Right Symmetric Models},''
  \href{http://dx.doi.org/10.1103/PhysRevD.94.095016}{{\em Phys. Rev. D}
  {\bfseries 94} no.~9, (2016) 095016},
  \href{http://arxiv.org/abs/1607.03504}{{\ttfamily arXiv:1607.03504
  [hep-ph]}}.

\bibitem{Nemevsek:2018bbt}
M.~Nemev\v{s}ek, F.~Nesti, and G.~Popara, ``{Keung-Senjanovi\'c process at the
  LHC: From lepton number violation to displaced vertices to invisible
  decays},'' \href{http://dx.doi.org/10.1103/PhysRevD.97.115018}{{\em Phys.
  Rev. D} {\bfseries 97} no.~11, (2018) 115018},
  \href{http://arxiv.org/abs/1801.05813}{{\ttfamily arXiv:1801.05813
  [hep-ph]}}.

\bibitem{Nemevsek:2016enw}
M.~Nemev\v{s}ek, F.~Nesti, and J.~C. Vasquez, ``{Majorana Higgses at
  colliders},'' \href{http://dx.doi.org/10.1007/JHEP04(2017)114}{{\em JHEP}
  {\bfseries 04} (2017) 114}, \href{http://arxiv.org/abs/1612.06840}{{\ttfamily
  arXiv:1612.06840 [hep-ph]}}.

\bibitem{Mandal:2019vhf}
S.~Mandal, {\em {Search for sterile neutrinos at colliders}}.
\newblock PhD thesis, HBNI, Mumbai, 2019.
\newblock
  \url{https://www.imsc.res.in/xmlui/bitstream/handle/123456789/445/HBNI%20Th157.pdf?sequence=1&isAllowed=y}.

\bibitem{Atre:2009rg}
A.~Atre, T.~Han, S.~Pascoli, and B.~Zhang, ``{The Search for Heavy Majorana
  Neutrinos},'' \href{http://dx.doi.org/10.1088/1126-6708/2009/05/030}{{\em
  JHEP} {\bfseries 05} (2009) 030},
  \href{http://arxiv.org/abs/0901.3589}{{\ttfamily arXiv:0901.3589 [hep-ph]}}.

\bibitem{Aaij:2014aba}
{\bfseries LHCb} Collaboration, R.~Aaij {\em et~al.}, ``{Search for Majorana
  neutrinos in $B^- \to \pi^+\mu^-\mu^-$ decays},''
  \href{http://dx.doi.org/10.1103/PhysRevLett.112.131802}{{\em Phys. Rev.
  Lett.} {\bfseries 112} no.~13, (2014) 131802},
  \href{http://arxiv.org/abs/1401.5361}{{\ttfamily arXiv:1401.5361 [hep-ex]}}.

\bibitem{Cvetic:2010rw}
G.~Cvetic, C.~Dib, S.~K. Kang, and C.~Kim, ``{Probing Majorana neutrinos in
  rare K and D, $D_s$, B, $B_c$ meson decays},''
  \href{http://dx.doi.org/10.1103/PhysRevD.82.053010}{{\em Phys. Rev. D}
  {\bfseries 82} (2010) 053010},
  \href{http://arxiv.org/abs/1005.4282}{{\ttfamily arXiv:1005.4282 [hep-ph]}}.

\bibitem{Cvetic:2020lyh}
G.~Cvetic, C.~Kim, S.~Mendizabal, and J.~Zamora-Saa, ``{Exploring CP-violation,
  via heavy neutrino oscillations, in rare B meson decays at Belle II},''
  \href{http://arxiv.org/abs/2007.04115}{{\ttfamily arXiv:2007.04115
  [hep-ph]}}.

\bibitem{Shuve:2016muy}
B.~Shuve and M.~E. Peskin, ``{Revision of the LHCb Limit on Majorana
  Neutrinos},'' \href{http://dx.doi.org/10.1103/PhysRevD.94.113007}{{\em Phys.
  Rev. D} {\bfseries 94} no.~11, (2016) 113007},
  \href{http://arxiv.org/abs/1607.04258}{{\ttfamily arXiv:1607.04258
  [hep-ph]}}.

\bibitem{Chun:2019nwi}
E.~J. Chun, A.~Das, S.~Mandal, M.~Mitra, and N.~Sinha, ``{Sensitivity of Lepton
  Number Violating Meson Decays in Different Experiments},''
  \href{http://dx.doi.org/10.1103/PhysRevD.100.095022}{{\em Phys. Rev. D}
  {\bfseries 100} no.~9, (2019) 095022},
  \href{http://arxiv.org/abs/1908.09562}{{\ttfamily arXiv:1908.09562
  [hep-ph]}}.

\bibitem{Mandal:2016hpr}
S.~Mandal and N.~Sinha, ``{Favoured $B_c$ Decay modes to search for a Majorana
  neutrino},'' \href{http://dx.doi.org/10.1103/PhysRevD.94.033001}{{\em Phys.
  Rev. D} {\bfseries 94} no.~3, (2016) 033001},
  \href{http://arxiv.org/abs/1602.09112}{{\ttfamily arXiv:1602.09112
  [hep-ph]}}.

\bibitem{Mandal:2017tab}
S.~Mandal, M.~Mitra, and N.~Sinha, ``{Constraining the right-handed gauge boson
  mass from lepton number violating meson decays in a low scale left-right
  model},'' \href{http://dx.doi.org/10.1103/PhysRevD.96.035023}{{\em Phys. Rev.
  D} {\bfseries 96} no.~3, (2017) 035023},
  \href{http://arxiv.org/abs/1705.01932}{{\ttfamily arXiv:1705.01932
  [hep-ph]}}.

\bibitem{Milanes:2016rzr}
D.~Milanes, N.~Quintero, and C.~E. Vera, ``{Sensitivity to Majorana neutrinos
  in $\Delta L=2$ decays of $B_c$ meson at LHCb},''
  \href{http://dx.doi.org/10.1103/PhysRevD.93.094026}{{\em Phys. Rev. D}
  {\bfseries 93} no.~9, (2016) 094026},
  \href{http://arxiv.org/abs/1604.03177}{{\ttfamily arXiv:1604.03177
  [hep-ph]}}.

\bibitem{Abada:2017jjx}
A.~Abada, V.~De~Romeri, M.~Lucente, A.~M. Teixeira, and T.~Toma, ``{Effective
  Majorana mass matrix from tau and pseudoscalar meson lepton number violating
  decays},'' \href{http://dx.doi.org/10.1007/JHEP02(2018)169}{{\em JHEP}
  {\bfseries 02} (2018) 169}, \href{http://arxiv.org/abs/1712.03984}{{\ttfamily
  arXiv:1712.03984 [hep-ph]}}.

\bibitem{Helo:2010cw}
J.~C. Helo, S.~Kovalenko, and I.~Schmidt, ``{Sterile neutrinos in lepton number
  and lepton flavor violating decays},''
  \href{http://dx.doi.org/10.1016/j.nuclphysb.2011.07.020}{{\em Nucl. Phys. B}
  {\bfseries 853} (2011) 80--104},
  \href{http://arxiv.org/abs/1005.1607}{{\ttfamily arXiv:1005.1607 [hep-ph]}}.

\bibitem{Abada:2019bac}
A.~Abada, C.~Hati, X.~Marcano, and A.~Teixeira, ``{Interference effects in LNV
  and LFV semileptonic decays: the Majorana hypothesis},''
  \href{http://dx.doi.org/10.1007/JHEP09(2019)017}{{\em JHEP} {\bfseries 09}
  (2019) 017}, \href{http://arxiv.org/abs/1904.05367}{{\ttfamily
  arXiv:1904.05367 [hep-ph]}}.

\bibitem{Aaboud:2018spl}
{\bfseries ATLAS} Collaboration, M.~Aaboud {\em et~al.}, ``{Search for heavy
  Majorana or Dirac neutrinos and right-handed $W$ gauge bosons in final states
  with two charged leptons and two jets at $ \sqrt{s}=13 $ TeV with the ATLAS
  detector},'' \href{http://dx.doi.org/10.1007/JHEP01(2019)016}{{\em JHEP}
  {\bfseries 01} (2019) 016}, \href{http://arxiv.org/abs/1809.11105}{{\ttfamily
  arXiv:1809.11105 [hep-ex]}}.

\bibitem{Sirunyan:2018pom}
{\bfseries CMS} Collaboration, A.~M. Sirunyan {\em et~al.}, ``{Search for a
  heavy right-handed W boson and a heavy neutrino in events with two
  same-flavor leptons and two jets at $\sqrt{s}=$ 13 TeV},''
  \href{http://dx.doi.org/10.1007/JHEP05(2018)148}{{\em JHEP} {\bfseries 05}
  (2018) 148}, \href{http://arxiv.org/abs/1803.11116}{{\ttfamily
  arXiv:1803.11116 [hep-ex]}}.

\bibitem{Aaboud:2019wfg}
{\bfseries ATLAS} Collaboration, M.~Aaboud {\em et~al.}, ``{Search for a
  right-handed gauge boson decaying into a high-momentum heavy neutrino and a
  charged lepton in $pp$ collisions with the ATLAS detector at $\sqrt{s}=13$
  TeV},'' \href{http://dx.doi.org/10.1016/j.physletb.2019.134942}{{\em Phys.
  Lett. B} {\bfseries 798} (2019) 134942},
  \href{http://arxiv.org/abs/1904.12679}{{\ttfamily arXiv:1904.12679
  [hep-ex]}}.

\bibitem{Ruiz:2013a}
T.~Han, I.~Lewis, R.~Ruiz, and Z.-g. Si, ``{Lepton number violation and
  ${W}^{\ensuremath{'}}$ chiral couplings at the LHC},''
  \href{http://dx.doi.org/10.1103/PhysRevD.87.035011}{{\em Phys. Rev. D}
  {\bfseries 87} (Feb, 2013) 035011}.
  \url{https://link.aps.org/doi/10.1103/PhysRevD.87.035011}.

\bibitem{Ruiz:2020b}
R.~Ruiz, ``{A quantitative study on helicity inversion in Majorana neutrino
  decays at the LHC},'' \href{http://arxiv.org/abs/2008.01092}{{\ttfamily
  arXiv:2008.01092 [hep-ph]}}.

\bibitem{Lindner:2016lpp}
M.~Lindner, F.~S. Queiroz, and W.~Rodejohann, ``{Dilepton bounds on left-right
  symmetry at the LHC run II and neutrinoless double beta decay},''
  \href{http://dx.doi.org/10.1016/j.physletb.2016.08.068}{{\em Phys. Lett. B}
  {\bfseries 762} (2016) 190--195},
  \href{http://arxiv.org/abs/1604.07419}{{\ttfamily arXiv:1604.07419
  [hep-ph]}}.

\bibitem{Hirsch:2020klk}
M.~Hirsch and Z.~S. Wang, ``{Heavy neutral leptons at ANUBIS},''
  \href{http://dx.doi.org/10.1103/PhysRevD.101.055034}{{\em Phys. Rev. D}
  {\bfseries 101} no.~5, (2020) 055034},
  \href{http://arxiv.org/abs/2001.04750}{{\ttfamily arXiv:2001.04750
  [hep-ph]}}.

\bibitem{Cvetic:2013eza}
G.~Cveti\v{c}, C.~S. Kim, and J.~Zamora-Sa\'a, ``{CP violations in $\pi^{\pm}$
  Meson Decay},'' \href{http://dx.doi.org/10.1088/0954-3899/41/7/075004}{{\em
  J. Phys. G} {\bfseries 41} (2014) 075004},
  \href{http://arxiv.org/abs/1311.7554}{{\ttfamily arXiv:1311.7554 [hep-ph]}}.

\bibitem{Cvetic_2015naa}
G.~Cvetic, C.~Dib, C.~S. Kim, and J.~Zamora-Saa, ``{Probing the Majorana
  neutrinos and their CP violation in decays of charged scalar mesons $\pi, K,
  D, D_s, B, B_c$},'' \href{http://dx.doi.org/10.3390/sym7020726}{{\em
  Symmetry} {\bfseries 7} (2015) 726--773},
  \href{http://arxiv.org/abs/1503.01358}{{\ttfamily arXiv:1503.01358
  [hep-ph]}}.

\bibitem{Cvetic_2015ura}
G.~Cvetic, C.~S. Kim, R.~Kogerler, and J.~Zamora-Saa, ``{Oscillation of heavy
  sterile neutrino in decay of $B \to \mu e \pi$},''
  \href{http://dx.doi.org/10.1103/PhysRevD.92.013015}{{\em Phys. Rev. D}
  {\bfseries 92} (2015) 013015},
  \href{http://arxiv.org/abs/1505.04749}{{\ttfamily arXiv:1505.04749
  [hep-ph]}}.

\bibitem{Moreno:2016cfz}
G.~Moreno and J.~Zamora-Saa, ``{Rare meson decays with three pairs of
  quasi-degenerate heavy neutrinos},''
  \href{http://dx.doi.org/10.1103/PhysRevD.94.093005}{{\em Phys. Rev. D}
  {\bfseries 94} no.~9, (2016) 093005},
  \href{http://arxiv.org/abs/1606.08820}{{\ttfamily arXiv:1606.08820
  [hep-ph]}}.

\bibitem{Zamora-Saa:2016ito}
J.~Zamora-Saa, ``{Resonant $CP$ violation in rare $\tau^{\pm}$ decays},''
  \href{http://dx.doi.org/10.1007/JHEP05(2017)110}{{\em JHEP} {\bfseries 05}
  (2017) 110}, \href{http://arxiv.org/abs/1612.07656}{{\ttfamily
  arXiv:1612.07656 [hep-ph]}}.

\bibitem{Cvetic:2018elt}
G.~Cveti\v{c}, A.~Das, and J.~Zamora-Sa\'a, ``{Probing heavy neutrino
  oscillations in rare $W$ boson decays},''
  \href{http://dx.doi.org/10.1088/1361-6471/ab1212}{{\em J. Phys. G} {\bfseries
  46} (2019) 075002}, \href{http://arxiv.org/abs/1805.00070}{{\ttfamily
  arXiv:1805.00070 [hep-ph]}}.

\bibitem{Cvetic:2019rms}
G.~Cveti\v{c}, A.~Das, S.~Tapia, and J.~Zamora-Sa\'a, ``{Measuring the heavy
  neutrino oscillations in rare W boson decays at the Large Hadron Collider},''
  \href{http://dx.doi.org/10.1088/1361-6471/ab4ee8}{{\em J. Phys. G} {\bfseries
  47} no.~1, (2020) 015001}, \href{http://arxiv.org/abs/1905.03097}{{\ttfamily
  arXiv:1905.03097 [hep-ph]}}.

\bibitem{Tapia:2019coy}
S.~Tapia and J.~Zamora-Sa\'a, ``{Exploring CP-Violating heavy neutrino
  oscillations in rare tau decays at Belle II},''
  \href{http://dx.doi.org/10.1016/j.nuclphysb.2020.114936}{{\em Nucl. Phys. B}
  {\bfseries 952} (2020) 114936},
  \href{http://arxiv.org/abs/1906.09470}{{\ttfamily arXiv:1906.09470
  [hep-ph]}}.

\bibitem{Gluza:2015goa}
J.~Gluza and T.~Jeli\'nski, ``{Heavy neutrinos and the pp$\rightarrow$lljj CMS
  data},'' \href{http://dx.doi.org/10.1016/j.physletb.2015.06.077}{{\em Phys.\
  Lett.\ B} {\bfseries 748} (2015) 125--131},
  \href{http://arxiv.org/abs/1504.05568}{{\ttfamily arXiv:1504.05568
  [hep-ph]}}.

\bibitem{Das:2017hmg}
A.~Das, P.~S.~B. Dev, and R.~N. Mohapatra, ``{Same Sign versus Opposite Sign
  Dileptons as a Probe of Low Scale Seesaw Mechanisms},''
  \href{http://dx.doi.org/10.1103/PhysRevD.97.015018}{{\em Phys. Rev. D}
  {\bfseries 97} no.~1, (2018) 015018},
  \href{http://arxiv.org/abs/1709.06553}{{\ttfamily arXiv:1709.06553
  [hep-ph]}}.

\bibitem{pilaftsis:2007sj}
S.~Bray, J.~S. Lee, and A.~Pilaftsis, ``Resonant cp violation due to heavy
  neutrinos at the lhc,''
  \href{http://dx.doi.org/https://doi.org/10.1016/j.nuclphysb.2007.07.002}{{\em
  Nuclear Physics B} {\bfseries 786} no.~1, (2007) 95 -- 118}.
  \url{http://www.sciencedirect.com/science/article/pii/S0550321307005329}.

\bibitem{pilaftsis:199761}
A.~Pilaftsis, ``Resonant cp violation induced by particle mixing in transition
  amplitudes,''
  \href{http://dx.doi.org/https://doi.org/10.1016/S0550-3213(97)00469-0}{{\em
  Nuclear Physics B} {\bfseries 504} no.~1, (1997) 61 -- 107}.
  \url{http://www.sciencedirect.com/science/article/pii/S0550321397004690}.

\bibitem{Gunion:1989ci}
J.~Gunion, R.~Vega, and J.~Wudka, ``{Higgs triplets in the standard model},''
  \href{http://dx.doi.org/10.1103/PhysRevD.42.1673}{{\em Phys. Rev. D}
  {\bfseries 42} (1990) 1673--1691}.

\bibitem{Maiezza:2010ic}
A.~Maiezza, M.~Nemevsek, F.~Nesti, and G.~Senjanovic, ``{Left-Right Symmetry at
  LHC},'' \href{http://dx.doi.org/10.1103/PhysRevD.82.055022}{{\em Phys. Rev.
  D} {\bfseries 82} (2010) 055022},
  \href{http://arxiv.org/abs/1005.5160}{{\ttfamily arXiv:1005.5160 [hep-ph]}}.

\bibitem{Senjanovic:2014pva}
G.~Senjanovi\'c and V.~Tello, ``{Right Handed Quark Mixing in Left-Right
  Symmetric Theory},''
  \href{http://dx.doi.org/10.1103/PhysRevLett.114.071801}{{\em Phys. Rev.
  Lett.} {\bfseries 114} no.~7, (2015) 071801},
  \href{http://arxiv.org/abs/1408.3835}{{\ttfamily arXiv:1408.3835 [hep-ph]}}.

\bibitem{Mohapatra:1977be}
R.~N. Mohapatra and D.~P. Sidhu, ``{Gauge Theories of Weak Interactions with
  Left-Right Symmetry and the Structure of Neutral Currents},''
\href{http://dx.doi.org/10.1103/PhysRevD.16.2843}{{\em Phys. Rev.} {\bfseries
  D16} (1977) 2843}.

\bibitem{Tello}
V.~Tello, {\em {Connections between the high and low energy violation of Lepton
  and Flavor numbers in the minimal left-right symmetric model}}.
\newblock PhD thesis, SISSA, Trieste, 2012.
\newblock \url{https://inspirehep.net/files/8d7885f38c339ea85b54ca6e4722c14d}.

\bibitem{Li:2014uha}
Z.-H. Li, Z.-G. Si, Y.~Wang, and N.~Zhu, ``{$B\to \pi\ell^{+}\ell^{-}$ decays
  revisited in the standard model},''
  \href{http://arxiv.org/abs/1411.0466}{{\ttfamily arXiv:1411.0466 [hep-ph]}}.

\bibitem{Littenberg:1993qv}
L.~Littenberg and G.~Valencia, ``{Rare and radiative kaon decays},''
  \href{http://dx.doi.org/10.1146/annurev.ns.43.120193.003501}{{\em Ann. Rev.
  Nucl. Part. Sci.} {\bfseries 43} (1993) 729--792},
  \href{http://arxiv.org/abs/hep-ph/9303225}{{\ttfamily arXiv:hep-ph/9303225}}.

\bibitem{Cirigliano:2011ny}
V.~Cirigliano, G.~Ecker, H.~Neufeld, A.~Pich, and J.~Portoles, ``{Kaon Decays
  in the Standard Model},''
  \href{http://dx.doi.org/10.1103/RevModPhys.84.399}{{\em Rev. Mod. Phys.}
  {\bfseries 84} (2012) 399}, \href{http://arxiv.org/abs/1107.6001}{{\ttfamily
  arXiv:1107.6001 [hep-ph]}}.

\bibitem{Dolgov:2000jw}
A.~Dolgov, S.~Hansen, G.~Raffelt, and D.~Semikoz, ``{Heavy sterile neutrinos:
  Bounds from big bang nucleosynthesis and SN1987A},''
  \href{http://dx.doi.org/10.1016/S0550-3213(00)00566-6}{{\em Nucl. Phys. B}
  {\bfseries 590} (2000) 562--574},
  \href{http://arxiv.org/abs/hep-ph/0008138}{{\ttfamily arXiv:hep-ph/0008138}}.

\bibitem{Kawasaki:2004qu}
M.~Kawasaki, K.~Kohri, and T.~Moroi, ``{Big-Bang nucleosynthesis and hadronic
  decay of long-lived massive particles},''
  \href{http://dx.doi.org/10.1103/PhysRevD.71.083502}{{\em Phys. Rev. D}
  {\bfseries 71} (2005) 083502},
  \href{http://arxiv.org/abs/astro-ph/0408426}{{\ttfamily
  arXiv:astro-ph/0408426}}.

\bibitem{Kawasaki:2017bqm}
M.~Kawasaki, K.~Kohri, T.~Moroi, and Y.~Takaesu, ``{Revisiting Big-Bang
  Nucleosynthesis Constraints on Long-Lived Decaying Particles},''
  \href{http://dx.doi.org/10.1103/PhysRevD.97.023502}{{\em Phys. Rev. D}
  {\bfseries 97} no.~2, (2018) 023502},
  \href{http://arxiv.org/abs/1709.01211}{{\ttfamily arXiv:1709.01211
  [hep-ph]}}.

\bibitem{Tanabashi:2018oca}
{\bfseries Particle Data Group} Collaboration, M.~Tanabashi {\em et~al.},
  ``{Review of Particle Physics},''
  \href{http://dx.doi.org/10.1103/PhysRevD.98.030001}{{\em Phys. Rev. D}
  {\bfseries 98} no.~3, (2018) 030001}.

\bibitem{deSalas:2020pgw}
P.~de~Salas, D.~Forero, S.~Gariazzo, P.~Mart\'{i}nez-Mirav\'{e}, O.~Mena,
  C.~Ternes, M.~T\'{o}rtola, and J.~Valle, ``{2020 Global reassessment of the
  neutrino oscillation picture},''
  \href{http://arxiv.org/abs/2006.11237}{{\ttfamily arXiv:2006.11237
  [hep-ph]}}.

\end{thebibliography}\endgroup
\bibliographystyle{utphys}
\end{document}